\documentclass{IEEEoj}

\usepackage[utf8]{inputenc}
\usepackage[T1]{fontenc}
\usepackage{amsthm}
\usepackage{amssymb,amsmath,amsfonts}
\usepackage{algorithm,algorithmic}
\usepackage{cite}
\usepackage{epsfig}
\usepackage{float}
\usepackage{graphicx,graphics}
\usepackage{multicol}
\usepackage{lipsum}
\usepackage{subfigure}
\usepackage{url}
\usepackage{verbatim}
\usepackage{xspace}
\usepackage{setspace}
\usepackage{soul}
\usepackage{dblfloatfix}

\usepackage{textcomp}
\def\BibTeX{{\rm B\kern-.05em{\sc i\kern-.025em b}\kern-.08em
    T\kern-.1667em\lower.7ex\hbox{E}\kern-.125emX}}
\AtBeginDocument{\definecolor{ojcolor}{cmyk}{0.93,0.59,0.15,0.02}}
\def\OJlogo{\vspace{-4pt}\hskip-4pt\includegraphics[height=18pt]{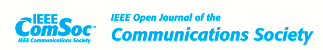}}

\newenvironment{list4}{
	\begin{list}{$\bullet$}{%
			\setlength{\itemsep}{0.05cm}
			\setlength{\labelsep}{0.2cm}
			\setlength{\labelwidth}{0.3cm}
			\setlength{\parsep}{0in} 
			\setlength{\parskip}{0in}
			\setlength{\topsep}{0in} 
			\setlength{\partopsep}{0in}
			\setlength{\leftmargin}{0.16in}}}
	{\end{list}}

\newenvironment{list4a}{
	\begin{list}{$\bullet$}{%
			\setlength{\itemsep}{0.05cm}
			\setlength{\labelsep}{0.2cm}
			\setlength{\labelwidth}{0.3cm}
			\setlength{\parsep}{0in} 
			\setlength{\parskip}{0in}
			\setlength{\topsep}{0in} 
			\setlength{\partopsep}{0in}
			\setlength{\leftmargin}{0.16in}}}
	{\end{list}}

\usepackage{url,changebar,bm,xspace,dsfont}
\let\mathbb=\mathds % I much prefer the dsfont over the bbfont
\DeclareMathOperator*{\argmax}{arg\,max}

\newtheorem{problem}{Problem}

\newtheorem{definition}{Definition}
\newtheoremstyle{remarkstyle}%
  {3pt}   % Space above
  {3pt}   % Space below
  {\itshape}  % Body font (italic)
  {}      % Indent amount
  {\bfseries} % Head font
  {.}     % Punctuation after head
  { }     % Space after head (INLINE!)
  {}      % Head spec

\theoremstyle{remarkstyle}
\newtheorem{remark}{Remark}

\makeatletter
\renewcommand{\subsubsection}{\@startsection
  {subsubsection} % name
  {3}             % level
  {0pt}           % indent
  {6pt}           % space before
  {2pt}           % space after
  {\normalfont\normalsize\itshape}} % style

\renewcommand{\subsection}{\@startsection
  {subsection} % name
  {2}             % level
  {-10pt}           % indent
  {8pt}   % space before
  {2pt}   % space after
  {\normalsize\bfseries}} % style

\makeatother
\begin{document}
\reviseddate{XX Month, XXXX}
\accepteddate{XX Month, XXXX}
\publisheddate{XX Month, XXXX}
\currentdate{11 January, 2024}
\doiinfo{OJCOMS.2024.011100}

\title{Toward Goal-Oriented Communication in Multi-Agent Systems: An overview}

\author{Themistoklis Charalambous\IEEEauthorrefmark{1,2,3} \IEEEmembership{(Senior~Member, IEEE)}, Nikolaos Pappas\IEEEauthorrefmark{4} \IEEEmembership{(Senior~Member, IEEE)}, Nikolaos Nomikos\IEEEauthorrefmark{5,6} \IEEEmembership{(Senior~Member, IEEE)}, Risto Wichman\IEEEauthorrefmark{2}
\IEEEmembership{(Senior Member, IEEE)}}
\affil{School of Engineering, University of Cyprus, Nicosia 1678, Cyprus}
\affil{School of Electrical Engineering, Aalto University, Espoo 02150, Finland}
\affil{FinEst Centre for Smart Cities, Ehitajate tee 5, 19086 Tallinn, Estonia}
\affil{Department of Computer and Information Science, Link\"oping University, Link\"oping SE-581 83, Sweden}
\affil{Department of Information and Communication Systems Engineering, University of the Aegean, Samos 83200, Greece}
\affil{Research and Development Department, Four Dot Infinity, 16777 Elliniko, Greece}
\corresp{CORRESPONDING AUTHOR: Themistoklis Charalambous (e-mail: charalambous.themistoklis@ucy.ac.cy).}
\authornote{The work of T. Charalambous was partly funded by MINERVA, a European Research Council (ERC) project funded under the European Union's Horizon 2022 research and innovation programme (Grant agreement No. 101044629).
Also, the works of T. Charalambous, N. Pappas, N. Nomikos, and R. Wichman were partly funded by 6G-LEADER, a project funded by the Smart Networks and Services Joint Undertaking (SNS JU) under the European Union’s Horizon Europe research and innovation programme (Grant agreement No. 101192080).
In addition, the work of R. Wichman was partially funded by Business Finland 6G-BRIDGE program. The work of N. Pappas has been supported in part by the Swedish Research Council (VR) and the ELLIIT.}
\markboth{Preparation of Papers for IEEE OPEN JOURNALS}{Charalambous \textit{et al.}}

% ========================================
%
% Abstract
%
% ========================================
\begin{abstract}
As multi-agent systems (MAS) become increasingly prevalent in autonomous systems, distributed control, and edge intelligence, efficient communication under resource constraints has emerged as a critical challenge. Traditional communication paradigms often emphasize message fidelity or bandwidth optimization, overlooking the task relevance of the exchanged information. In contrast, goal-oriented communication prioritizes the importance of information with respect to the agents’ shared objectives. This review provides a comprehensive survey of goal-oriented communication in MAS, bridging perspectives from information theory, communication theory, and machine learning. We examine foundational concepts alongside learning-based approaches and emergent protocols. Special attention is given to coordination under communication constraints, as well as applications in domains such as swarm robotics, federated learning, and edge computing. The paper concludes with a discussion of open challenges and future research directions at the intersection of communication theory, machine learning, and multi-agent decision making.
\end{abstract}


\maketitle
% ========================================
%
% Introduction
%
% ========================================
\section{Introduction}\label{sec:intro}

%\TC{Things to add:
%\begin{itemize}
%    \item \url{https://www.mdpi.com/1099-4300/25/5/728}
%    \item Learning Task-Oriented Channel Allocation for Multi-Agent Communication \url{https://ieeexplore.ieee.org/document/9847027}
%\end{itemize}
%}

\IEEEPARstart{I}{n recent years}, there has been a growing interest toward employing networks of autonomous agents to solve complex tasks collaboratively. These systems, known as multi-agent systems (MASs), achieve global objectives through localized interactions and decentralized decision-making among individual agents~\cite{8352646,SYS-019,e24121734}. Modern MASs are increasingly deployed in dynamic, distributed, and communication-constrained environments. Applications such as autonomous vehicle coordination~\cite{Hult2016,machines2017,Li_Hoang_Lin_Vu_Koenig_2023}, distributed sensing and estimation~\cite{Bechlioulis2020,Reza2022,Fioravanti2024}, and swarm robotics~\cite{mostofi2021,Pollyanna2021,e25111536,appliedmath2024} demand not only local intelligence but also effective communication among agents. 

Communication plays a critical role in enabling coordination and information sharing across agents, particularly when local observations are limited or partial. However, several fundamental challenges hinder effective communication in MASs. The major challenges are listed below. 
\begin{list4}
\item \emph{Bandwidth limitations} restrict how much information can be exchanged in real-time, especially over wireless links. Classical communication approaches in MAS often rely on predefined protocols or fully connected message passing structures that assume unrestricted communication. These models struggle in resource-constrained settings. Moreover, communication protocols in MAS have traditionally focused on information fidelity or throughput. However, in many real-world scenarios, these metrics fall short of what is truly needed: relevance to the task at hand, i.e., they rarely consider the \emph{value} or \emph{necessity} of a specific piece of information to the global task treating all information equally, leading to inefficient use of communication bandwidth. Agents must therefore prioritize what to communicate, potentially compressing or omitting less relevant data. 
\item The \emph{distributed architecture} of MAS means that there is often no centralized controller with global knowledge. Each agent must make decisions based on partial, noisy, or outdated information. Communication becomes a tool not only for sharing state but for aligning beliefs and coordinating actions under uncertainty.
\item \emph{Partial observability} further complicates message design. When agents have incomplete or private observations of the environment, they must infer what others know, anticipate their actions, and tailor messages accordingly. This creates a form of nested reasoning that is both computationally and communicatively demanding.
\item Another limitation lies in the \emph{separation of communication and decision-making modules}. In many standard architectures, agents first communicate and then make decisions, regardless of whether the exchanged information is critical for the decision-making process. Such decoupling fails to exploit the potential of jointly optimizing messaging and action policies in service of a common task. 
\item Finally, with increasing \emph{heterogeneity of agents} (differing in sensors, objectives, or capabilities), fixed communication topologies or uniform message policies can result in redundant information sharing, overload, or irrelevance, thus deteriorating coordination. Such heterogeneity necessitates flexible, adaptive protocols rather than static message formats. 
\end{list4}
These constraints motivate a rethinking of communication strategies: not simply how to convey information accurately, but what is worth communicating, given the task and context.

The paradigm of \emph{goal-oriented communication} (often referred to as \emph{task-driven messaging}\footnote{Task-driven messaging refers to the design and use of communication protocols where information is exchanged solely based on its utility toward accomplishing a specific goal. Unlike traditional paradigms that emphasize information fidelity or throughput, task-driven communication prioritizes relevance, efficiency, and the impact of coordination.}) shifts the focus from transmitting information for its own sake to transmitting only what is useful for achieving a specific task~\cite{GOcomms_first}. This shift is particularly vital in cases where bandwidth, energy, or latency constraints are present, as excessive or irrelevant communication can degrade system performance or even render coordination infeasible.
From this perspective, the insight is that the utility of a message should be assessed by its contribution to achieving a collective objective. Agents must therefore learn or infer what to communicate, when to communicate, and with whom to communicate, in a manner that maximizes task performance under real-world constraints. 
In practice, goal-oriented messaging may manifest as sparse message passing, event-triggered communication, or the emergence of protocol learning in multi-agent reinforcement learning settings. These approaches all share a common goal: aligning communication behavior with task-specific objectives and constraints.

% ========================================
% Scope and Contributions of the Review
% ========================================
\subsection{Relation between goal-oriented versus semantic communication}

Before delving more into goal-oriented communication for MASs, we should clarify that goal-oriented communication and semantic communication represent two distinct yet interconnected paradigms in next-generation wireless systems. Semantic communication primarily focuses on understanding and transmitting the semantics of information rather than raw bits. The framework extracts semantic content elements from raw data streams, employs causal reasoning and contextual awareness, and emphasizes the properties of minimalism, generalizability, and efficiency in its representations.

In contrast, goal-oriented communication emphasizes optimizing communication strategies to achieve specific tasks or objectives, particularly in MASs. This paradigm tailors communication protocols to maximize task performance and coordination efficiency among distributed agents. Goal-oriented communication explicitly ties all messaging decisions to task success metrics, carefully considers resource constraints and utility functions, and emphasizes coordination and cooperation among multiple agents working toward shared or complementary objectives. The success of such systems is measured not by the fidelity of information transmission, but by how effectively communication serves the overall system goals and task completion.

However, these approaches are not mutually exclusive but rather complementary frameworks that can work synergistically. 
The goal of the considered communication problem provides the necessary context for the appropriate definition of the semantics of information. This becomes relevant since it can allow for efficient prioritization of information and efficient resource allocation leading to significantly reducing the wastage of resources~\cite{Commag21,IoTM24,OJCOMS24,2024:SurveysnTutorials,2024:IEEEWCmagazine}.
Goal-oriented communication is often considered the first major application of semantic communication, as goal-oriented systems frequently require semantic representations to achieve their objectives efficiently~\cite{2024:SurveysnTutorials}.

% ========================================
% Scope and Contributions of the Review
% ========================================
\subsection{Scope and contributions of the overview}

Communication is a fundamental enabler of coordination, perception, and control in MASs. From autonomous driving to robotic swarms, collaborative agents must frequently exchange information to form consistent world models, synchronize behaviors, and jointly complete tasks. Traditionally, communication frameworks have focused on reconstructing the transmitted signals or states with high fidelity, following principles laid out by classical Shannon theory. However, in many real-world tasks, perfect reconstruction is not necessary, what matters is whether the received message enables the receiver to act effectively toward its goal. This shift from \emph{transmission fidelity} to \emph{goal relevance} has sparked growing interest in goal-oriented communication; see, for example, \cite{Commag21, OJCOMS24, IoTM24} and references therein.

This overview offers a comprehensive and integrative overview of goal-oriented communication in MASs, with a focus on information-theoretic models and learning-based approaches. Specifically, we:
\begin{list4}
    \item Recap classical communication theory) and discuss its limitations for intelligent, goal-oriented communications (Section~\ref{sec:foundations}).

    \item Describe theoretical frameworks of wireless networked control systems (WNCSs) and discuss issues related to control systhesis and remote estimator design (Section~\ref{sec:foundations}).

    \item Present a systematic overview of key information-theoretic tools, such as the Information Bottleneck (IB), semantic rate-distortion (SRD) theory, and G theory, that quantify and guide purposeful communication. Also, we discuss how these tools can be conceptually integrated into a unified framework (Section~\ref{sec:GOIT}).

    \item Examine how goal-oriented principles reshape coordination and control in MASs when communication is costly or constrained. We build on models from wireless networked control (Section~\ref{sec:foundations}) and introduces priority metrics (such as Cost of Information Loss, Value of Information, and Age of Information), and analyzes the impact of shared and unreliable channels (Section~\ref{sec:WNCSs}).
        
    \item Cover approaches where communication strategies are learned, such as multi-agent reinforcement learning (MARL), sparsity/attention mechanisms, and semantic representation learning. Additionally, we aim at highlighting connections and synergies between learning, information theory, and semantic communication (Section~\ref{sec:learning_based}).

    \item Illustrate the practical significance of goal-oriented communication through selected applications: cooperative autonomous vehicles, distributed SLAM, federated learning, and edge intelligence. Each example demonstrates unique challenges and solutions enabled by goal-oriented design (Section~\ref{sec:applications}).
    
    \item Identify open research directions, including hierarchical representations, causal semantic compression, and goal-oriented coordination under uncertainty (Section~\ref{sec:challenges}).

    \item Summarize the key insights, reiterate the contribution of the overview, and encourage further research at the intersection of communication theory, semantics, and intelligent systems (Section~\ref{sec:conclusions}).
\end{list4}

We aim to bridge the conceptual foundations of information theory with emerging practices in distributed learning, control, and perception, establishing a common language for researchers and practitioners working on semantic and goal-oriented communication. By placing goal-oriented messaging at the intersection of theory and application, we seek to accelerate the development of scalable, interpretable, and context-aware communication strategies, thus enabling agents to exchange information not merely to share data, but to advance shared objectives. This overview thus provides both conceptual clarity and practical guidance for designing intelligent, purpose-driven communication protocols in multi-agent environments.

% ========================================
% Prior works and relevance to this overview
% ========================================
\subsection{Prior works}

While several contributions have advanced the field of goal-oriented communications from multiple angles, ranging from distributed simultaneous localization and mapping (SLAM) and decentralized control to semantic coding and IB theory, research remains fragmented across disciplines. Recently, several survey and tutorial papers have emerged addressing goal-oriented communication, reflecting the growing interest in communication efficiency for intelligent systems; see, for example, \cite{Commag21, IoTM24, e26020102,2024:SurveysnTutorials,2024:IEEEWCmagazine,OJCOMS24,2024:Oulu} and references therein. 

The survey and tutorial paper~\cite{2024:SurveysnTutorials} presents a visionary, tutorial-style framework for evolving from data-centric to knowledge- and reasoning-driven semantic communication systems and provides a comprehensive roadmap for scalable, reasoning-driven, semantic communication networks in large-scale cellular environments. Finally, they discuss what challenges still lie ahead for plugging these ideas into large-scale 6G networks. Overall, the paper provides a foundational, holistic vision of semantic communication for 6G, for the fundamental transformation of communication into a knowledge and reasoning-driven paradigm, while our work provides an in-depth exploration of a specific, critical application of this vision, namely goal-oriented communication within the complex domain of MASs.

Kountouris and Pappas in~\cite{Commag21} introduce a multiscale semantic definition spanning source-level event importance, link-level attributes (freshness, timeliness), and system-level impact on application objectives, enabling smart devices to perform active, semantics-aware sampling that reduces redundancy and resource waste. Through a real-time remote actuation case study, they demonstrate that semantics-empowered policies achieve fewer uninformative samples, lower reconstruction error, and reduced costly actuation mistakes compared to classical approaches. The work identifies key open challenges including concrete semantic metrics, goal-oriented resource orchestration, and scalable optimization frameworks, positioning semantic communication as the foundation for next-generation networked intelligence.

The paper by Zhou \emph{et al.}~\cite{IoTM24} presents a comprehensive goal-oriented semantic communication framework specifically designed for 6G networks that goes beyond traditional semantic approaches by incorporating both semantic-level information extraction and effectiveness-aware performance metrics tailored to diverse application requirements. The framework’s practical effectiveness is demonstrated through a unmanned aerial vehicle (UAV) control case study, where {Age of Information (AoI)} and {Value of Information (VoI)} metrics enable intelligent filtering of redundant command signals, achieving over 60\% reduction in data transmission while maintaining control effectiveness. This paper is systems-oriented and implementation-focused, providing operational solutions for specific use cases. Our overview is more foundational and mathematical, aiming to address the fundamental theoretical gaps. Hence, they are essentially complementary perspectives on the same communication paradigm.

Trevlakis \emph{et al.}~\cite{OJCOMS24} surveys and categorizes a wide range of recent works on semantic and goal-oriented communication, including fundamental theories, architectural models, enabling techniques, deep learning, knowledge graphs, and practical implementations. Additionally, it introduces a semantic networking architecture that evolves traditional point-to-point communication into a distributed, multi-user, edge-to-cloud paradigm. This architecture enables the extraction and filtering of goal-specific semantic information at the source and post-processing/decoding at the destination. While it includes goal-oriented communications within its scope, it does not provide the in-depth survey of MASs' communication challenges and approaches offered herein.

The paper by Li \emph{et al.}~\cite{2024:IEEEWCmagazine} introduces the Goal-Oriented Tensor (GoT) as a unified metric for semantic communication in 6G networks, capturing both the semantic significance of information and its impact on task effectiveness. The authors propose a holistic architecture where semantic quantization, sparse semantics-aware sampling, goal-oriented channel coding, and intelligent decision making collaborate to minimize real-world costs. They validate GoT through a fire monitoring and rescue case study, demonstrating its superiority in selective sampling and cost reduction. While this paper focuses on a practical metric unification and system design, providing an applied, tensor-based metric and architecture tailored for real-world 6G scenarios, our overview lays the mathematical groundwork for goal-oriented communications, emphasizing the theoretical foundations.

Xin \emph{et al.}~\cite{e26020102} explores semantic information theory, covering major topics and discusses how these concepts extend and differ from classical Shannon information theory, emphasizing the urgent need for a unified theoretical framework for semantic communications. However, the paper focuses solely on semantic information theory and reviews several mathematical theories and tools and evaluates their applicability in the context of semantic communication.

The paper by Getu \emph{et al.}~\cite{2024:Oulu} provides a comprehensive review of existing metrics for evaluating both semantic communication and specifically goal-oriented communication across wireless, optical, and quantum domains. It addresses the need for unified performance metrics and covers topics like semantic metrics for text, speech, image, video, resource allocation, and system throughput. The survey highlights research directions for 6G and beyond. Nevertheless, MAS scenarios require an additional higher-level integration step to couple such metrics with decision impact and coordination efficiency in decentralized settings.

% ========================================
% How they differ to this overview
% ========================================
\subsection{How this work differs}

Despite the breadth of recent work, the literature on semantic and goal-oriented communication remains fragmented across multiple research communities, including information theory, control and robotics, distributed AI, and networking, each advancing its own set of models, metrics, and application scenarios. While these studies have produced important theoretical insights and compelling application-specific demonstrations, they often lack a unifying framework that connects foundational principles to the specific coordination and decision-making challenges encountered in MASs. In particular, few works integrate communication efficiency, semantic relevance, and task performance within a single end-to-end perspective, or address the nuanced requirements of safety-critical and human-centric environments. This gap motivates the present overview.

While recent surveys and tutorials on semantic and goal-oriented communication have introduced valuable concepts and demonstrated isolated case studies, they have not provided a \emph{comprehensive, MAS-centric synthesis} that unifies theoretical foundations with practical deployment considerations. Existing works often examine semantic signaling in isolation from resource constraints, or focus on AI-driven link-level solutions without establishing a rigorous connection to end-to-end multi-agent objectives. Moreover, safety-critical and human-centric dimensions are frequently treated as peripheral topics rather than core design drivers. In contrast, this paper delivers a unified, conceptually rigorous framework for goal-oriented communication in MASs, distinguished by the following key contributions:
\begin{list4a}
  \item It integrates information-theoretic principles with distributed control/estimation and robotics models, as well as emerging learning-based architectures, aiming at creating a shared formalism applicable across heterogeneous MASs domains.
  \item It emphasizes messaging whose primary purpose is to advance joint agent objectives, ensuring \emph{efficiency} (minimizing communication overhead) and \emph{effectiveness} (maximizing task performance) under real-world constraints.
  \item It bridges theory and practice by aligning mathematical principles with concrete MASs use cases, enabling the design of flexible, goal-oriented protocols that emphasize actionable information instead of simply preserving raw data, exploring how these models can be combined in MASs with real-world constraints (e.g., limited observability, decentralized optimization).
\end{list4a}
Rather than merely cataloging existing contributions, our work distills their essential mechanisms and organizes them into a cohesive framework tailored for MASs applications, guiding the design of communications that are both resource-aware and tightly coupled to agent goals. This transition is critical for the next decade’s communication and networking needs, especially given the demands of emerging applications (e.g., cooperative autonomous systems and distributed AI/ML).

% ========================================
%
% Foundations of Goal-Oriented Communication
%
% ========================================
\section{Foundations of Goal-Oriented Communication}
\label{sec:foundations}

% ========================================
% Classical Communication Theory
% ========================================
\subsection{Classical Communication Theory}

Shannon’s classical information theory~\cite{shannon1948mathematical,cover2006elements} provides the foundational framework for modeling and analyzing communication systems. In its canonical form, communication is viewed as the process of mapping symbols from a source $X \in \mathcal{X}$ through an encoder, transmitting them over a possibly noisy channel characterized by a conditional distribution $p(y|x)= \mathbb{P}(Y = y|X=x)$, and producing an output $Y \in \mathcal{Y}$ at the receiver. The primary objective of such a system is to ensure that the receiver obtains sufficient information to accurately infer or reconstruct the transmitted data, subject to channel limitations. The theory abstracts away semantic communication, focusing instead on the \emph{statistical} relationship between $X$ and $Y$.

\subsubsection{Entropy}
Let $X$ be a discrete random variable with alphabet $\mathcal{X}$ and probability mass function  $p(x) \equiv p_X(x) = \mathbb{P}(X = x)$. The \emph{Shannon entropy} of $X$ is defined as:
\begin{equation}
H(X) = -\sum_{x \in \mathcal{X}} p(x) \log_2 p(x),
\end{equation}
where the logarithm is taken base $2$ and the entropy is measured in bits. The entropy quantifies the average uncertainty or information content in the random variable $X$. It reaches its maximum when $X$ is uniformly distributed over $\mathcal{X}$, and equals zero when $X$ is deterministic.

\subsubsection{Mutual Information}

A central quantity in Shannon's theory is the \emph{mutual information} \( I(X; Y) \), which quantifies the amount of information that the output \( Y \) conveys about the input \( X \). It is defined as the reduction in uncertainty about \( X \) after observing \( Y \), and can be expressed in terms of the entropy function \( H(\cdot) \) as:
\begin{equation}
I(X; Y) = H(X) - H(X|Y),
\end{equation}
where \( H(X) \) is the entropy of the source and \( H(X|Y) \) is the conditional entropy of \( X \) given \( Y \), given by: 
\begin{equation}
H(X | Y) = -\sum_{y \in \mathcal{Y}} p(y) \sum_{x \in \mathcal{X}} p(x|y) \log_2 p(x | y).
\end{equation}
Equivalently, it can be written using the joint distribution as:
\begin{equation}
H(X | Y) = -\sum_{x \in \mathcal{X}} \sum_{y \in \mathcal{Y}} p(x, y) \log_2 p(x | y).
\end{equation}
The conditional entropy quantifies the average uncertainty remaining about $X$ after observing $Y$. Therefore, the mutual information can also be written as:
\begin{equation}
I(X; Y) = \sum_{x \in \mathcal{X}} \sum_{y \in \mathcal{Y}} p(x, y) \log \frac{p(x, y)}{p(x) p(y)},
\end{equation}
where \( p(x, y) \) is the joint probability distribution of \( X \) and \( Y \), and \( p(x) \), \( p(y) \) are the marginal distributions. This measure captures the statistical dependence between the transmitted and received signals. For discrete random variables $X$, $Y$, and $Z$ with joint probability mass function $p(x,y,z)$, the conditional mutual information of $X$ and $Y$ given $Z$ is defined
by 
\begin{align}\label{eq:conditionalMI}
I(X;Y | Z) &\triangleq H(X| Z) - H(X | Y,Z) \\
&= \sum_{x \in \mathcal{X}} \sum_{y \in \mathcal{Y}} \sum_{z \in \mathcal{Z}}p(x, y, z) \log \frac{p(x, y|z)}{p(x|z) p(y|z)}. \nonumber    
\end{align}
Making use of~\eqref{eq:conditionalMI}, for multiple random variables, the following result holds:
\begin{align}
I(X_1,X_2,\ldots,X_n; Y) \!=\! \sum_{i=1}^n I(X_i; Y | X_{i-1}, \ldots, X_1 ).    
\end{align}
Shannon also introduced the notion of \emph{channel capacity} \( C \), defined as the maximum mutual information achievable over the channel:
\begin{equation}
C = \max_{p(x)} I(X; Y),
\end{equation}
which represents the highest rate at which information can be reliably transmitted across the channel. Together, these concepts form the basis for understanding the limits of communication under noise and bandwidth constraints.

When exact reconstruction of a source is not necessary or feasible (such as in image compression, video streaming, or real-time sensing), \emph{rate-distortion theory} provides a framework to study lossy source compression~\cite{Berger:1971BOOK}. Instead of requiring perfect fidelity, the goal is to find an optimal trade-off between the number of bits used for encoding and the quality of reconstruction, as measured by a distortion metric. Distortion is measured in terms of signal fidelity; e.g., mean squared error or Hamming distance.

Consider a memoryless source $X \in \mathcal{X}$ with known distribution $p(x)$, and a reproduction alphabet $\hat{\mathcal{X}}$. Let $d(x, \hat{x}) \geq 0$ be a distortion function that quantifies how ``close'' a reconstruction $\hat{x} \in \hat{\mathcal{X}}$ is to the true source value $x \in \mathcal{X}$. The encoder maps each source symbol to a codeword, and the decoder reconstructs a symbol $ \hat{x}$. The pair \( (X, \hat{X}) \) then follows a joint distribution \( p(x, \hat{x}) = p(x)p(\hat{x}|x) \), where \( p(\hat{x}|x) \) defines a (possibly stochastic) reconstruction rule. The expected distortion is given by:
\begin{equation}
\mathbb{E}[d(X, \hat{X})] = \sum_{x \in \mathcal{X}} \sum_{\hat{x} \in \hat{\mathcal{X}}} p(x) p(\hat{x}|x) d(x, \hat{x}).
\end{equation}

\subsubsection{Rate distortion theory}

Rate-distortion theory, introduced by Shannon in 1959 \cite{shannon1959coding}, provides the fundamental theoretical framework for understanding the trade-off between compression efficiency and reconstruction fidelity in lossy data compression. This theory addresses the question: \emph{What is the minimum amount of information (rate) required to represent a source such that it can be reconstructed with a specified level of distortion?} Unlike lossless compression, which seeks perfect reconstruction, rate-distortion theory acknowledges that in many practical scenarios, perfect fidelity is neither necessary nor achievable under bandwidth constraints, making controlled distortion an acceptable compromise.

The cornerstone of rate-distortion theory is the \emph{rate-distortion function}, which characterizes the minimum achievable rate $R$ (measured in bits per symbol) required to transmit a source under a distortion constraint $D$. Formally, it is defined as:
\begin{equation}
R(D) = \min_{p(\hat{x}|x): \mathbb{E}[d(X, \hat{X})] \leq D} I(X; \hat{X}),
\end{equation}
where $I(X; \hat{X})$ represents the mutual information between the source $X$ and its reconstruction $\hat{X}$, and $d(X, \hat{X})$ is a distortion measure that quantifies the difference between the original and reconstructed symbols. The minimization is performed over all conditional probability distributions $p(\hat{x}|x)$ (representing different encoding strategies) that satisfy the average distortion constraint $\mathbb{E}[d(X, \hat{X})] \leq D$.
The rate-distortion function $R(D)$ is a convex, non-increasing function of $D$ that describes the fundamental trade-off between compression rate and reconstruction quality.

The choice of distortion measure $d(X, \hat{X})$ significantly impacts the rate-distortion function and reflects the application-specific notion of reconstruction quality. Common distortion measures include:
\begin{list4}
    \item \emph{Squared Error Distortion}: $d(x, \hat{x}) = (x - \hat{x})^2$, widely used for continuous sources;
    \item \emph{Hamming Distortion}: $d(x, \hat{x}) = \mathbf{1}_{x \neq \hat{x}}$, appropriate for discrete sources;
    \item \emph{Absolute Error}: $d(x, \hat{x}) = |x - \hat{x}|$, used in robust applications.
\end{list4}
For specific source distributions and distortion measures, closed-form expressions for $R(D)$ can be derived. A fundamental result is the rate-distortion function for a Gaussian source with squared error distortion:
\begin{equation}
R(D) = \begin{cases}
\frac{1}{2} \log_2 \frac{\sigma^2}{D}, & \text{if } D \leq \sigma^2 \\
0, & \text{if } D > \sigma^2
\end{cases}
\end{equation}
where $\sigma^2$ is the source variance. This result demonstrates the logarithmic relationship between rate and distortion for Gaussian sources, a relationship that appears frequently in practical compression systems.

The rate-distortion function provides both a lower bound on compression performance and a roadmap for optimal encoder design. The \emph{rate-distortion theorem} establishes that for any rate $R > R(D)$, there exist encoding and decoding schemes that achieve average distortion arbitrarily close to $D$ with sufficiently long block lengths. Conversely, no encoding scheme can achieve average distortion $D$ with rate less than $R(D)$. This operational interpretation is realized through the concept of \emph{typical sequences} and \emph{jointly typical encoding/decoding}. The optimal encoder partitions the source space into regions, assigning codewords to representative points (centroids) within each region. The decoder maps received codewords back to these centroids, achieving the minimum distortion for a given rate.
Rate-distortion theory underlies the design of virtually all modern compression algorithms. In practice, this theory underlies the design of compression algorithms (e.g., JPEG, MP3, H.264) by quantifying the best possible trade-off between rate and quality. For a comprehensive treatment of rate–distortion theory, the reader is also referred to~\cite{Berger:1971BOOK}.

\subsubsection{Limitations of classical communication theory}

While classical information theory offers a mathematically elegant and universally applicable framework, its standard distortion measures are fundamentally \emph{task-agnostic}: i.e., they treat all deviations equally, regardless of their semantic or functional significance. For example, an agent performing classification may tolerate significant pixel-level noise, provided that the semantic content (e.g., object category) is preserved. This mismatch between \emph{fidelity} (accuracy of signal reconstruction) and \emph{utility} (contribution to the intended task) becomes critical in modern multi-agent and cyber–physical systems, where communication is increasingly embedded within decision-making loops. In such contexts, transmitting every detail of the source signal is not only unnecessary but often leads to overall performance deterioration. These observations motivate a paradigm shift toward \emph{goal-oriented communication}, in which compression, coding, and transmission are explicitly optimized to preserve information that is \emph{relevant} for achieving the system’s objective. By aligning communication metrics with task-specific utility, such approaches can dramatically improve efficiency while maintaining or even enhancing decision quality. We explore this extension in Section~\ref{sec:GOIT}.

% ========================================
% Classical Control and Estimation Theory
% ========================================
\subsection{Wireless networked control systems (WNCSs)}\label{subsec:WNCS}

A WNCS is a feedback control architecture where sensors, controllers, and actuators (may) communicate over wireless links. In such systems, deciding when to transmit sensor data is critical due to shared channel access, power limitations, and real-time performance demands. This decision is typically governed by a transmission policy that initiates communication when new information significantly improves estimation or control and suppresses communication when data is redundant or contributes little to task performance. Such policies can be event-triggered, self-triggered, or based on value-aware decision metrics, and are designed to minimize communication load while maintaining control quality.

\begin{figure} [h]
	\centering
	\includegraphics[width=0.99\columnwidth]{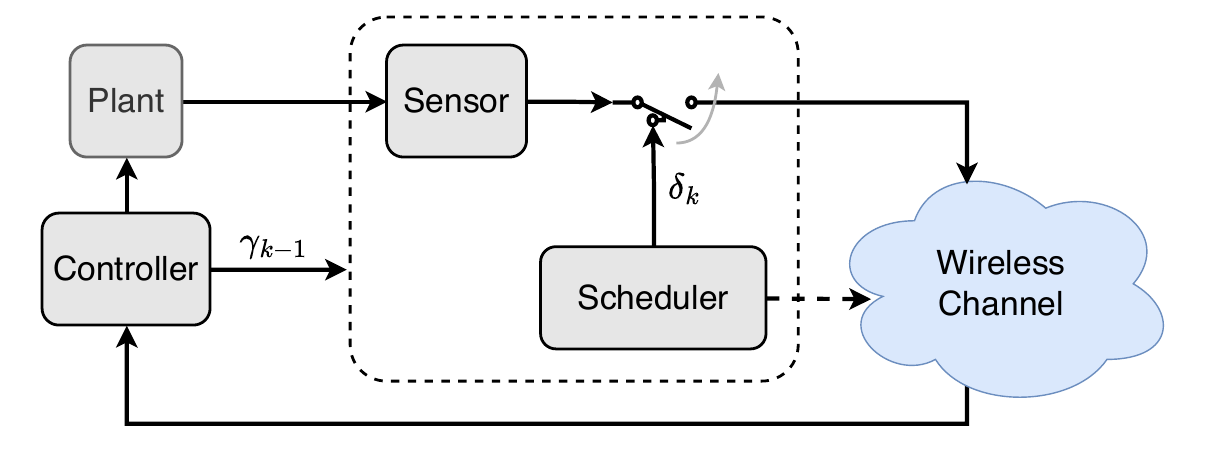}
	\caption{A system consisting of a plant, a sensor and a controller closes the loop through a network. The channel access decision is determined by the (goal-oriented) scheduler block.}
	\label{fig:NCS}
\end{figure}

Consider the simplified model of a system that closes its loop over a network as depicted in Figure~\ref{fig:NCS}. 

\noindent Without getting into the specifics yet, we assume that the transmission policy is captured by the scheduler block. This allows us to, without loss of generality, represent the fundamentals with simplified notation. The plant dynamics are described by a linear time-invariant (LTI) discrete-time system:
\begin{align}\label{eq:process}
	x_{k+1} = Ax_k + Bu_k + w_k,
\end{align}
where \( x_k \in \mathbb{R}^n \) and \( u_k \in \mathbb{R}^m \) represent the state and control input at time \( k \), respectively. The matrices \( A \in \mathbb{R}^{n \times n} \) and \( B \in \mathbb{R}^{n \times m} \) are time-invariant and of appropriate dimensions. {The process noise \( w_k \in \mathbb{R}^n \) is a zero-mean Gaussian random variable with covariance \( W \)}, and the initial state \( x_0 \) is Gaussian with mean \( \bar{x}_0 \) and covariance \( X_0 \). The sensor \( \mathcal{S} \) observes the system through a linear measurement model:
\begin{align} \label{eq:output}
	y_k = Cx_k + v_k,
\end{align}
where \( y_k \in \mathbb{R}^p \) is the measurement, \( C \in \mathbb{R}^{p \times n} \) is the output matrix, and {\( v_k \in \mathbb{R}^p \) is a zero-mean Gaussian noise term with covariance \( V \). The random variables \( x_0 \), \( w_k \), and \( v_k \) are assumed to be mutually independent.} Upon obtaining a measurement, the sensor constructs a data packet. Whether this packet is transmitted to the controller \( \mathcal{C} \) is determined by a binary channel access decision variable \( \delta_k \in \{0, 1\} \), defined as:
\begin{align} \label{eq:deltach1}
	\delta_k = \begin{cases}
		1, & \text{if the sensor transmits at time } k, \\
		0, & \text{otherwise}.
	\end{cases}
\end{align}
In case of unreliable communication channels packet dropouts are possible. We consider the case for which instantaneous packet ack\-nowledge\-ments\-/neg\-ative-\-acknow\-ledge\-ments (ACK/NACKs) are available through an error-free feedback channel and define
\begin{align}
	\gamma_{k}\!=\! \begin{cases}
		1, & \text{if $\delta_{k}\!=\!1$ and packet successfully received},\\
		0, & \text{otherwise}.
	\end{cases}
\end{align}

First, we focus on a setting where the network is modeled as a single ideal communication channel. As a result, when channel access is granted (i.e., \( \delta_k = 1 \)), the corresponding data packet is successfully received by the controller without loss or interference.
In addition, we define 
\begin{align} \label{eq:t}
	t_k=\min\{\kappa\geq0:\delta_{{k-\kappa}}=1\}
\end{align}
as the time elapsed since the most recent successful packet reception at the controller. 
Based on these definitions, the information set available at the sensor at time $k$ is denoted by $\mathcal{I}_k^s = \{\bm{y}^k, \bm{u}^{k-1}, \bm{\delta}^{k-1} \}$, where boldface notation (e.g., $\bm{\delta}^{k-1} = \{\delta_0, \delta_1, \dots, \delta_{k-1}\}$) is used to represent sequences of variables up to time $k-1$. The controller's information set is given by $\mathcal{I}_k^c = \{\bm{y}^{k - t_k}, \bm{u}^{k-1}, \bm{\delta}^k \}$, reflecting the sensing and estimation architecture, which will be elaborated in the subsequent section. Additionally, the information set available at the scheduler is denoted by $\mathcal{I}_k^\tau$. While its structure will be described in detail later, for now we assume that it satisfies $\mathcal{I}_{k-1}^c \subseteq \mathcal{I}_k^\tau$.

Let the \emph{channel access policy} $\bm{d}=\{d_0,d_1,\dots,d_{K-1}\}$ and \emph{control policy} $\bm{f}=\{f_0,f_1,\dots,f_{K-1}\}$ denote the admissible policies for the finite horizon $K$ with
\begin{align} 
	\delta_k&=d_k(\mathcal{I}_k^\tau),\\
	u_k&=f_k(\mathcal{I}_k^c),
\end{align}
where the mappings $d_k$ and $f_k$ are measurable mappings of their available information. 

The objective is to find the optimal channel access and control policy that minimize the standard quadratic cost over the infinite horizon, i.e.,
\begin{align} \label{eq:optdf}
	\min_{\bm{d},\bm{f}} J_{0:\infty},
\end{align}
where
\begin{align}  \label{eq:Jinfty}
	J_{0:\infty} = \lim\limits_{K \to \infty}\frac{1}{K}\mathbb{E}\left\{\sum_{k=0}^{K-1}\left(x^T_{k}Qx_{k}+u^T_{k}R u_{k}\right)\right\},
\end{align}
is the quadratic cost penalizing the deviation of the states and inputs from their reference values with respective positive semi-definite and positive definite weighting matrices $Q$ and $R$.

The simplistic \emph{always transmit} policy, where the sensor sends its measurement at every time step regardless of content or context, is suboptimal in many practical scenarios. While this strategy ensures that the controller receives the freshest possible data, it completely disregards communication costs, bandwidth constraints, and the potential redundancy of the transmitted information. In resource-limited systems such as {WNCSs}, constant transmission leads to channel congestion, increased energy consumption, and inefficient use of communication resources.

Moreover, in goal-oriented settings, not all information updates contribute equally to task performance. For example, when the system state evolves slowly or remains within predictable bounds, transmitting every measurement offers little improvement in estimation or control accuracy. Instead, selective transmission policies (e.g., based on event-triggering, prediction error, or task-relevance metrics) can reduce communication load while maintaining or even improving overall system performance \cite{SECON23Polina, CSCN24, TNET25Polina}.

% ========================================
% Separation of channel access scheme and controller design
% ========================================
\subsubsection{Separation of channel access scheme and controller design} 
\label{sec:CE}

Despite the complexities involved in the joint optimization of channel access and control policies, under certain conditions, their design can be separated in the sense that \eqref{eq:optdf} can be rewritten as
\begin{align} \label{eq:eqoptdf}
\min_{\bm{d}}\left\{\min_{\bm{f}} J_{0:\infty}\right\}.
\end{align}
In this regard, we first review the conditions on the channel access policy which ensure that the \emph{certainty equivalence principle} holds. As a consequence, the classical results in stochastic control theory can be adopted to design the optimal controller. We then present the detailed model of the sensing and estimation architecture and finally derive an equivalent problem formulation for designing the channel access policy.

The deterministic optimal controller refers to the control policy that minimizes the cost function in~\eqref{eq:Jinfty} under the assumption of perfect information, i.e., full knowledge of the exact realizations of all involved random variables. Let \( \bar{u}_k \) denote the optimal control input at time \( k \) in this idealized setting. Importantly, under the \emph{certainty equivalence principle}, there is no loss in optimality when the controller instead applies the conditional expectation \( \mathbb{E}\{\bar{u}_k | \mathcal{I}_k^c\} \), computed using the available (partial and noisy) information set \( \mathcal{I}_k^c \). This principle, as formalized in~\cite{BarShalom:1974}, ensures that the separation between estimation and control remains optimal in linear-quadratic-Gaussian (LQG) settings.
\begin{definition}[Certainty equivalence principle]
The certainty equivalence principle holds if the closed-loop optimal controller has the same form as the certainty equivalent controller.	
\end{definition}
\begin{definition}[Certainty equivalent controller]
A certainty equivalent controller uses the deterministic optimal controller, with the state $x_{i,k}$ replaced by the estimate $\hat{x}_{k|k} = \mathbb{E}\{x_{k}|\mathcal{I}_{k}^c\}$.	
\end{definition}

% ========================================
% Controller synthesis
% ========================================
\subsubsection{Controller synthesis}

The case under consideration in non-classical and, as a result, the applicability of the certainty equivalence principle is no longer guaranteed by default. However, it has been shown that certainty equivalence continues to hold under specific structural conditions; that is, when the channel access decisions are independent of the control actions. This requirement is satisfied if the scheduling policy is solely a function of the system's primitive random variables. Formally, the channel access decision at time \( k \) must take the form \( \delta_k = d_k(x_0, \bm{w}^{k-1}, \bm{v}^{k-1}) \) for all \( k \in \{0, 1, \dots, K-1\} \)~\cite{Ramesh:2014Thesis}. Under this assumption, the control and estimation problems decouple, allowing for the use of a certainty equivalent controller. By employing a dynamic programming framework, one can derive the optimal control law that minimizes the expected cost given partial information. The resulting certainty equivalent controller corresponds to the standard LQG form, as detailed in~\cite[Chapter~8]{Astrom:2006}, given by
\begin{align} \label{eq:u}
	u_{k} = L_{\infty}\hat{x}_{k|k},
\end{align}
where the optimal feedback gain $L_{\infty}$ is given by
\begin{align} \label{eq:L}
	L_{\infty} = -(B^T\Pi_{\infty} B + R)^{-1}B^T\Pi_{\infty} A,
\end{align}
and $\Pi_{\infty}$ is the positive semi-definite solution of the following discrete-time algebraic Riccati equation (DARE)
\begin{align} \label{eq:DARE}
	\Pi_{\infty}=A^T \Pi_{\infty} A + Q - L_{\infty}^T(B^T \Pi_{\infty} B+R)L_{\infty}.
\end{align}
We assume that the pair $(A,B)$ and $(A,Q^{1/2})$ are controllable and observable, respectively, to guarantee that $\Pi_{\infty}$ always exists.

It is important to emphasize that, while restricting the channel access policy to exclude dependence on past control inputs guarantees the validity of the certainty equivalence principle, this restriction does not, in general, yield the optimal solution. Nevertheless, such a constraint simplifies the optimization problem by making the formulation in~\eqref{eq:eqoptdf} equivalent to the decoupled control problem in~\eqref{eq:optdf}. This significantly facilitates the design of the channel access policy, as the optimal control law is known \emph{a priori} and corresponds to the classical certainty equivalent controller given by~\eqref{eq:u}.
However, allowing the channel access decisions to depend on past control inputs can, in general, improve system performance. In such cases, the certainty equivalent controller may no longer be optimal, since the control actions influence both the system state and the information available to the controller. As a result, the joint optimization problem described in~\eqref{eq:optdf} must be solved in its full generality~\cite{Ramesh:2014Thesis}.

% ========================================
% Estimator design
% ========================================
\subsubsection{Estimator design}

 The control input in~\eqref{eq:u} is computed using the state estimate \( \hat{x}_{k|k} = \mathbb{E}\{x_k | \mathcal{I}_k^c\} \), which is based on the information available to the controller at time \( k \). To formalize the structure of \( \mathcal{I}_k^c \) and the computation of \( \hat{x}_{k|k} \), we begin by describing the sensor's operation and the contents of the data packet it generates.

We assume that the systems is equipped with smart sensors, meaning that each sensor possesses local computational capabilities. This enables the sensors to autonomously assess the necessity of transmission and actively participate in channel access decisions. More importantly, the use of smart sensors, as opposed to primitive, computation-limited ones, allows for local preprocessing of raw measurements prior to transmission, which has been shown to significantly enhance estimation quality at the receiver~\cite{Schenato:2007}.

At the beginning of transmission frame \( k \), the information available to the sensor is given by \( \mathcal{I}_k^s = \{\bm{y}^k, \bm{u}^{k-1}, \bm{\delta}^{k-1} \} \). Notably, knowledge of past channel access decisions \( \bm{\delta}^{k-1} \) allows the sensor to infer the past control inputs \( \bm{u}^{k-1} \) without requiring explicit communication from the controller.

We define the \emph{a priori} and \emph{a posteriori} state estimates at the sensor as
\begin{align*}
	\hat{x}_{k|k-1}^s &\triangleq \mathbb{E}\{x_k | \mathcal{I}_{k-1}^s\},\\
	\hat{x}_{k|k}^s &\triangleq \mathbb{E}\{x_k | \mathcal{I}_k^s\},
\end{align*}
with their corresponding estimation error covariances given by
\begin{align*}
	P_{k|k-1}^s &\triangleq \mathbb{E}\big\{(x_k - \hat{x}_{k|k-1}^s)(x_k - \hat{x}_{k|k-1}^s)^T | \mathcal{I}_{k-1}^s\big\},\\
	P_{k|k}^s &\triangleq \mathbb{E}\big\{(x_k - \hat{x}_{k|k}^s)(x_k - \hat{x}_{k|k}^s)^T | \mathcal{I}_k^s\big\}.
\end{align*}

The sensor runs a local Kalman filter to recursively compute the state estimate $\hat{x}_{k|k}^s$. Let $h,\,g: \mathbb{S}_+^n \rightarrow \mathbb{S}_+^n$ be defined as
\begin{align}
	h(X) &\triangleq A X A^T + W, \label{eq:h}\\
	g(X) &\triangleq X - X C^T (C X C^T + V)^{-1} C X. \label{eq:g}
\end{align}
Then, the recursive Kalman filtering equations at the sensor are given by:
\begin{subequations} \label{eq:TVkalman}
	\begin{align}
		\hat{x}_{k|k}^s &= \hat{x}_{k|k-1}^s + K_k(y_k - C \hat{x}_{k|k-1}^s),\\
		\hat{x}_{k+1|k}^s &= A \hat{x}_{k|k}^s + B u_k,\\
		P_{k|k}^s &= g \circ h(P_{k-1|k-1}^s), \label{eq:PposS}\\
		P_{k+1|k}^s &= h(P_{k|k}^s),\\
		K_{k+1} &= P_{k+1|k}^s C^T \big(C P_{k+1|k}^s C^T + V\big)^{-1}, \label{eq:TVgain}
	\end{align}
\end{subequations}
where $g \circ h(\cdot) \triangleq g(h(\cdot))$ denotes function composition, and the initial conditions are $\hat{x}_{0|-1}^s = 0$ and $P_{0|-1}^s = X_0$.

Since the sensor has uninterrupted access to the measurements, the estimation error covariance converges exponentially fast~\cite{BrianD.O.Anderson:2012}. Assuming that the pair $(A, C)$ is observable and $(A, W^{1/2})$ is controllable, the fixed-point equation $g \circ h(X) = X$ admits a unique positive semi-definite solution, denoted by $\bar{P}$~\cite{BrianD.O.Anderson:2012}. {At this point, we explicitly switch from the time-varying Kalman filter to its steady-state form. Specifically, once the filter has reached steady state, the covariance matrices and gains converge, i.e., $P_{{k|k}}^s\to\bar{P}$ and $K_{k}\to K$, for all $k$ beyond a transient period.} Hence, we assume the filter has reached steady state, so that $P_{k|k}^s = \bar{P}$ for all $k$, which is a standard assumption in the literature~\cite{Han:2017,Leong:2017a,Wu:2019,Wu:2020,Leong:2020,Iwaki:2021}. Under this assumption, the filtering equations simplify to the time-invariant form:
\begin{align}
\hat{x}_{k|k}^s &= \hat{x}_{k|k-1}^s + K (y_k - C \hat{x}_{k|k-1}^s), \label{eq:hatxsTIV_single}\\
\hat{x}_{k+1|k}^s &= A \hat{x}_{k|k}^s + B u_k,	
\end{align}
where the steady-state Kalman gain is given by:
\begin{align} \label{eq:TIVgain}
	K = h(\bar{P}) C^T \big(C h(\bar{P}) C^T + V \big)^{-1}.
\end{align}

\noindent If the sensor gains access to the channel at time $k$, it transmits a data packet containing $\hat{x}_{k|k}^s$ from~\eqref{eq:hatxsTIV_single}, which implicitly conveys information about the sequence of recent measurements $\{y_{k - t_{k-1} + 1}, \dots, y_k\}$ to the controller~\cite{Leong:2017}. Consequently, the information set available to the controller becomes $\mathcal{I}_k^c = \{\bm{y}^{k - t_k}, \bm{u}^{k-1}, \bm{\delta}^k \}$. {Let $h^{t}(X)$, $t\in\mathbb{N}$, denote the  $t$-fold composition of $h$, i.e.,  
\begin{align*}
    h^{t}(X) = \underbrace{h\circ h\circ \ldots \circ h(X)}_{t \mathrm{~times}},
\end{align*}
with $h^0(X) \triangleq X$. Hence,} the state estimate and corresponding error covariance {of the controller} are given by:
\begin{align}
	\hat{x}_{k|k} &= (A + B L_\infty)^{t_k} \hat{x}_{k - t_k|k - t_k}^s,\\
	P_{k|k} &= h^{t_k}(\bar{P}). \label{eq:Pctrl}
\end{align}
%where $h^0(X) \triangleq X$ and $h^{t}(X)$ denotes the  $t$-fold composition of $h$.

% ========================================
% From Single to Multiple Wireless Networked Control Systems: Limitations and Challenges
% ========================================
\subsubsection{From Single to Multiple Wireless Networked Control Systems: Limitations and Challenges}
\label{subsec:wncs_challenges}

WNCSs integrate control loops with wireless communication, enabling flexible automation without rigid wiring. In a single WNCS, the control loop typically involves a sensor, controller, and actuator that communicate over a shared wireless link. While this architecture is attractive for cost-effective and modular system design, it also introduces several limitations tightly coupled to the quality of its communication link. Key limitations include:
\begin{list4}
    \item {Packet loss and delay:} Wireless links are inherently unreliable due to fading, interference, and congestion, which degrade estimation and control performance.
    \item {Energy constraints:} In systems with battery-powered sensors or actuators, excessive communication reduces operational lifespan.
    \item {Trade-off between communication rate and control accuracy:} Frequent updates improve control quality but consume more bandwidth and energy.
    \item {Design complexity:} Control and communication layers are traditionally designed separately, ignoring cross-layer dependencies that impact system stability and robustness.
\end{list4}
Even under idealized assumptions (e.g., memoryless channels, perfect synchronization), designing event-triggered or state-aware scheduling policies that optimize control performance while respecting communication constraints is a challenging task. When scaling from a single WNCS to a network of multiple WNCSs, e.g., in a MAS where each agent is a WNCS subsystem, the challenges compound significantly. These include:
\begin{list4}
    \item {Shared wireless medium:} Multiple WNCSs must contend for access to a common communication channel, introducing collisions, interference, and scheduling complexity.
    \item {Coupled control objectives:} In MASs, agents often share a global objective or must coordinate actions. This leads to coupling not only in control dynamics but also in communication dependencies.
    \item {Scalability of scheduling:} Centralized scheduling becomes infeasible with many agents. Distributed or priority-based policies must be employed. Designing such mechanisms with performance guarantees remains an open problem.
    \item {Asynchronous and heterogeneous dynamics:} Agents may operate at different time scales or use different sensing modalities, further complicating coordination.
    \item {Information bottlenecks and redundancy:} As more agents participate, ensuring that only task-relevant and non-redundant information is communicated becomes essential to avoid congestion.
\end{list4}

These challenges highlight the need for goal-oriented communication strategies in multi-agent WNCSs, where decisions about \emph{what}, \emph{when}, and \emph{to whom} to communicate are made with the global task in mind \cite{ICAS21Nikos}. We believe that by embedding goal-oriented relevance metrics into scheduling and encoding policies, MASs can overcome the limitations of traditional WNCS frameworks and scale effectively in dynamic, resource-constrained environments.

\begin{remark}
Some initial studies that deal with the aforementioned topics when a receiver tracks remotely DTMCs with the purpose of remote actuation can be found in \cite{ICAS21Nikos, TCOM24Mehrdad, TCOM25Jiping, CL23Manos, TIT25Jiping}. In these studies relevant metrics such as the \textit{Cost of Actuation Error} and \textit{Age of Consecutive Error} have been defined and studied under different systems.    
\end{remark}

% ========================================
%
% Information-Theoretic Models of Goal-Oriented Messaging
%
% ========================================
\section{Information-Theoretic Models of Goal-Oriented Messaging}\label{sec:GOIT}

In many real-world systems, particularly in multi-agent and robotic scenarios, exact reconstruction of the source is not necessary. Instead, what matters is whether the received information enables the receiver (e.g., an agent, controller, or decision module) to achieve its intended goal.

% ========================================
% IB
% ========================================
\subsection{Information bottleneck (IB)}\label{subsec:IB}

The IB principle~\cite{tishby2000informationbottleneckmethod} provides a theoretical framework for extracting relevant information from a source variable $X$ with respect to a target variable $Y$. Unlike traditional compression approaches that aim to preserve all information about $X$, the IB framework selectively retains only the components of $X$ that are informative about $Y$, thus enabling purposeful and task-driven representation. Formally, the IB method seeks a stochastic encoder $p(t | x)$ that maps input data $X$ to a compressed representation $T$, such that $T$ retains maximal mutual information about the task-relevant variable $Y$, while being minimally dependent on the full input, i.e., 
\begin{problem} \label{problem:IB}
\begin{equation}
    \min_{p(t|x)} I(X; T) \quad \text{subject to} \quad I(T; Y) \geq \beta,
\end{equation}
\end{problem}
\noindent where $I(\cdot ; \cdot)$ denotes mutual information, and $\beta$ is a tunable parameter that balances compression and relevance. 
\noindent Equivalently, the objective can be formulated in Lagrangian form as:
\begin{equation}
    \mathcal{L}_{\text{IB}} = I(X; T) - \beta I(T; Y).
\end{equation}

\noindent This formulation defines a trade-off: reducing $I(X; T)$ compresses the representation, while maximizing $I(T; Y)$ ensures task-relevant information is retained. The optimal point on this trade-off curve depends on the complexity of the task and the acceptable information loss. Here, $Y$ may represent labels, control actions, or decisions. The IB principle basically formalizes the trade-off between compression (removal of irrelevant information) and informativeness (preservation of task-relevant content).

Practical implementation of the IB principle has been enabled by \emph{variational approximations}, most notably the Deep Variational IB (VIB)~\cite{alemi2017deep}. In these models, the encoder $p(t|x)$ and decoder $p(y|t)$ are parameterized using deep neural networks, and mutual information terms are approximated using tractable variational bounds:
\begin{equation}
\mathcal{L}_{\text{VIB}} = \mathbb{E}_{p(x,y)} \big[ \mathbb{E}_{p(t|x)} [ -\log p(y|t) ] \big] + \beta D_{\mathrm{KL}}(p(t|x) \| r(t)),
\end{equation}
where $r(t)$ is a prior distribution (e.g., standard Gaussian) and $D_{\mathrm{KL}}$ is the Kullback–Leibler divergence \cite{cover2006elements}. This formulation closely resembles a regularized autoencoder with a task-aware bottleneck.

% ========================================
% Implications of IB for MASs
% ========================================
\subsubsection{Implications of IB for MASs} \label{subsubsec:implicationsIB}

The IB principle offers a compelling lens for designing \emph{communication protocols} and \emph{internal representations} that are both efficient and effective. Each agent may observe high-dimensional, partially observable environments and must communicate compressed messages to other agents or to a centralized controller. Using IB-based objectives, agents can learn to encode only the aspects of their observations that are relevant to shared goals, or global rewards. For instance, in distributed cooperative tasks, agent $i$ may share embeddings $T_i$ derived from its local observations $X_i$, with the goal of maximizing performance on a joint task $Y$. By applying the IB principle, each agent can learn a mapping that discards irrelevant or redundant features while preserving only what is needed for team-level coordination. This reduces communication overhead and supports robustness to noise or spurious correlations in the inputs.

% ========================================
% Semantic rate-distortion theory
% ========================================
\subsection{Semantic rate distortion (SRD): Adapting for goal-oriented communication}
\label{subsec:semantic_rd}

Traditionally, distortion is measured in terms of signal fidelity (e.g., mean squared error or Hamming distance) but the metric is task-agnostic: it does not account for how distortion affects a downstream objective. This limitation has motivated a growing interest in extending the notion of rate-distortion theory to incorporate \emph{task-aware} distortion measures; see, for example, \cite{JSAC:2023Gunduz, 2024:ToMC_Gunduz,2025:SemanticMATHEORY} and references therein. In this broader perspective, mutual information is not the final goal but a proxy that must be evaluated in the context of how the information contributes to a specific task. That is, high mutual information does not necessarily imply high task performance. This divergence is a proxy to goal-oriented formulations that treat communication as purposeful, where relevance to a task supersedes transmission fidelity.

{SRD} theory extends Shannon’s classical framework by prioritizing \emph{task relevance} over mere signal fidelity. Rather than minimizing reconstruction error across all data dimensions, semantic extensions optimize communication to preserve the information that matters most for downstream objectives. This shift enables communication systems to become both efficient and goal-oriented, especially in settings with strict bandwidth or energy constraints~\cite{JSAC:2023Gunduz}.

% ========================================
% Goal-oriented distortion/utility function
% ========================================
\subsubsection{Goal-oriented distortion/utility function}\label{subsubsec:GORD}

In semantic communication, the distortion function $d(x, \hat{x})$ is replaced with a task-aware cost metric. Let $f: \mathcal{X} \to \mathcal{A}$ denote a decision task, such as classification, estimation, or control, and let $\ell(\cdot,\cdot)$ measure task loss. The goal-oriented distortion is defined as:
\begin{align}
d_s(x, \hat{x}) = \ell\bigl(f(x), f(\hat{x})\bigr).    
\end{align}
For instance, in object detection, $d_s$ could represent alignment or miss rates of bounding boxes, rather than pixel-level error. This criterion ensures communication resources focus on transmitting information that directly influences decision quality, while allowing irrelevant details to be compressed or dropped~\cite{JSAC:2023Gunduz}. An alternative formulation uses a \emph{goal-oriented utility} function $u(x,\hat{x})$, capturing the effectiveness of reconstructed data in achieving the task. We then define a rate–utility function:
\begin{align}
R(U) = \min_{p(\hat{x} | x):\, \mathbb{E}[u(X,\hat{X})] \ge U} I(X; \hat{X}),
\end{align}
which reflects the minimal rate required to ensure expected utility $U$. This formulation aligns naturally with semantic sufficiency and relevance criteria.

% ========================================
% Indirect semantic rate‑distortion for intrinsic‑extrinsic source models
% ========================================
\subsubsection{Indirect semantic rate‑distortion for intrinsic‑extrinsic source models}
\label{subsec:indirect_rd}

\begin{figure} [h]
	\centering
	\includegraphics[width=0.99\columnwidth]{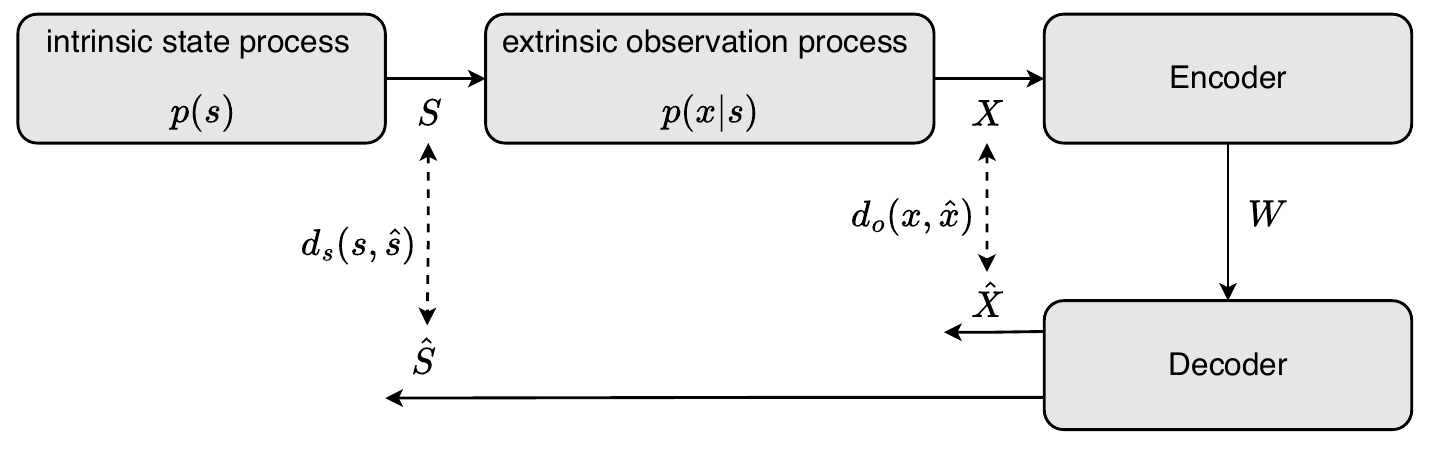}
	\caption{Illustration of a semantic source and its lossy compression (redrawn from~\cite{TCOM:2022RD}).}
	\label{fig:semanticRD}
\end{figure}

A different extension to classical rate-distortion theory addresses scenarios where the \emph{semantic state} is not directly observable, but only indirectly inferred through noisy observations~\cite{2021:SRDF-ISIT,TCOM:2022RD,isit:2022Photis,2023:TCOM_Photis}. In~\cite{2021:SRDF-ISIT,TCOM:2022RD}, the authors propose a two-tier source model comprising an \emph{intrinsic (semantic) state} $S$ representing the true "meaning" or latent feature of the source and an \emph{extrinsic observation} $X$, which is stochastically related to $S$ via a known conditional distribution $p(x|s)$. The encoder observes only $X$, and the goal is to compress it into a codeword that allows (\emph{i}) reconstruction $\hat{S}$ of the intrinsic state, and (\emph{ii}) reproduction  $\hat{X}$ of the observation itself; see Figure~\ref{fig:semanticRD}. 

Distortion metrics $d_s\bigl(s, \hat{s}\bigr)$ and $d_x\bigl(x, \hat{x}\bigr)$ measure the fidelity of the semantic reconstruction and the syntactic fidelity of the extrinsic signal, respectively. The joint rate-distortion problem is stated as:
\begin{problem}\label{problem:semanticRD}
\begin{subequations}
\begin{align}
&R(D_s, D_x) = \min_{p(\hat{s},\hat{x} | x)} I(X; \hat{S}, \hat{X}) \\
\mathrm{s.t.} \quad 
&\mathbb{E}\bigl[d_s(S,\hat{S})\bigr]\le D_s,\\ 
&\mathbb{E}\bigl[d_x(X,\hat{X})\bigr]\le D_x.
\end{align}
\end{subequations}
\end{problem}
Here, $D_s$ and $D_x$ are the allowable semantic and extrinsic distortions, and $I(X;\hat{S},\hat{X})$ quantifies the required encoding rate. Since $S$ is not observed directly, this is an instance of the \emph{indirect (or remote) rate-distortion} problem~\cite{Berger:1971BOOK}. The decoder jointly reconstructs $(\hat{S},\hat{X})$ based solely on the compressed $\hat{X}$. 

Focusing on a tractable yet revealing example, the paper assumes:
\begin{align}
X = A S + W,    
\end{align}
where $S$ and $W$ are zero-mean Gaussian random vectors, i.e., $S \sim \mathcal{N}(0, \Sigma_S)$, $W \sim \mathcal{N}(0, \Sigma_W)$, and $A$ is a known matrix. Distortions are quadratic, i.e., $d_s(s, \hat{s}) = \|s - \hat{s}\|^2$ and $d_x(x, \hat{x}) = \|x - \hat{x}\|^2$. Under these conditions, the rate-distortion problem reduces to a convex optimization over the error covariance matrix of $\hat{S}$. Specifically, one minimizes the mutual information $I(X; \hat{S}, \hat{X})$ subject to covariance constraints linked to $D_s$ and $D_x$. When $\Sigma_S$, $\Sigma_W$, and $A$ are simultaneously diagonalizable, the optimal solution admits a \emph{water-filling} structure, allocating bit-rate to dimensions according to their signal-to-noise and semantic importance.

Building on the semantic source model discussed above, where the communication goal is to preserve task-relevant information about an intrinsic semantic state, another significant extension is offered in~\cite{isit:2022Photis}, which adapts the {SRD} paradigm to \emph{discrete, finite-alphabet settings}, motivated by robust and task-aligned communication for symbolic or logical data (e.g., decisions, classifications, or control actions). With a structure similar to the Gaussian model above, the authors in~\cite{isit:2022Photis} formulated the general rate-distortion problem as follows.
\begin{problem}
\begin{subequations}
\begin{align}
    & R(D_s, D_x) = \min_{p(u|x),\ \phi(u),\ \psi(u)} I(X; U) \\
    \mathrm{s.t.} \quad & \mathbb{E}[d_s(S, \phi(U))] \leq D_s \\
    &\mathbb{E}[d_x(X, \psi(U))] \leq D_x.
\end{align}    
\end{subequations}
\end{problem}
Here, \( \phi \) and \( \psi \) are decoders for semantic and syntactic reconstructions, respectively. 

To provide analytical insights, \cite{isit:2022Photis} presents detailed solutions for the case of binary source and goal variables under standard distortion metrics; namely:
\begin{list4}
    \item Case 1: Binary source and semantic variables under Hamming distortions. A closed-form expression for \( R(D_s, D_x) \) is derived using convexity and symmetry arguments.
    \item Case 2: Binary variables with Hamming and erasure distortion. This illustrates how the choice of distortion affects achievable rates, emphasizing that semantic fidelity depends not only on joint distributions but also on the \emph{operational meaning} of distortion.
\end{list4}

A major conclusion from this study is that \emph{the nature of the distortion constraint critically shapes the encoding strategy}. For example, when erasure distortion is allowed, the encoder may choose to remain silent under high uncertainty, thereby improving average semantic fidelity under limited bandwidth. This aligns with broader trends in goal-oriented communication: not all bits are equally valuable, and the value of information depends on context and task.

Together with the Gaussian analysis of indirect {SRD}, this work completes the picture by offering a rigorous, finite-alphabet formulation with operational relevance to digital systems. It supports the design of practical encoding schemes for real-world deployments, where semantic variables are not observed directly, channel rates are constrained, and multiple fidelity objectives must be satisfied concurrently.
This {SRD} model highlights key principles: latent semantic states can be integrated into source models; multiple distortion metrics can be managed jointly; and optimal encoding requires balancing both semantic and extrinsic goals. Extensions of this framework can be invaluable in designing goal-oriented communication for multi-agent systems and semantic-aware control architectures.

% ========================================
% Implications of semantic rate distortion for MASs
% ========================================
\subsubsection{Implications of semantic rate distortion for MASs} \label{subsubsec:implicationSRDF}

In MASs, {SRD} offers a theoretical foundation for designing \emph{goal-oriented} and \emph{resource-aware} communication policies. Agents can learn to encode and transmit only information that is causally or statistically relevant to shared objectives, such as cooperative localization, distributed mapping, or team-based control. This is particularly valuable in bandwidth-constrained and decentralized settings, where excessive communication may lead to congestion, energy drain, or poor scalability~\cite{JSAC:2023Gunduz}. Furthermore, agents may adapt their compression behavior dynamically, depending on context or uncertainty. For instance, in high-stakes or high-uncertainty regions of the task space, agents can increase communication fidelity, while in stable or low-impact settings, they can reduce or suppress transmission altogether. These context-sensitive behaviors align with the {SRD} paradigm. It is deduced that {SRD} facilitates the emergence of intelligent communication protocols that optimize not for data preservation, but for decision utility, advancing the vision of multi-agent systems that communicate with purpose.

% ========================================
% Semantic information theory and G theory
% ========================================
\subsection{Semantic information theory and G theory}
\label{subsec:GTheory}

A key innovation in Lu's \emph{semantic information theory}, or \emph{G theory}~\cite{e27050461}, lies in redefining the notion of information through a semantic lens. Rather than relying solely on Shannon’s probabilistic framework, which quantifies information as the reduction in uncertainty about a source variable, G theory proposes measuring the \textit{semantic alignment} between messages and the real-world states they describe. This is especially useful in contexts such as human-robot interaction and cooperative multi-agent systems, where communication is not only about statistical inference but also about conveying and interpreting the significance of the information. 

At the heart of G theory is the \emph{truth function}, denoted $T_\theta(y | x)$, which measures the semantic compatibility between a message $y$ and a source state $x$, as perceived by a receiver with interpretation model $\theta$. Here, $\theta$ parametrizes the receiver’s internal semantic model, allowing for non-symmetric interpretations of messages across different receivers or tasks. Based on this framework, the \emph{semantic distortion} is defined as:
\begin{equation}
d(y | x) = -\log T_\theta(y | x),
\end{equation}
which penalizes semantic mismatch between the message and the ground truth. The greater the semantic misalignment, the higher the distortion. Building upon this, G theory defines the \emph{semantic mutual information} between the source variable $X$ and the interpreted message $Y_\theta$ as:
\begin{equation}
I(X; Y_\theta) = \sum_{x, y} p(x, y) \log \frac{T_\theta(y | x)}{T_\theta(y)},
\end{equation}
where
\begin{equation}
T_\theta(y) = \sum_{x} p(x) T_\theta(y | x)
\end{equation}
is the marginal semantic alignment of the message $y$ over the distribution of possible source states. 
This formulation captures the semantic contribution of a message to the receiver’s understanding of the world. Specifically, it quantifies how much the message $y$ helps the receiver, modeled by $\theta$, improve its belief about the true state $x$, not in terms of probabilistic prediction but in terms of semantic correctness. Notably, this measure is generally \emph{asymmetric}, i.e., $I(X; Y_\theta) \neq I({X_\theta; Y})$, reflecting the asymmetry of interpretation inherent to communication with semantics.
Unlike Shannon's mutual information $I(X; Y)$, which is bounded by the channel capacity and independent of task or context, the semantic mutual information $I(X; Y_\theta)$ directly depends on the receiver's interpretive model $\theta$. This opens the door to goal- and context-dependent communication systems where the significance/relevance of a message is judged by its utility to the recipient.

{Building upon this semantic foundation, recent work has introduced algorithmic and optimization principles that enable the practical use of G theory in inference and learning systems. In particular, the \emph{semantic variational Bayes} (SVB) framework and the associated \emph{maximum information efficiency} (MIE) principle~\cite{2025:Gtheory_FEP} extend classical variational inference by explicitly incorporating semantic information into the optimization process. Instead of minimizing the standard variational free energy, SVB relies on the analysis of the semantic rate--fidelity function and seeks to maximize the ratio between semantic information and communication rate, i.e., the information efficiency. This leads to optimization procedures in which the semantic channel, characterized by the truth function $T_\theta(y|x)$, is iteratively aligned with the statistical channel, resulting in improved convergence properties and a more interpretable objective grounded in semantic consistency.}

{From this perspective, G theory provides the underlying representation of semantic information, while the MIE principle offers a way to optimize it under resource constraints. This formulation highlights that efficient communication is not solely about maximizing information transmission, but rather about maximizing \emph{useful} or \emph{goal-relevant} information. Nevertheless, both G theory and its MIE-based extensions primarily focus on representation and inference aspects, and do not explicitly address how semantic information should be generated, selected, or exchanged in multi-agent systems with task-driven objectives.}

{The practical relevance of this framework is illustrated through several representative applications in~\cite{2025:Gtheory_FEP}. For example, in latent-variable models such as mixture models, the MIE principle leads to a reinterpretation of the expectation–maximization (EM) algorithm, where convergence is more accurately characterized by the decrease of the \emph{information difference}, defined as the gap between the total transmitted information $R$ and the semantic (task-relevant) information $G$. Intuitively, this quantity measures how much of the communicated information is \emph{not} useful for the task at hand, and its reduction indicates that the representation becomes increasingly efficient in capturing relevant semantics. This perspective explains why the classical variational free energy may exhibit non-monotonic behavior during learning, while the information difference provides a more consistent indicator of convergence. 
In the context of active inference and control, the framework extends naturally to goal-oriented settings by introducing \emph{purposeful information}, which quantifies how well actions align the system state with desired objectives. Furthermore, in data compression, semantic constraints enable efficient encoding schemes that preserve task-relevant information, achieving near-optimal trade-offs between compression rate and semantic fidelity. These examples highlight the versatility of the MIE-based formulation across inference, control, and communication problems, while reinforcing the central role of semantic relevance in guiding efficient information processing.}

\begin{remark}
% From channel-aware to goal-aware communication
{It is worth noting that several extensions of classical information theory (see, e.g.,~\cite{e25050728}) have already introduced forms of context-aware adaptation, particularly in the study of fading channels with channel state information (CSI). In these settings, transmission strategies are adapted to the instantaneous channel conditions through mechanisms such as adaptive codewords, power control, and generalized mutual information (GMI)–based decoding. These approaches effectively exploit side information to improve reliability and spectral efficiency, and can be interpreted as early instances of context-aware communication. However, despite this increased level of adaptivity, such frameworks remain fundamentally data-centric. The objective is still to maximize achievable rates or minimize decoding error, and the notion of relevance is tied to channel conditions rather than to the underlying task or decision-making objective. In particular, while techniques such as output partitioning and mismatched decoding introduce implicit forms of importance weighting, they do not explicitly quantify the impact of transmitted information on downstream actions or system performance.}

{From a goal-oriented perspective, this highlights a key limitation: adapting communication to the channel is not sufficient when the ultimate objective is to support control, estimation, or coordination tasks. Instead, communication strategies must be designed to prioritize information based on its contribution to task success. This perspective naturally motivates the introduction of task-aware metrics that explicitly capture the utility of information, which will be discussed in Section~\ref{sec:WNCSs}.\ref{subsec:metrics}. These metrics provide a way to quantify how communication decisions affect system performance, enabling a shift from purely information-centric to decision-centric design. This transition marks a fundamental shift from communication systems that are optimized for information transfer to systems that are optimized for decision-making and action.}
\end{remark}

\subsection{Toward a unified framework: integrating theoretical models}
\label{subsec:unified_theory}

The development of goal-oriented communication  has drawn from multiple theoretical traditions, each offering unique strengths. However, these models often appear disjoint in formulation and application. Herein, we discuss how foundational models (namely, the IB principle, {SRD} theory, and G theory) can be conceptually integrated into a unified framework. At the core of all three approaches is the idea that communication should serve a \textit{purpose}, such as improving task performance, reducing uncertainty about relevant variables, or aligning interpretations. Each model formalizes this idea differently:
\begin{list4}
    \item IB frames the trade-off between compressing raw observations and retaining task-relevant content. It is especially suited for learning agent representations under uncertainty and bandwidth constraints. This is often achieved through variational inference or neural encoders.
    \item {SRD} generalizes classical rate-distortion theory by replacing syntactic error measures with task-aware or utility-driven metrics. It enables communication that focuses on decision utility rather than full-state reconstruction. More specifically, communication channels are evaluated based on how well compressed representations support specific decision outcomes, such as control accuracy, detection rate, or reward maximization. This leads to problem-specific distortion or utility functions.
    \item G Theory redefines information through semantic alignment between messages and states. Beyond task metrics, the "meaning" of a message is modeled explicitly via truth functions $T_\theta(y | x)$ that reflect semantic compatibility. It accounts for interpretation asymmetry and receiver-specific beliefs, making it suitable for scenarios involving heterogeneity, abstraction, or symbolic reasoning.
\end{list4}
Essentially, these approaches are not mutually exclusive. For example, in the broader landscape of goal-oriented communication, G theory provides a complement to task-based rate-distortion approaches. Whereas task-oriented models such as the one in~\cite{TCOM:2022RD} formulate the problem through classical information-theoretic lenses by defining distortions in terms of task accuracy or utility (e.g., control error, estimation loss), G theory further generalizes this idea by redefining information and distortion in terms of semantics, thus offering a \emph{semantic abstraction} by evaluating messages through shared truth models. When task distortion functions are interpreted through semantic alignment or truth functions, both paradigms can be unified under a more general framework for \emph{goal-oriented communication}.

It is envisioned that such methods will be integrated and goal-oriented communication will be treated as a multi-layer optimization. For example, a message may be produced via an IB encoder, optimized via a semantic rate distortion utility objective, and evaluated semantically via a G-theoretic truth function. In the realm of MASs, each agent may adapt its messaging behavior based on local context, the task at hand, and the semantic expectations of its neighboring agents. Such a unified perspective can guide the design of practical architectures that combine representation learning, rate-aware communication, and semantic reasoning, enabling MASs that are not only efficient and distributed—but also interpretable, adaptive, and aligned with mission goals.

\subsection{Distributed information-theoretic goal-oriented communication in MASs}
\label{subsec:distributed_GOIT}

The theoretical foundations laid by the IB principle provide a powerful tool for designing MASs, in which each agent may observe high-dimensional, noisy data and must compress its observations into low-bandwidth messages. The IB objective offers a way to learn representations that retain only information relevant to a shared decision variable (e.g., the global reward or consensus), allowing agents to communicate concisely yet effectively. By minimizing $I(X; T)$ while maximizing $I(T; Y)$, agents discard nuisance factors while preserving task-critical content. Variational extensions, such as the deep VIB framework, make such representations learnable via gradient-based optimization. Thus, in order to enable scalable learning of communication policies even in such MAS environments, distributed learning methods can be invoked.

{SRD} theory offers a complementary perspective by redefining distortion not in terms of reconstruction fidelity, but in terms of the effectiveness of the transmitted message in supporting a goal. In distributed MAS settings, this leads to communication protocols that minimize rate while guaranteeing acceptable levels of task performance. For instance, agents can use goal-oriented distortion metrics such as classification error, control deviation, or belief inconsistency, rather than raw signal deviation. When semantic representations are latent or indirectly observed, recent advances in \emph{indirect {SRD}} provide optimization formulations that jointly minimize the rate needed to support both semantic accuracy and syntactic fidelity.

Rate-utility formulations are particularly useful in heterogeneous MASs, where agents may possess different sensing modalities or decision roles, and thus require asymmetric or adaptive communication strategies. For example, one agent may prioritize transmission of localization cues while another may share semantic object descriptors. Goal-oriented utility metrics allow for such diversity while maintaining coordination efficiency.

Going further, Lu’s \emph{G theory} adds a semantic abstraction layer by replacing task-based distortion metrics with \emph{truth functions} that model semantic alignment between messages and the world. The concept of \emph{semantic mutual information}, $I(X; Y_\theta)$, introduces an information measure that accounts for interpretability and "meaning" from the perspective of the receiver. This is especially relevant in human-agent or robot-robot interaction scenarios where message interpretation may differ between entities. G theory thus enables the design of agents that not only learn what to communicate but also how their messages will be interpreted, potentially adjusting messages for different agents or roles in a task.

These approaches are especially valuable in real-world applications like robot teams, distributed sensing, and collaborative decision-making, where communication must be both efficient and aligned with shared goals. Looking ahead, these ideas can be extended to handle large teams, time-varying tasks, and complex reasoning by incorporating hierarchical structures, memory over time, and causal understanding, paving the way for scalable, intelligent, and goal-driven communication.

% ========================================
%
% Coordination of WNCSs under Communication Constraints
%
% ========================================
\section{Coordination of WNCSs under Communication Constraints}
\label{sec:WNCSs}

A central challenge in designing and benchmarking goal-oriented communication systems is selecting appropriate metrics that reflect both communication efficiency and task effectiveness. Unlike traditional communication systems where metrics like throughput or bit error rate suffice, in MASs, coordination is essential for achieving shared goals. However, this coordination is increasingly constrained by communication limitations, such as limited bandwidth, energy constraints, and latency. In goal-oriented MASs, agents must strike a balance between communication cost and coordination accuracy. Achieving effective coordination under such conditions requires intelligent decision-making regarding when, what, and with whom to communicate. To explore these trade-offs, we focus on estimation and control problems in the context of WNCSs. A MAS may consist of multiple WNCSs, either independent or coupled, potentially sharing a common wireless network. This shared medium introduces contention, interference, and scheduling complexity—making communication-aware coordination even more critical.

\begin{figure} [h]
	\centering
	\includegraphics[width=0.99\columnwidth]{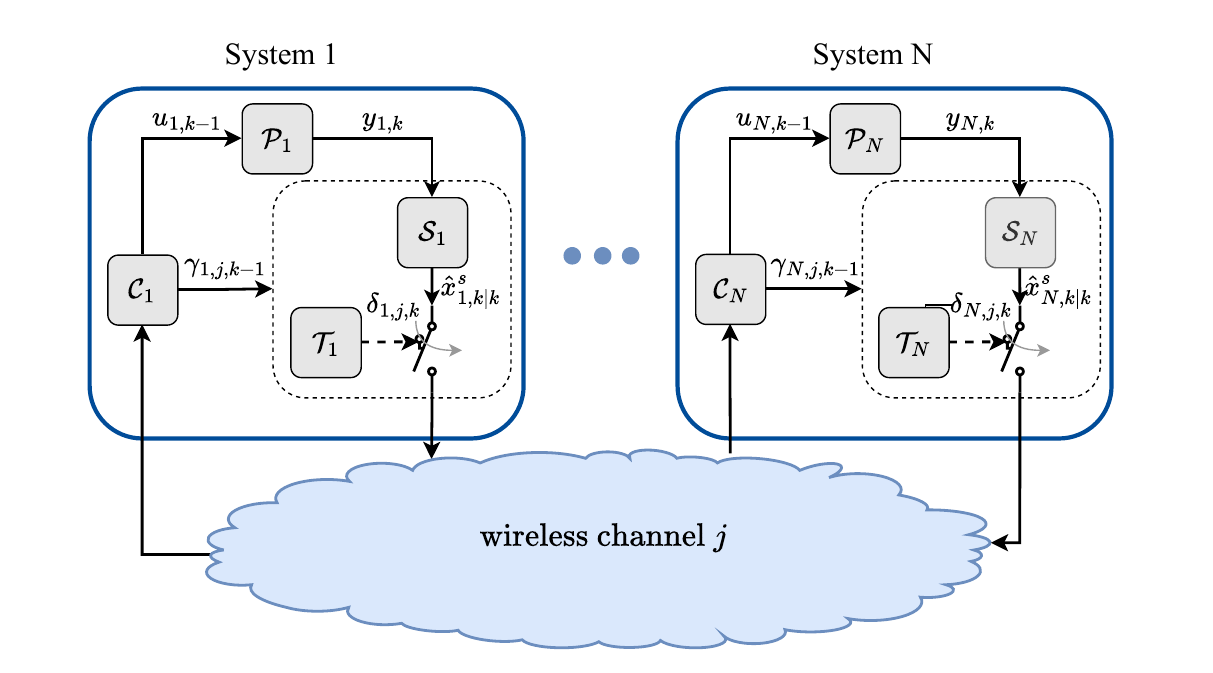}
	\caption{Example of a MAS layout where $N$ WNCSs that exchange no explicit information compete to access channel $j\in\mathcal{M}$. Each subsystem $i\in\mathcal{N}$ consists of a plant ($\mathcal{P}_i$), sensor ($\mathcal{S}_i$), controller ($\mathcal{C}_i$), and the channel access decisions are determined by $\mathcal{T}_i$ which is embedded in the sensor block.}
	\label{fig:WNCS}
\end{figure}

To ground our discussion of goal-oriented communication metrics, we first present a representative model of a MAS consisting of multiple WNCSs. Using this model as a foundation, we introduce and analyze key metrics that quantify the importance/significance of information for communication scheduling and control performance. Therefore, we extend the system presented in Section~\ref{subsec:WNCS} to account of a MAS consisting of multiple WNCSs. For simplicity of exposition, we consider the case for which the dynamics are decoupled, as depicted in Figure~\ref{fig:WNCS}.

\noindent {Since there are $N$ WNCSs , to represent system $i$, $i\in\{1, \ldots, N\}$, we use the same notation as for the system presented in Section~\ref{sec:foundations}.\ref{subsec:WNCS}, with the difference that now we use subscript $i$. If $N=1$, the MAS reduces to the single WNCS introduced in Section~\ref{sec:foundations}.\ref{subsec:WNCS}. Hence,} each plant $\mathcal{P}_i$ is modeled by the following LTI difference equation
\begin{align}\label{eq:xCh4}
	x_{i,k+1} = A_{i}x_{i,k}+B_{i}u_{i,k}+w_{i,k},
\end{align}
where $x_{i,k} \in \mathbb{R}^{n_{i}}$ and $u_{i,k} \in \mathbb{R}^{m_{i}}$ are the states and inputs of subsystem $i$ at time step $k$, respectively. The system and input matrices are denoted by $A_{i}$ and $B_{i}$, respectively, and we assume that $\sigma_{\max}(A_{i})>1$ for all $i$. The initial state $x_{i,0}$ is a Gaussian random variable with mean $\bar{x}_{i,0}$ and covariance $X_{i,0}$ and the process disturbance $w_{i,k}$ is zero-mean Gaussian with covariance $W_i$. The measurement taken by the sensor $\mathcal{S}_i$ at $k$ is given by
\begin{align}
	y_{i,k} = C_{i}x_{i,k}+v_{i,k},
\end{align}
where $C_{i}\in\mathbb{R}^{p_i\times n_i}$ is the output matrix and $v_{i,k}$ is zero-mean Gaussian with covariance $V_i$. We assume the $v_{i,k}$, $w_{i,k}$ and $x_{i,0}$ are mutually independent. The sensor has access to the past control inputs and the entire measurement history which is utilized to determine the \emph{a posteriori} state estimate by running the following local Kalman filter 
\begin{subequations} \label{eq:TIVkalmanCh4}
	\begin{align}
		\hat{x}_{i,k|k}^s &= \hat{x}_{i,k|k-1}^s + K_i(y_{i,k}-C_i\hat{x}_{i,k|k-1}^s), \label{eq:hatxsTIVch4}\\		
		\hat{x}_{i,k+1|k}^s &= A_i\hat{x}_{i,k|k}^s+B_iu_{i,k},
	\end{align}
\end{subequations}
where 
\begin{align}\label{eq:KgainCh4}
	K_i = h_i(\bar{P}_i) C_i^T( C_ih_i(\bar{P}_i) C_i^T + V_i )^{-1},
\end{align}
is the steady-state filter gain, $h_i(\cdot)$ is the Lyapunov-like function defined as $h_{{i}}(X) \triangleq A_{{i}} X A_{{i}}^T + W_{{i}}$ and $\bar{P}_i$ is the steady-state \emph{a posteriori} error covariance {of system $i$ (cf. with $h(X)$ and $\bar{P}$ in Section~\ref{sec:foundations}.\ref{subsec:WNCS})}.

\subsubsection{Shared communication resources}

The wireless network consists of $M$ ($|\mathcal{M}|=M$) wireless channels and the bandwidth limitation is such that $M<N$. The decision variable $\delta_{i,j,k}\in\{0,1\}$ represents whether subsystem $i$ transmits on channel $j$ in frame $k$ as follows. 
\begin{align}
	\delta_{i,j,k} = \begin{cases}
		1, & \mathcal{S}_i \text{ transmits } \hat{x}_{i,k|k}^s \text{ on channel } j \text{ to } \mathcal{C}_i, \\
		0, & \text{otherwise}.
	\end{cases}
\end{align}
We assume that instantaneous packet ack\-nowledge\-ments\-/neg\-ative-\-acknow\-ledge\-ments (ACK/NACKs) are available through an error-free feedback channel and define
\begin{align}
	\gamma_{i,j,k} \!=\! \begin{cases}
		1, & \text{if $\delta_{i,j,k}\!=\!1$ and packet successfully received},\\
		0, & \text{otherwise}.
	\end{cases}
\end{align}
We consider the case in which $\mathbb{P}\{\gamma_{i,j,k}=1|\delta_{i,j,k}=1\}\leq 1$ as opposed to the case of reliable channels where $\mathbb{P}\{\gamma_{i,j,k}=1|\delta_{i,j,k}=1\}= 1$. For clarity of exposition, we assume memoryless wireless channels, corresponding to multipath-dominated environments where channel gain variations are uncorrelated across time and space. This assumption is widely adopted in the design and analysis of WNCSs~\cite{Sinopoli:2004, Schenato:2007, Shi:2010, SECON23Polina}. In this setting, packet dropouts are modeled as a Bernoulli process with a fixed success probability \( \bar{q}_{i,j} \in (0,1] \). Specifically, the probability of a successful transmission from agent \( i \) to agent \( j \) at time \( k \), conditioned on a channel access attempt, is given by:
\begin{align} \label{eq:qbar}
	\mathbb{P}\{\gamma_{i,j,k} = 1 | \delta_{i,j,k} = 1\} = \bar{q}_{i,j}.
\end{align}

\noindent  Additionally, we let
\begin{align}
	\theta_{i,k} = \begin{cases}
		1, & \text{if }\sum_{j=1}^M\gamma_{i,j,k}=1, \\
		0, & \text{otherwise},
	\end{cases}
\end{align}
represent whether $\mathcal{C}_i$ receives a data packet at $k$. Consequently, the time elapsed since the last successful packet reception at $\mathcal{C}_i$ is given by
\begin{align} \label{eq:tch4}
	t_{i,k}=\min\{\kappa\geq0:\theta_{i,{k-\kappa}}=1\}.
\end{align}

\noindent We consider a communication model where each frame consists of a single time-slot, and channel access decisions must adhere to certain constraints. In particular, to avoid interference and ensure collision-free transmissions, the following condition is imposed.
\begin{align} \label{eq:cnstrnt1Ch4}
	\sum_{i=1}^N\delta_{i,j,k} \leq 1, \quad \forall j\in\mathcal{M},\, \forall k\in\mathbb{Z}_{\geq 0}.
\end{align}	
Second, we assume no diversity scheme is used and that each sensor may only occupy one channel at a time, i.e.,
\begin{align} \label{eq:cnstrnt2Ch4}
	\sum_{j=1}^M\delta_{i,j,k} \leq 1, \quad \forall i\in\mathcal{N},\, \forall k\in\mathbb{Z}_{\geq 0}.
\end{align}

\subsubsection{Control and estimation}

The objective of the controller design is to minimize the standard quadratic cost of the entire system which is given by	
\begin{align}\label{eq:JinftyCh4}
	J_{0:\infty} = \lim\limits_{K \to \infty}\frac{1}{K}\mathbb{E}\left\{\sum_{k=0}^{K-1}\sum_{i=1}^N\left(x^T_{i,k}Q_ix_{i,k}+u^T_{i,k}R_i u_{i,k}\right)\right\},
\end{align}
where $Q_i$ and $R_i$ are positive semi-definite and positive definite weighting matrices of appropriate dimensions, respectively. We assume that the pairs $(A_i,B_i)$ and $(A_i,Q_i^{1/2})$ are controllable and observable, respectively. 
The channel access schemes proposed in the subsequent sections result in the independence of channel access decisions and past control actions. Therefore, the certainty equivalence principle holds and the optimal controller is 
\begin{align} \label{eq:uCh4}
	u_{i,k} = L_{i,\infty}\hat{x}_{{i,}k|k},
\end{align}
where $L_{i,\infty}$ is the optimal feedback gain given by 
\begin{align}
	L_{i,\infty} = -(B^T_i\Pi_{i,\infty} B_i + R_i)^{-1}B^T_i\Pi_{i,\infty} A_i,
\end{align}
and $\Pi_{i,\infty}$ is the unique positive semi-definite solution of the following DARE
\begin{align} \label{eq:DARE}
	\Pi_{i,\infty}=A_i^T \Pi_{i,\infty} A_i + Q_i - L_{i,\infty}^T(B_i^T \Pi_{i,\infty} B_i+R_i)L_{i,\infty}.
\end{align}
Moreover, $\hat{x}_{i,k|k} \triangleq \mathbb{E}\{x_{i,k}|\mathcal{I}_{i,k}^c\}$ is the state estimate calculated by the controller {with the information set available to controller $i$ being $\mathcal{I}_{i,k}^c$ (note that in the single-agent case the information set available to the controller was denoted by $\mathcal{I}_{k}^c$).} 
To describe $\mathcal{I}_{i,k}^c$, we first discuss the information available at $\mathcal{S}_i$. At the beginning of $k$, the sensor is aware of the latest measurement and the entire history of past measurements, control inputs, channel access decisions on all channels, and their outcome, i.e.,
\begin{align} \label{eq:IsCh4}
	\mathcal{I}_{i,k}^s=\{\bm{y}_i^k,\bm{u}_i^{k-1},\bm{\delta}_{i,1}^{k-1},\bm{\gamma}_{i,1}^{k-1},\dots,\bm{\delta}_{i,M}^{k-1},\bm{\gamma}_{i,M}^{k-1} \},
\end{align}
where we use bold face to denote a sequence of variables such as 
$$
\bm{\delta}_{i,j}^{k-1}=\{\delta_{i,j,0},\delta_{i,j,1},\dots,\delta_{i,j,k-1}\}. 
$$
Recall that the data packet contains the sensor's state estimate and thus a successful transmission at $k$ provides $\hat{x}_{i,k|k}^s$ to the controller. Therefore, the information available at the controller is given by
\begin{align} \label{eq:IcCh4}
	\mathcal{I}_{i,k}^c=\{\bm{\theta}_i^{k}\odot\bm{\hat{x}}_i^{s^k},\bm{u}_i^{k-1},\bm{\delta}_{i,1}^{k},\bm{\gamma}_{i,1}^{k},\dots,\bm{\delta}_{i,M}^{k},\bm{\gamma}_{i,M}^{k}\},
\end{align}
where $\bm{\theta}_i^{k}$ is inferred from the ACK/NACK history and $\odot$ denotes the Hadamard product. Based on this information, the state estimate and error covariance at the controller are obtained by
\begin{align}
	\hat{x}_{i,k|k}&= (A_i+B_iL_{i,\infty})^{t_{i,k}}\hat{x}_{i,k-t_{k}|k-t_{i,k}}^s,\label{eq:hatxnnCh4}\\
	P_{i,k|k}&= h^{t_{i,k}}(\bar{P}_i), \label{eq:PctrlCh4}
\end{align}
where $h_i^{0}(X)\triangleq X$ and $t_{i,k}$ is given in \eqref{eq:tch4}.

Recall that the information available to sensor $i$ at the beginning of transmission frame $k$ is given in \eqref{eq:IsCh4}. Note that knowledge of the past channel access decisions is sufficient to infer the past control actions without requiring explicit communication from the controller. Define
\begin{align}
	\hat{x}_{i,k|k-1}^s&\triangleq \mathbb{E}\{x_{k} | \mathcal{I}_{i,k-1}^s\},\\
	\hat{x}_{i,k|k}^s&\triangleq \mathbb{E}\{x_{k} | \mathcal{I}_{i,k}^s\},
\end{align}
as the \emph{a priori} and \emph{a posteriori} state estimates at the sensor, respectively, with the respective error covariances defined as
\begin{align*}	
	P_{i,k|k-1}^s&\triangleq\mathbb{E}\{(x_{i,k}-\hat{x}_{i,k|k-1}^s)(x_{i,k}-\hat{x}_{i,k|k-1}^s)^T | \mathcal{I}_{i,k-1}^s\},\\
	P_{i,k|k}^s&\triangleq\mathbb{E}\{(x_{i,k}-\hat{x}_{i,k|k}^s)(x_{i,k}-\hat{x}_{i,k|k}^s)^T | \mathcal{I}_{i,k}^s\}.
\end{align*}
By running a local Kalman filter, the sensor can calculate $\hat{x}_{i,k|k}^s$ recursively. Let functions $h_{{i}},\,g_{{i}}: \mathbb{S}_+^n \to \mathbb{S}_+^n$ be defined as
\begin{align}
	h_{{i}}(X) &\triangleq A_{{i}} X A_{{i}}^T + W_{{i}}, \label{eq:h}\\
	g_{{i}}(X) &\triangleq X - X C_{{i}}^T (C_{{i}} X C_{{i}}^T + V_{{i}})^{-1} C_{{i}} X. \label{eq:g}
\end{align}
The recursive Kalman filtering equations for sensor $i$ are given by
\begin{subequations} \label{eq:TVkalman}
	\begin{align}
		\hat{x}_{i,k|k}^s &= \hat{x}_{i,k|k-1}^s + K_{i,k}(y_{i,k}-C_i\hat{x}_{i,k|k-1}^s),\\		
		\hat{x}_{i,k+1|k}^s &= A_i\hat{x}_{i,k|k}^s+B_i u_{i,k},\\
		P_{{i,k|k}}^s &= g_{{i}}\circ h_{{i}}(P_{{i,k-1|k-1}}^s), \label{eq:PposS}\\
		P_{{i,k+1|k}}^s &= h(P_{{i,k|k}}^s),\\
		K_{i,k+1} &= P_{{i,k+1|k}}^s C_i^T( C_i P_{{i,k+1|k}}^s C_i^T + V_i )^{-1}, \label{eq:TVgain}		
	\end{align}
\end{subequations}
where $ g_{{i}}\circ h_{{i}}(\cdot)\triangleq g_{{i}}(h_{{i}}(\cdot))$ denotes the function composition and the initial conditions are $\hat{x}_{i,0|-1}^s=0$ and $P_{{i,0|-1}}^s=X_{i,0}$. 
Since the measurements are available uninterruptedly at the sensor, the error covariance converges exponentially fast \cite{BrianD.O.Anderson:2012}. Based on the assumption that the pairs $(A_i,C_i)$ and $(A_i,W_i^{1/2})$ are observable and controllable, respectively, $g_{{i}}\circ h_{{i}}(X)=X$ has a unique positive semi-definite solution which we denote by $\bar{P}_{{i}}$ \cite{BrianD.O.Anderson:2012}. 
{At this point, we explicitly switch from the time-varying Kalman filter to its steady-state form. Specifically, once the filter has reached steady state, the covariance matrices and gains converge, i.e., $P_{{i,k|k}}^s\to\bar{P}_i$ and $K_{i,k}\to K_i$, for all $k$ beyond a transient period. Hence,} for simplicity of exposition, we will assume that the filter has already entered steady state and $P_{{i,k|k}}^s=\bar{P}_i$ for all $k$, which is a common assumption in similar settings \cite{Han:2017,Leong:2017a, Wu:2019, Wu:2020, Leong:2020, Iwaki:2021}. As a result, the filtering equations can be simplified as
\begin{subequations} \label{eq:TIVkalman}
	\begin{align}
		\hat{x}_{i,k|k}^s &= \hat{x}_{i,k|k-1}^s + K_i(y_{i,k}-C_i\hat{x}_{i,k|k-1}^s), \label{eq:hatxsTIV}\\		
		\hat{x}_{i,k+1|k}^s &= A_i\hat{x}_{i,k|k}^s+B_i u_{i,k},
	\end{align}
\end{subequations}
where the time-invariant Kalman filter gain is given by
\begin{align}\label{eq:TIVgain}
		K_i = h(\bar{P}_i) C_i^T( C_i h_{{i}}(\bar{P}_i) C_i^T + V_i )^{-1}.
\end{align}
If channel access is granted $k$, the sensor transmits a data packet which contains $\hat{x}_{i,k|k}^s$ in \eqref{eq:hatxsTIV} which conveys information about the missing measurements to the the controller, i.e., $\{y_{i,k-t_{k-1}+1},\dots, y_{i,k}\}$ \cite{Leong:2017}. Consequently, the information available at the controller is given by $\mathcal{I}_{i,k}^c=\{\bm{y}^{i,k-t_k},\bm{u}^{i,k-1},\bm{\delta}^{i,k}\}$. {Similarly to Section~\ref{subsec:WNCS}, $h_i^{t}(X)$, $t\in\mathbb{N}$, denote the  $t$-fold composition of $h_i$ with $h_i^0(X) \triangleq X$. Hence,} the state estimate and the corresponding error covariance {of the controller} become
\begin{align}
	\hat{x}_{i,k|k}&= (A_i+B_iL_{i,\infty})^{t_{k}}\hat{x}_{i,k-t_{k}|k-t_{k}}^s,\\
	P_{i,k|k}&= h_{{i}}^{t_{k}}(\bar{P}_i). \label{eq:Pctrl}
\end{align}
%where $h_{{i}}^{0}(X)\triangleq X$.

\subsection{Metrics for Information Importance/Significance in WNCSs and relevant optimization problems}
\label{subsec:metrics}

The cost should be defined in a way that it can be evaluated based on the local information of each decision-maker, i.e., $\mathcal{I}_{i,k-1}^c$. Using the law of total expectation, \eqref{eq:JinftyCh4} can be written as
\begin{align} \label{eq:Jktot}
	J_{0:\infty} =&\lim\limits_{K \to \infty}\frac{1}{K}\mathbb{E}\left\{\sum_{k=0}^{K-1}\sum_{i=1}^N\mathbb{E}\left\{J_{i,k}|\mathcal{I}_{i,k}^c\right\} \right\} \notag \\
    &+\sum_{i=1}^N \mathrm{tr}(\Pi_{i,\infty}W_i),
\end{align}
where
\begin{align} \label{eq:Jik}
	J_{i,k}\triangleq\mathrm{tr}\left(\Gamma_{i,\infty}(x_{i,k}-\hat{x}_{i,k|k})(x_{i,k}-\hat{x}_{i,k|k})^T\right).
\end{align}

To illustrate the intuition behind prioritized channel access and derive the priority measures, we begin by letting $\mathcal{F}_k\triangleq \{i:i\in\mathcal{N},\delta_{i,k}=1\}$ denote the subset of subsystems that transmit at $k$ and define $\bar{\mathcal{F}}_k\triangleq \mathcal{N}\setminus \mathcal{F}_k$. Moreover, we denote the cost incurred by subsystem $i$ at $k$ by $J_{i,k}$. The cost of the entire system at $k$ can be written as in~\eqref{eq:Jk2Ch4}.
\begin{figure*}
\begin{align}\label{eq:Jk2Ch4}
	\sum_{i\in\mathcal{N}}\mathbb{E}\left\{J_{i,k}|\mathcal{I}_{i,k}^c\right\}&=\sum_{i\in \bar{\mathcal{F}}_k}\mathbb{E}\left\{J_{i,k}|\mathcal{I}_{i,k}^\tau,\theta_{i,k}=0\right\}  + \sum_{i\in \mathcal{F}_k}\mathbb{P}\left\{\gamma_{i,j,k}=1|\delta_{i,j,k}=1\right\}\mathbb{E}\left\{J_{i,k}|\mathcal{I}_{i,k}^\tau,\theta_{i,k}=1\right\}\notag\\
	&\quad + \sum_{i\in \mathcal{F}_k}\mathbb{P}\left\{\gamma_{i,j,k}=0|\delta_{i,j,k}=1\right\}\mathbb{E}\left\{J_{i,k}|\mathcal{I}_{i,k}^\tau,\theta_{i,k}=0\right\}\notag\\
	&= \sum_{i\in \mathcal{N}}\mathbb{E}\left\{J_{i,k}|\mathcal{I}_{i,k}^\tau,\theta_{i,k}=0\right\} -\sum_{i\in \mathcal{F}_k}\left(\mathbb{E}\left\{J_{i,k}|\mathcal{I}_{i,k}^\tau,\theta_{i,k}=0\right\}- \mathbb{E}\left\{J_{i,k}|\mathcal{I}_{i,k}^\tau,\theta_{i,k}=1\right\}\right)\bar{q}_{i,j}.
\end{align}
\end{figure*}
Recall that $\mathcal{I}_{i,k}^\tau$ denotes the information available to Scheduler block and our only assumption so far is that $\mathcal{I}_{i,k-1}^c\subseteq \mathcal{I}_{i,k}^\tau$. The last equality in \eqref{eq:Jk2Ch4} reveals that the channel access decisions at $k$ only influence the last summation. Furthermore, due to the lack of constraints on energy consumption, any transmission is always preferred. As a result, minimizing the cost of the entire system at $k$ subject to \eqref{eq:cnstrnt1Ch4} and \eqref{eq:cnstrnt2Ch4} is equivalent to
\begin{align} \label{eq:FkgenCh3}
\mathcal{F}_k=\argmax_{i\in\mathcal{N}} \quad m_{i,k},
\end{align}
where
\begin{align}
m_{i,k}=&\sum_{i\in \mathcal{N}}\mathbb{E}\left\{J_{i,k}|\mathcal{I}_{i,k}^\tau,\theta_{i,k}=0\right\} \notag \\
    &-\sum_{i\in \mathcal{F}_k}\Big(\mathbb{E}\left\{J_{i,k}|\mathcal{I}_{i,k}^\tau,\theta_{i,k}=0\right\} \notag \\
    &- \mathbb{E}\left\{J_{i,k}|\mathcal{I}_{i,k}^\tau,\theta_{i,k}=1\right\}\Big)\bar{q}_{i,j}.
\end{align}
denotes the priority measure. The impact of the choice of $J_{i,k}$ and $\mathcal{I}_{i,k}^\tau$ on the resulting priority measure are discussed next. 

% ========================================
% Cost of information loss (CoIL)
% ========================================
\subsubsection{Cost of information loss (CoIL)}

The term Cost of Information Loss (CoIL) was introduced in \cite{Charalambous:2017} to quantify the performance degradation caused by missing information in networked control systems. A key property of CoIL is that it depends on the statistical properties, such as the variance of the process and measurement noise, rather than the actual sensor readings. This enables preemptive scheduling decisions to be made at the controller side without requiring real-time measurements, significantly reducing communication overhead in centralized systems. In distributed scenarios, CoIL can still be computed based solely on timing information, which allows control units to prioritize channel access independently of the sensor's computational capacity. This makes CoIL particularly attractive for systems with low-cost or passive sensors. Such architectures motivate variance-based triggering laws (e.g., \cite{Trimpe:2014, Leong:2017a}) and enable controller-side negotiation of access priorities, as further developed in follow-up works \cite{Farjam:2022, Farjam:2023}. 

Within this metric, we solve the following problem on a single-step horizon, i.e.,
\begin{problem} \label{problem:p2Ch4}
	\begin{align}
		\begin{split}
			\min_{\Delta_k} & \quad \sum_{i\in\mathcal{N}}\mathbb{E}\left\{J_{i,k}|\mathcal{I}_{i,k}^c\right\}, \\
			\text{s. t.}&\quad \eqref{eq:cnstrnt1Ch4}, \eqref{eq:cnstrnt2Ch4},
		\end{split}
	\end{align}
	where $J_{i,k}$ is given in \eqref{eq:Jik}
    and $\Delta_{k}$ is a binary matrix that includes all the optimization variables at time $k$, i.e.,
	\begin{align} \label{eq:Delta}
		\Delta_{k} \triangleq 
		\begin{bmatrix}
			\delta_{1,1,k} & \ldots& \delta_{N,1,k} \\
			\delta_{1,2,k} & \ldots& \delta_{N,2,k} \\
			\vdots & & \vdots \\
			\delta_{1,M,k} & \ldots& \delta_{N,M,k} 
		\end{bmatrix}.
	\end{align}
\end{problem}
 
\noindent The aim is to obtain a goal-oriented (prioritization) scheme whose implementation accomplishes the objective of  Problem~\ref{problem:p2Ch4}.

Note that the estimation at the controller side at the end of frame $k$ only concerns $\theta_{i,k}$ through $t_{i,k}$, i.e., whether a data packet has been received at $k$ regardless of the selected channel. In case a sensor is not granted channel access, i.e., $i\in\bar{\mathcal{F}}_k$, then certainly $\theta_{i,k}=0$. However, for $i\in\mathcal{F}_k$, the quality of the assigned wireless link influences the outcome of $\theta_{i,k}$ and thus $\sum_{i\in \mathcal{N}}\mathbb{E}\left\{J_{i,k}|\mathcal{I}_{i,k}^c\right\}$.

Considering that the information utilized for decision-making is the controller's information, i.e., $\mathcal{I}_{i,k}^\tau=\mathcal{I}_{i,k-1}^c$, substituting \eqref{eq:Jik} in \eqref{eq:Jk2Ch4} yields
\begin{align}\label{eq:Jk3Ch4}
\sum_{i\in\mathcal{N}}\mathbb{E}\left\{J_{i,k}|\mathcal{I}_{i,k}^c\right\}
=&\sum_{i\in \mathcal{N}}\mathbb{E}\left\{J_{i,k}|\mathcal{I}_{i,k-1}^c,\theta_{i,k}=0\right\} \notag \\
&-\sum_{i\in \mathcal{F}_k}{\mathrm{CoIL}}_{i,k}\bar{q}_{i,j},
\end{align}
where ${\mathrm{CoIL}}_{i,k}$ is given by
\begin{align} \label{eq:CoILch4}
	{\mathrm{CoIL}}_{i,k}&\triangleq \mathbb{E}\left\{J_{i,k}|\mathcal{I}_{i,k-1}^c,\theta_{i,k}=0\right\} \notag \\ 
    & \quad - \mathbb{E}\left\{J_{i,k}|\mathcal{I}_{i,k-1}^c,\theta_{i,k}=1\right\} \notag\\
	&= \mathrm{tr}\left(\Gamma_{i,\infty}\left[h_i^{t_{i,k-1}+1}(\bar{P}_i)-\bar{P}_i\right]\right).
\end{align}
Due to the independence of the first term on the right-hand side of \eqref{eq:Jk3Ch4} from the channel access decisions at $k$, minimizing \eqref{eq:Jk3Ch4} is equivalent to maximizing the last summation. Consequently, by reintroducing the constraints, Problem~\ref{problem:p2Ch4} can equivalently be formulated as follows.
\begin{problem} \label{problem:p3Ch4}
\begin{align} \label{eq:maximization}
	\begin{split}
		\max_{\Delta_{k}} &\quad \sum_{i\in\mathcal{N}}\sum_{j\in\mathcal{M}}{\mathrm{CoIL}}_{i,k}\bar{q}_{i,j}\delta_{i,j,k}, \\
		\text{subject to}&\quad \eqref{eq:cnstrnt1Ch4}, \eqref{eq:cnstrnt2Ch4},
	\end{split}
\end{align}
\end{problem}

\noindent The probability of successful transmission over each link is assumed to be known~\cite{Charalambous:2017,Farjam:2018} or it can be learned~\cite{Farjam:2022} and local information is sufficient for evaluating ${\mathrm{CoIL}}_{i,k}$ and thus the local cost can be defined as $m_{i,j,k}={\mathrm{CoIL}}_{i,k}\bar{q}_{i,j}$. 

% ========================================
% Value of Information (VoI)
% ========================================
\subsubsection{Value of Information (VoI)}

The concept of Value of Information (VoI) quantifies the price a decision‑maker is willing to pay to access certain information before making decisions. In its foundational treatment, VoI was defined as the expected increase in utility obtained by reducing uncertainty prior to selecting an action. This notion has since been adopted across diverse disciplines. In information economics, VoI captures the economic worth of information in markets and mechanism-design frameworks, linking information asymmetry to decision efficiency~\cite{Bikhchandani:2013}. In risk management, VoI has informed decisions about investing in information gathering for uncertain systems and environments~\cite{Gould:1974}. In shortest path optimization, VoI is used to determine whether acquiring additional data (e.g., traffic or environmental conditions) justifies the computational or communication cost, as in \cite{Rinehart:2011}. 

In the context of networked control systems, a specific variant of VoI is adopted in \cite{Molin:2015,Molin:2019}. The main distinguishing feature of this variant of VoI is that the actual measurements are taken into consideration for evaluating it. To be more specific, instead of using the information available at the controller for priority assignment as in CoIL, VoI utilizes the sensor's information, i.e., $\mathcal{I}_{i,k}^\tau=\mathcal{I}_{i,k}^s$. Subsequently, we define
\begin{align} \label{eq:VoIdefch3}
	{\mathrm{VoI}}_{i,k}\triangleq \mathbb{E}\left\{J_{i,k}|\mathcal{I}_{i,k}^s,\theta_{i,k}=0\right\} - \mathbb{E}\left\{J_{i,k}|\mathcal{I}_{i,k}^s,\theta_{i,k}=1\right\},
\end{align}
where $J_{i,k}$ is defined as \eqref{eq:Jik}. Let $e_{i,k|k}^s\triangleq x_{i,k}-\hat{x}_{i,k|k}^s$ and $\check{e}_{i,k|k}\triangleq\hat{x}_{i,k|k}^s-\hat{x}_{i,k|k}$. Since 
$$
e_{i,k|k}= x_{i,k}-\hat{x}_{i,k|k}=e_{i,k|k}^s+\check{e}_{i,k|k}, 
$$
we can write
\begin{align} \label{eq:Jik0voi}
&\mathbb{E}\{{J}_{i,k}|{\mathcal{I}}_{i,k}^s,\theta_{i,k}{=}0\} \notag \\ 
&= \mathrm{tr}(\Gamma_{i,\infty}\mathbb{E}\{(e_{i,k|k}^s+\check{e}_{i,k|k})(e_{i,k|k}^s+\check{e}_{i,k|k})^T|{\mathcal{I}}_{i,k}^s,\theta_{i,k}{=}0\} \notag\\
	&= \mathrm{tr}(\Gamma_{i,\infty}\mathbb{E}\{e_{i,k|k}^s{e_{i,k|k}^s}^T|{\mathcal{I}}_{i,k}^s,\theta_{i,k}{=}0\}){+}\mathrm{tr}(\Gamma_{i,\infty}\check{e}_{i,k|k}{\check{e}_{i,k|k}}^T) \notag \\
	&= \mathrm{tr}(\Gamma_{i,\infty}\bar{P}_i)+\mathrm{tr}(\Gamma_{i,\infty}\check{e}_{i,k|k}{\check{e}_{i,k|k}}^T),
\end{align}
where the second equality follows from the facts that $\mathbb{E}\{e_{i,k|k}^s|{\mathcal{I}}_{i,k}^s,\delta_{i,k}{=}0\}=0$ and $\check{e}_{i,k|k}$ is deterministically given by
\begin{align} \label{eq:hate}
	\check{e}_{i,k|k}=\sum_{\check{t}=k-t_{i,k-1}+1}^k A_i^{k-\check{t}}K_i(y_{i,\check{t}}-C_i\hat{x}_{\check{t}|\check{t}-1}^s),
\end{align}
where $K_i$ in this case is the invariant Kalman gain in \eqref{eq:TIVgain}. Furthermore, absence of packet dropouts means that $\hat{x}_{i,k|k}=\hat{x}_{i,k|k}^s$. Therefore, by following the same steps as in \eqref{eq:Jik0voi} we obtain
\begin{align} \label{eq:Jik1voi}
	\mathbb{E}\{{J}_{i,k}|{\mathcal{I}}_{i,k}^s,\theta_{i,k}{=}1\} =\mathrm{tr}(\Gamma_{i,\infty}\bar{P}_i).
\end{align}
Finally, substituting \eqref{eq:Jik0voi} and \eqref{eq:Jik1voi} in \eqref{eq:VoIdefch3} yields
\begin{align} \label{eq:VoIch3}
	{\mathrm{VoI}}_{i,k}= \mathrm{tr}(\Gamma_{i,\infty}\check{e}_{i,k|k}{\check{e}_{i,k|k}}^T).
\end{align}

A key limitation of employing VoI as the priority metric in \eqref{eq:FkgenCh3} is that its implementation typically requires smart sensors with sufficient computational capabilities. However, since VoI incorporates more information than CoIL in its computation, it is generally expected to yield improved performance, as supported by empirical results in \cite{Trimpe:2015} and \cite{Farjam:CDC2021}. This holds true even though CoIL-based scheduling has been shown to result in lower estimation error covariance at the controller, as reported in \cite{Farjam:2019ECC}. An illustrative example demonstrating how adopting each priority
measure influences the channel access decisions and performance is provided in~\cite[Example 3.3.1]{2022:TahmooresThesis}.

% ========================================
% Age of Information (AoI)
% ========================================
\subsubsection{Age of Information (AoI)}
\label{subsubsec:AoI}

Age of Information (AoI) \cite{kosta2017age, 2021:Yates_AoI, pappas2023age,2023:FountoulakisTCOM} is a performance metric that quantifies the timeliness or freshness of information in communication systems. Unlike traditional metrics such as latency or throughput, which measure packet delivery delay or volume, AoI captures the age of the most fresh received update at a decision-maker. Formally, at any time $t$, the AoI is defined as the time elapsed since the generation of the most fresh successfully received status update, i.e., 
$$
\Delta(t) = t - u(t), 
$$
where $u(t)$ denotes the generation time of the most fresh received packet by time $t$.

Originally introduced to address the demands of real-time status monitoring, AoI has become a critical metric in systems where the significance of information diminishes over time. Intuitively understood as a measure of \emph{information freshness}, AoI was designed to capture strict timeliness requirements in machine-type communications, including intelligent transportation systems, industrial Internet of Things (IIoT), and other ultra-reliable low-latency (URLLC) applications. In such environments, ensuring that the receiver has access to the most recent data is often more important than maximizing data throughput or minimizing the delay. The primary appeal of AoI lies in its ability to capture the temporal relevance of information. Traditional latency metrics only account for individual packet delays, whereas AoI reflects how stale the receiver’s view of the system is. This makes AoI particularly well-suited for systems with tight feedback loops, where outdated data can lead to poor control decisions or system instability. The introduction of AoI has led to the development of a rich body of theoretical and applied research. Among the topics explored are sampling and scheduling policies designed to minimize AoI under bandwidth and energy constraints, the analysis of AoI dynamics in queuing systems and wireless multi-access networks, and the design of medium access control (MAC) protocols that are explicitly AoI-aware. More recently, AoI has been adapted to more specialized forms. Two key AoI-based performance measures have been extensively studied: 
\begin{list4}
    \item[i)] \emph{Average AoI (AAoI)}: the long-term time average of the AoI process,
    \begin{equation}
        \bar{\Delta} = \lim_{T \to \infty} \frac{1}{T} \int_0^T \Delta(t) \, dt,
    \end{equation}
    which can also be computed as the ratio of the expected area under the AoI curve $Q_n$ to the expected inter-update duration $Y_n$~\cite{2021:Yates_AoI}:
    \begin{equation}
        \bar{\Delta} = \frac{\mathbb{E}[Q_n]}{\mathbb{E}[Y_n]}.
    \end{equation}

    \item[ii)] \emph{Peak AoI (PAoI)}: the age value just before a new update is received. If $A_n$ denotes the age just before the $n$-th update arrives, then the long-run average peak age is
    \begin{align}
        \Delta^{(p)} &= \lim_{N \to \infty} \frac{1}{N} \sum_{n=1}^N A_n \notag \\
        &= \mathbb{E}[A_n] = \mathbb{E}[Y_n] + \mathbb{E}[T_n],
    \end{align}
    where $T_n$ is the random delay associated with delivering the $n$-th packet~\cite{2021:Yates_AoI}.
\end{list4}

\subsubsection{{Version Age of Information (VAoI)}}

{While AoI captures the timeliness of information it does not explicitly account for whether the underlying information source has actually evolved. In many practical systems, especially those with event-driven or state-dependent dynamics, information may remain unchanged for extended periods. In such cases, AoI may not accurately reflect the effective staleness of information. This observation motivates the introduction of the Version Age of Information (VAoI) \cite{YatesISIT21}.% \todo{ADD REFS HERE.}}

{VAoI associates each update generated at the source with a version index that reflects the evolution of the underlying information. Let $N_t^{\mathrm{S}}$ and $N_t^{\mathrm{R}}$ denote the version indices at the source and the receiver, respectively, at time $t$. The VAoI is defined as
\begin{equation}
\Delta_v(t) = N_t^{\mathrm{S}} - N_t^{\mathrm{R}}.
\end{equation}
This metric quantifies how many updates the receiver is lagging behind the source. Unlike AoI, which increases continuously with time, VAoI increases only when a new version is generated at the source and the destination is not informed, otherwise remains constant. Thus, VAoI captures the evolution of information content rather than purely temporal staleness. VAoI can be viewed as a content-aware extension of AoI that aligns freshness with actual information changes.}

{From the perspective of goal-oriented communication, VAoI naturally reflects the relevance of information. Since not all updates contribute equally to task performance, tracking the number of missed updates (or versions) is often more informative than measuring elapsed time. In particular, VAoI avoids unnecessary transmissions when the system state remains unchanged, while prioritizing communication when new and potentially important information is generated. This makes VAoI especially suitable for resource-constrained MASs, where communication decisions must balance efficiency and task performance.}

{Incorporating VAoI into scheduling and control policies leads to event-driven communication strategies. For instance, transmissions can be triggered when the version gap $\Delta_v(t)$ exceeds a threshold, indicating that the receiver has missed a sufficient number of updates. Compared to AoI-based policies, such strategies reduce redundant transmissions and improve resource utilization, especially in systems with slowly varying dynamics. Moreover, since VAoI does not rely on timestamps, it is robust to clock synchronization issues, which are common in large-scale distributed systems.}

{Several extensions of VAoI have been proposed to capture system dynamics and semantic relevance better \cite{TCOM25Mehrdad}. For example, the \emph{Version Innovation Age} (VIA) accounts for missed state transitions in Markovian sources, increasing only when the source changes but the update is not successfully delivered. Similarly, the \emph{Age of Incorrect Versions} (AoIV) \cite{TCOM25Mehrdad} focuses on outdated information, specifically when the receiver's estimate is incorrect, refining the notion of semantic staleness. These variants further highlight that freshness metrics can be adapted to capture not only when information is outdated, but also whether it is relevant or erroneous.}

{Overall, VAoI provides a complementary perspective to AoI by incorporating the evolution of the information source into the notion of freshness. This makes it a valuable tool for designing goal-oriented communication strategies that prioritize relevant updates and avoid unnecessary use of communication resources \cite{CL25Erfan, COMMAGErfan, WiOpt24Erfan, TCOM25Erfan}.}

\begin{remark}
The success of AoI has spawned numerous extensions and variants tailored to specific application requirements. Notable developments include Age of Synchronization (AoS), which measures temporal misalignment between distributed nodes, and Age of Incorrect Information (AoII), which accounts for both the duration and magnitude of estimation errors. Recent innovations have extended AoI beyond traditional timeliness considerations to encompass the complete information lifecycle. The concept of \textit{Age of Actuation}, introduced in \cite{AoIWS23Ali, AliTCOM26}, focuses on the timeliness of control actions rather than merely information reception. Building upon this foundation, \textit{Age of Actuated Information} \cite{GC24Ali} provides a more comprehensive framework that considers both the timeliness of information generation and its subsequent utilization in decision-making processes. These metrics recognize that in many systems, the ultimate value of information lies not in its reception but in its effective translation into timely actions.
\end{remark}

In the context of multi-agent systems and goal-oriented communication, AoI serves as both a foundational metric and a building block for more sophisticated information valuation frameworks. While AoI provides essential insights into information freshness, intelligent agents operating in complex environments require additional criteria to determine not only whether information is fresh, but also whether it is worth communicating given its expected contribution to task performance and coordination objectives. This evolution has led to the development of hybrid metrics that combine AoI with task-relevance measures and semantic importance indicators. In such systems, communication decisions are driven by multi-objective optimization problems that balance information freshness against communication costs, energy consumption, and task performance requirements; see, for example, \cite{tzortzis2025remote}. The integration of AoI with goal-oriented frameworks enables the design of intelligent communication protocols that can adaptively prioritize updates based on their temporal relevance and decision-making impact \cite{SECON23Polina, TCOM24Erfan, TIT25Jiping, ToC_Pouya:2025}.

% ========================================
% Goal-oriented communications for heterogeneous spatially-distributed MASs
% ========================================
\subsection{Goal-oriented communications for heterogeneous spatially-distributed MASs} 
\label{sec:scheduling_routing}

As multi-agent systems scale to hundreds or thousands of agents, such as in swarm robotics, sensor networks, and distributed AI, the efficiency of communication becomes increasingly critical. In such systems, centralized scheduling and naive all-to-all communication quickly become infeasible due to congestion, delay, and energy limitations. Therefore, mechanisms, such as scheduling, routing, and prioritization, must be carefully designed to allocate bandwidth and attention to the most task-relevant agents, messages, and locations. This section discusses key strategies to address these challenges, focusing on \emph{role-aware}, \emph{spatially-aware}, and \emph{graph-structured} communication paradigms.

% ========================================
% Role-based and spatially-aware communication
% ========================================
\subsubsection{Role-based and spatially-aware communication}
\label{subsec:role-based}

In cooperative MASs, agents often assume distinct roles or operate in spatial sub-regions of the environment. These roles may be predefined, such as sensor, relay, or decision-maker, or may emerge dynamically through task requirements, for instance in leader-follower coordination or exploration–exploitation patterns. \emph{Role-based communication} leverages these structures by prioritizing transmissions from agents whose information is deemed most critical to global task performance. For example, in multi-robot exploration scenarios, certain agents may be assigned temporary relay roles to preserve network connectivity, while others focus on sensing or actuation. One such implementation is the Connectivity-Aware Relay Algorithm (CARA), which dynamically assigns communication roles based on real-time signal quality and link stability, significantly improving mission reliability and transmission efficiency~\cite{CARA:2022}.

In parallel, \emph{spatial awareness} plays a crucial role in guiding communication decisions, especially in tasks like coverage control or environmental monitoring. Agents that are situated near task-relevant features—such as event hotspots, boundaries of explored regions, or dynamic fields—may be given higher priority for channel access. A representative approach is Where2comm~\cite{Where2comm:2022}, which uses spatial confidence maps to quantify regions of perceptual importance. Agents near uncertain or unobserved zones are prioritized in the communication schedule, ensuring effective collaborative perception while significantly reducing data transmission overhead. This form of prioritization is particularly useful when dealing with redundant observations from densely clustered agents. Moreover, spatially-aware communication can be implemented through proximity-triggered transmission policies, where agents initiate communication only when certain geometric thresholds are exceeded—such as separation from neighbors or movement into new regions. These threshold-based mechanisms reduce unnecessary communication within tightly clustered groups while still allowing for adaptive updates when an agent transitions into a previously uncovered area. Such techniques are widely used in multi-robot exploration and coverage control tasks to balance coordination fidelity and network load.

Together, role-based and spatially-aware communication strategies provide scalable and efficient frameworks for message prioritization in large MASs. By allocating communication resources to agents based on contextual relevance, either through task role or spatial significance, these approaches align naturally with goal-oriented metrics, complementing more formal information-theoretic scheduling strategies like VoI and CoIL.

% ========================================
% Attention-based message prioritization
% ========================================
\subsubsection{Attention-based message prioritization}

Inspired by neural attention mechanisms, \emph{attention-based communication} enables agents to selectively attend to the most informative peers—those whose data is likely to enhance performance or reduce uncertainty. In the context of goal-oriented multi-agent reinforcement learning, attention scores are typically computed from features like task relevance, prediction error, or belief divergence.

For example, in Message-Aware Graph Attention Networks (MAGAT)~\cite{RAE:2021}, agents apply a key–query mechanism to dynamically weight incoming messages based on their relevance to cooperative path planning. Importantly, MAGAT demonstrates significantly improved success rates under communication constraints compared to models with uniform aggregation. Similarly, sparsity-promoting algorithms such as those described in~\cite{Sun:2020IROS} learn an adaptive communication graph via attention, allowing agents to communicate only with the neighbors that matter most and resulting in enhanced scalability and efficiency.

Another foundational study~\cite{jiang:2018attentional}, equips agents with a gating policy that determines on-the-fly whether and with whom to communicate. This gating is based on local observations and predicted impact on the shared task. By combining this with message aggregation, the Attention-based Communication (ATOC) can significantly reduce irrelevant traffic in sizable agent populations.

These methods support end-to-end learning of communication policies in which attention weights are optimized jointly with task execution strategies. The result is sparse, dynamic, and \emph{task-sensitive} message routing, delivering both communication efficiency and high coordination quality.

% ========================================
% Graph-Based approaches
% ========================================
\subsection{Graph-based approaches}\label{subsec:graph}

In MASs, agents often need to coordinate decisions based on partial observations and limited communication bandwidth. To address this challenge, \emph{graph neural networks (GNNs)} and \emph{message passing algorithms} have emerged as foundational tools for learning structured representations and policies over dynamic agent networks. These methods model agents as nodes in a graph and their communication links as edges, allowing information to be exchanged via iterative message aggregation from local neighborhoods. The aggregation functions may be learned end-to-end or defined heuristically, and are often conditioned on edge features such as spatial proximity, sensing overlap, trust metrics, or semantic relevance.

Recent research has begun to integrate \emph{goal-oriented metrics} into graph-based message-passing frameworks. These methods enable agents to dynamically reshape the communication graph to prioritize task‑relevant links and optimize coordination under resource constraints.
For example, Graph Neural Networks (GNNs) have recently gained popularity for learning communication policies in decentralized multi-agent systems. Standard GNN architectures often employ basic message aggregation schemes, which limit the agents' ability to distinguish and prioritize task-relevant information. Such limitations ultimately affect the performance of solutions found and the overall resilience of the MAS. New trends of research focus on communication-aware techniques that enable agents to dynamically adapt the communication topology according to task requirements. 

% ========================================
% Graph-Based Message Passing
% ========================================
\subsubsection{Graph-Based Message Passing}
\label{subsubsec:MAGAT}

Dobbe \emph{et al.}~\cite{NIPS2017_8bb88f80} presents an information-theoretic framework based on rate-distortion theory for fully decentralized policies in multi-agent systems. They treat decentralization as a compression problem and use classical rate-distortion results to analyze performance limits of distributed communication. More specifically, the authors formulate the design of fully decentralized control policies as a distributed rate–distortion problem.  Let $u_i^*$ denote agent~$i$’s optimal action in the centralized solution to the following problem.
\begin{problem}\label{problem:u}
\begin{align*}
  \min_{u} \;& f_o(x,u) \\
  \quad\text{s.t.}\quad & g(x,u)=0, \\
  & h(x,u)\le0,
\end{align*}
where $x=(x_1,\dots,x_N)$ is the global state and only a subset of agents $i\in\mathcal{C}$ control components $u_i$. 
\end{problem}
In the decentralized setting, each agent observes only its local state $x_i$ and implements a learned policy 
\begin{equation}
  \hat u_i \;=\;\hat\pi_i(x_i),
\end{equation}
which approximates $u_i^*$ but incurs distortion 
\begin{align}
d\bigl(\hat u_i,\,u_i^*\bigr)\;=\;\bigl\|\hat u_i - u_i^*\bigr\|^2. 
\end{align}
By interpreting the local policy as a compression of the full state $x$ into the single observation $x_i$, the expected distortion admits a lower bound via rate–distortion theory.  Specifically, for each agent~$i$ one considers the optimization
\begin{align}
D_i^* \;=\;&\min_{p(\hat u_i\,|\,u_i^*)} 
               \;\mathbb{E}\bigl[d(\hat u_i,u_i^*)\bigr]\notag\\
  &\text{s.t.}\quad
     I\bigl(\hat u_i;\,u_i^*\bigr)\;\le\;I\bigl(x_i;\,u_i^*\bigr),
\end{align}
where $I(\cdot;\cdot)$ denotes the mutual information.  The constraint follows from the data processing inequality, since $\hat u_i$ depends on $u_i^*$ only through $x_i$.  Classical rate–distortion theory then implies that the optimal distortion $D_i^*$ decreases monotonically as the mutual information $I(x_i;u_i^*)$ increases, yielding a fundamental performance limit for any regression‐based policy. The same framework naturally extends to limited communication among agents.  If agent~$i$ may also observe up to $k$ additional local states $\{x_j:j\in S\}$, then one maximizes
\begin{equation}
  S_i=\arg\max_{S\subseteq\mathcal{N}\setminus\{i\},\,|S|=k}
    I\bigl(u_i^*;\,x_i,\{x_j:j\in S\}\bigr),
\end{equation}
in order to minimize the corresponding rate–distortion lower bound. In the jointly Gaussian, mean‐squared error case, this selection criterion admits a closed‐form solution based on squared correlations, and the resulting decentralized policies closely approach the centralized optimum.% in both synthetic examples and in an optimal power flow case study.

Another particularly notable example in this domain is Multi-Agent Graph Attention Transformer (MAGAT)~\cite{RAE:2021}, which integrates attention-based GNNs with robotic path planning. In MAGAT, each agent constructs its local observation graph and learns to selectively attend to neighboring agents based on task-relevant features. The model employs a key-query attention mechanism, where the importance of each neighbor is computed through the similarity between a query vector (representing the agent’s internal state or goal) and key vectors (representing neighbors' encoded messages). This attention-weighted message passing scheme allows agents to dynamically focus their communication on the most informative peers, thereby reducing redundancy and enhancing coordination.
MAGAT is trained end-to-end to jointly learn both the communication policy (who to attend to) and the control policy (how to act), optimizing for global task performance in scenarios such as cooperative navigation and obstacle avoidance. One of its key advantages is the ability to adaptively reshape the communication topology in response to the environment and team dynamics, enabling robust scalability in heterogeneous or partially connected systems. Furthermore, the use of attention mechanisms allows for interpretable communication flows, which can reveal the emergent structure of decentralized collaboration in MASs.
The success of MAGAT demonstrates the potential of GNN-based architectures not only to serve as function approximators for distributed control but also to encode flexible, goal-oriented communication strategies that generalize across tasks and team configurations.

% ========================================
% Graph Information Bottleneck
% ========================================
\subsubsection{Graph Information Bottleneck (GIB)}
\label{subsubsec:graph_ib}

In many real-world scenarios, including distributed sensing, swarm robotics, and cooperative control, agents must extract, share, and utilize information that is not only relevant to their own goals but also influenced by their neighbors. Classical IB does not account for such relational dependencies. The \emph{Graph IB} (GIB)~\cite{2020:GIB} extends the classical IB principle to data structured as graphs. While the original IB method focuses on learning compressed representations $T$ of an input $X$ that preserve relevant information about a target variable $Y$, GIB seeks to perform this compression in the context of \emph{nodes} embedded within a \emph{graph topology}. This extension is particularly powerful for learning structured representations in MASs, where each agent corresponds to a node in a communication graph, and interactions are encoded via edges. This is particularly beneficial in scenarios where agents must maintain a shared understanding of the environment (e.g., distributed SLAM, coverage control, or joint planning) without overwhelming the network.

Let $G = (\mathcal{V}, \mathcal{E})$ denote a graph with node set $\mathcal{V}$, edge set $\mathcal{E}$, and node features $\{X_v\}_{v \in \mathcal{V}}$. The objective is to learn node-level representations $\{T_v\}$ such that each $T_v$ retains information relevant to a task variable $Y_v$ (e.g., a class label, policy action, or control input), the representation is compact with respect to the input $X_v$ and its neighborhood $\mathcal{N}(v)$, and the learned representation incorporates and respects the graph structure. This leads to a localized bottleneck objective of the form:
\begin{equation}
    \mathcal{L}_{\text{GIB}} = \sum_{v \in \mathcal{V}} \left[ I(T_v; X_{\mathcal{N}(v)}) - \beta I(T_v; Y_v) \right],
\end{equation}
where $X_{\mathcal{N}(v)}$ denotes the features of node $v \in \mathcal{V}$ and its neighbors, and $\beta$ is a trade-off parameter.

In practice, the GIB framework is implemented using \emph{Graph Neural Networks} (GNNs). The encoder network \( p_\phi(T_v | X_{\mathcal{N}(v)}) \) maps node features to latent embeddings through message passing and neighborhood aggregation:
\begin{equation}
    T_v = \text{GNN}_\phi(X_v, \{X_u : u \in \mathcal{N}(v)\}),
\end{equation}
while the decoder \( q_\psi(Y_v | T_v) \) predicts the downstream variable \( Y_v \). To approximate the mutual information terms, variational upper and lower bounds are employed, typically using techniques from the VIB~\cite{alemi2017deep} and Deep Graph Infomax~\cite{DGI:2019}. GIB thus provides a theoretically grounded and practically scalable framework for task-aware communication and representation learning over structured multi-agent networks.

% ========================================
%
% Learning-Based Approaches to Goal-Oriented Messaging
%
% ========================================
\section{Learning-based approaches to goal-oriented communications for MASs}
\label{sec:learning_based}

In this section, the integration of learning-based methods with goal-oriented semantic communication concepts will be investigated. Already it is shown in some works that the \emph{joint} design (semantic content selection together with intelligent scheduling) can dramatically improve control accuracy. For example, Wu \emph{et al.}~\cite{2024:TWC_GOSC} introduced a goal-oriented semantic communication framework for remote robot control. In their UAV example, the controller is learned by deep reinforcement learning (DRL) to send only commands that directly affect the task outcome. The UAV receiver uses an AoI/VoI policy to queue messages: more urgent semantic packets override stale ones. In simulations, their framework reduced trajectory tracking error by approximately 91.5\% compared to naive periodic control, showing the power of sending only the right information at the right time to achieve goals.

Learning-based communication strategies have emerged as powerful tools for enabling agents in MASs to adaptively determine \emph{what}, \emph{when}, and \emph{with whom} to communicate. Unlike classical methods that rely on predefined communication protocols or centralized control, these approaches empower agents to autonomously discover efficient communication policies that are tightly coupled with task performance. Through experience and task-driven feedback, agents can learn to exchange only those messages that are necessary for achieving collective goals. This section provides an overview of key developments in learning-based goal-oriented communication, with emphasis on MARL, attention-based sparsity mechanisms, and semantic representation learning.

\subsection{Communication learning via MARL}
\label{subsec:comm_learning}

MARL has emerged as a cornerstone framework for developing learning-based communication strategies in MASs~\cite{e24040470,e25020221,e25020299,e27010004}. In MARL, agents are trained to optimize long-term cumulative rewards that depend not only on their own actions, but also on the actions and messages of other agents in the system. This provides a natural mechanism for agents to jointly learn communication and control policies, particularly under decentralized or partially observable settings.

Agents can be endowed with explicit communication channels, enabling them to send discrete or continuous messages to other agents. These messages can be jointly optimized along with control actions using gradient-based methods, often through centralized training with decentralized execution (CTDE)~\cite{amato2024CIDE}. Architectures such as CommNet and DIAL\footnote{CommNet and DIAL are architectures for multi-agent communication and learning, particularly in the context of deep reinforcement learning. CommNet, or Communication Network, allows agents to communicate with each other by sharing a single, learned communication vector. DIAL, or DeepInfluence Architecture, is a more complex architecture where agents can influence each other's policies through a dedicated communication channel.} have demonstrated that differentiable communication modules can facilitate the emergence of efficient, interpretable communication patterns. More recent work explores message gating, dropout, or hard attention to further constrain and refine message exchange during policy learning. These techniques help to reduce communication overhead by ensuring that only messages that improve team-level performance are transmitted.

Moreover, the integration of the IB principle into MARL has led to further innovations in compact and purposeful communication~\cite{alemi2017deep}. By regularizing the mutual information between the agent’s observation and its outgoing message, agents are encouraged to compress their messages to include only task-relevant content. This not only improves generalization but also aligns naturally with semantic encoder-decoder architectures commonly used in cooperative MASs. Wang \emph{et al.}~\cite{pmlr-v119-wang20i} developed Informative Multi-Agent Communication (IMAC) using information bottleneck principles to learn efficient communication protocols under bandwidth constraints. They prove that limited bandwidth requires low-entropy messages and use information theory to optimize message scheduling.

{He~\emph{et al.}~\cite{He:2022} has extended this paradigm by jointly optimizing communication policies and underlying wireless resource allocation. In particular, task-oriented channel allocation has been investigated within MARL frameworks, where agents not only learn what information to communicate, but also how to allocate limited communication resources based on the task relevance of the transmitted messages. For instance, the authors propose a task-oriented communication framework in which agents extract task-relevant features using a VIB~\cite{alemi2017deep} and a centralized entity allocates communication channels by maximizing the expected contribution of each message to the overall task performance. This approach highlights the importance of coupling communication learning with resource allocation, showing that prioritizing goal-oriented information can significantly improve coordination efficiency compared to rate-maximization or random allocation strategies. These results illustrate a broader trend toward integrating communication learning with system-level constraints, where bandwidth, interference, and channel variability are explicitly accounted for within the learning process. From a goal-oriented perspective, such approaches bring MARL-based communication closer to practical deployment in wireless multi-agent systems. At the same time, it is important to note that in most MARL-based approaches, including the one above, the notion of task relevance is learned implicitly through reward signals, rather than being explicitly defined through information-theoretic or semantic metrics. This can limit interpretability and generalization, and highlights the need for frameworks that combine learning-based communication with explicit measures of information value.}

Real-world applications, such as wireless multi-UAV networks, illustrate the utility of these approaches under communication-constrained and dynamic settings~\cite{Riku2023}. Agents in such domains learn to weigh communication costs against potential improvements in navigation, mapping, or coverage, achieving robust coordination despite limited and noisy channels. These demonstrations highlight the potential of MARL to enable adaptive, self-organizing communication behavior in large-scale and realistic MAS deployments.

\subsection{Sparsity and Attention Mechanisms}
\label{subsec:sparsity_attention}

In MASs, where agents operate in bandwidth-limited or latency-sensitive environments, communication efficiency becomes essential. Learning-based approaches have increasingly focused on strategies that induce sparsity and selectivity in inter-agent communication, often guided by learned attention mechanisms.

Attention mechanisms, originally popularized in natural language processing, have proven to be powerful tools for prioritizing information flow among agents~\cite{jiang:2018attentional,HU:2024-128015,2025:Younas}. In the context of MASs, attention allows each agent to dynamically evaluate the relevance of messages from its neighbors. Agents learn to assign attention weights based on spatial proximity, uncertainty, prediction divergence, or shared task context. As a result, attention mechanisms enable selective aggregation of messages, ensuring that only the most informative inputs are integrated into decision-making processes.

To promote sparsity~\cite{2011:SparseMARL}, various regularization techniques have been introduced~\cite{LI:2023-120517}. These include $L_1$ penalties on communication magnitudes, entropy-based constraints that limit the spread of attention weights~\cite{2025:TAC_Entropybased}, and explicit communication budgets that cap the number of messages an agent can send or receive. By treating communication as a costly action, these mechanisms incentivize agents to develop parsimonious messaging strategies, transmitting only when the expected task benefit outweighs the cost.

Overall, the combination of sparsity and attention mechanisms enables scalable communication in MASs by filtering unnecessary exchanges and focusing inter-agent dialogue on task-critical content. This results in systems that are not only more efficient but also more robust to communication failure, interference, and delays.

\subsection{Semantic Representation Learning for Message Generation}
\label{subsec:semantic_rep}

Learning what to communicate is as important as learning when and to whom. Semantic representation learning focuses on developing communication strategies where agents generate messages that are not raw data, but compressed representations optimized for downstream utility. These representations encode latent variables, such as beliefs, predictions, or summaries of local observations, that are directly relevant to the shared task.

End-to-end architectures, often inspired by encoder-decoder models, are commonly used for semantic communication learning. Here, the encoder maps the agent's observation into a compact latent message, while the decoder uses this message to guide decision-making. The parameters of both encoder and decoder are trained jointly, using gradients from a shared objective, such as classification accuracy, control performance, or accumulated task reward.

One major advantage of semantic communication is that it enables abstraction. By discarding task-irrelevant details, agents can generalize their communication behavior across different scenarios or environments. For instance, instead of transmitting full state vectors, agents may learn to share predictions about other agents' intent, likelihood of collisions, or the presence of strategic opportunities, information that is more useful for coordination than raw sensor data.

Recent works further refine this paradigm by integrating semantic loss functions into the learning process. These include task-specific metrics such as classification margin, trajectory divergence, or goal achievement rates. By explicitly linking message quality to task outcome, agents learn to encode the most relevant aspects of their observations in a compact and efficient manner.

\subsection{Outlook of learning-based goal-oriented communication for MASs}

Learning-based goal-oriented communication is poised to transform how MASs coordinate and reason in complex, distributed settings. These methods offer scalable, decentralized, and adaptive mechanisms for enabling communication protocols that are aligned with both environmental dynamics and task objectives. The ability of agents to learn when, what, and how to communicate through interaction and feedback allows MASs to move beyond rigid, handcrafted protocols toward flexible and robust cooperation.

Despite significant progress, several challenges remain. Interpreting the emergent communication policies is often difficult, making it hard to verify safety or correctness. Moreover, the learned protocols may be brittle under non-stationary conditions, such as changing team composition or adversarial interference. Finally, while current methods are largely empirical, stronger theoretical foundations are needed to ensure that the developed approaches will work across a wide range of tasks and network topologies.

% ========================================
%
% Application of Goal-Oriented Communication in Real-World Systems
%
% ========================================
\section{Application of Goal-Oriented Communication in Real-World Systems}
\label{sec:applications}

% ========================================
% Semantic Networks for CAVs.
% ========================================
\subsection{Semantic networks for cooperative autonomous vehicles.}
\label{subsec:CAVs}

The Internet of Vehicles (IoV) connects cars, infrastructure, and users to support intelligent transportation. However, modern IoV applications (e.g. autonomous driving, traffic management, vehicle surveillance) generate massive amounts of data that must be sent to servers under strict latency constraints~\cite{2020:ProceedingsIEEE_IoV}. Traditional radio networks are already pushed to their Shannon limits, so the IoV’s spectrum resources are quickly becoming inadequate. In addition, nearby vehicles often sense correlated or overlapping information (e.g. multiple cameras see the same scene), so naively sending all raw data wastes bandwidth. The redundancy among the transmitted data dramatically deteriorates the spectrum efficiency and irrelevant content can cause severe network congestion. As a consequence, IoV faces an exploding data load and spectrum scarcity that conventional (symbol-level) communications cannot efficiently handle~\cite{2023:gowc_agv_petar}.

To address these challenges, Xu \emph{et al.}~\cite{2023:VTM_IoV} propose shifting focus from bits to task-relevant semantics, i.e., instead of insisting on delivering every source symbol perfectly, the goal is to extract and transmit only the task-relevant semantics of the data. For example, if multiple vehicles capture images of the same car, they need only send semantic features (like an ID or key descriptors) rather than the full raw images. This semantic-aware approach promises to drastically cut down the transmitted data, alleviating IoV spectrum bottlenecks and congestion. To implement this vision, the paper proposes a Cooperative Semantic Communication (Co-SC) framework for multi-vehicle IoV. The core idea is that multiple nearby vehicles jointly encode their sensed data into shared semantic information. The architecture involves the following key components:
\begin{list4}
    \item Semantic encoders (at vehicles): Each vehicle (user) processes its raw data through a semantic encoder, which extracts a high-dimensional semantic feature that captures the task-relevance of the data (e.g. object identity, scene description).
    \item Joint Source-Channel (JSC) encoders (at vehicles): The semantic features are then passed through a deep JSC encoder that maps them into channel symbols. Unlike conventional channel coding, this JSC encoding is semantically aware. Additionally, it learns to protect more important features by allocating more symbols, and it is trained jointly with the semantic encoder.
    \item Shared channel transmission: Vehicles share a common wireless channel (e.g., uplink to a roadside server) to send their semantic symbols. Since the semantic encoder has removed overlap, the data from different cars is more compact.
    \item Cooperative JSC decoder (at server): At the server or edge host, a cooperative JSC decoder jointly processes the received symbols from all vehicles. Because the vehicles’ data is correlated, this decoder can use one user’s symbols to help decode another’s, narrowing the uncertainty space and correcting errors.
    \item Semantic decoder and task modules (at server): The recovered semantic features are then fed into a semantic decoder or directly into task-specific modules. In the paper’s example (an image-based ID retrieval task), the server uses the combined semantic features to identify vehicles in a database. The framework can also simply feed semantic features to deep learning models for tasks like detection or localization.
\end{list4}
The proposed framework, Co-SC, is a novel multiuser cooperative architecture. It integrates semantic encoders/decoders and joint source-channel coding for each vehicle, plus a shared knowledge base, so that multiple vehicles’ data is compressed and decoded jointly. By jointly designing the semantic encoders and decoders across users, the system learns shared representations: shared (common) semantics and distinct (unique) semantics are separated. The cooperative decoders exploit these correlations to compress redundancies and improve decoding. 

Eldeeb \emph{et al.}~\cite{2024:CAVs_WCL} developed a multi-task semantic communication framework among connected autonomous vehicles (CAVs). They focus on sharing traffic sign information: one vehicle’s camera captures a sign, a convolutional autoencoder extracts its semantic features, and the compact encoding is sent to other CAVs over a network. At the receiver, one decoder reconstructs the image (for visualization) and another classifies the sign. In simulations (e.g., in foggy conditions), this task-oriented scheme outperforms conventional approaches in terms of classification accuracy and image similarity while using fewer bits. In other words, CAVs share only goal-relevant semantic data (the label of the sign), enhancing bandwidth efficiency without sacrificing safety-critical understanding.

The multiuser cooperative designs~\cite{2023:VTM_IoV,2024:CAVs_WCL} highlight how cooperation in MASs is key. Enabling cooperative, semantic-rich transmissions could transform IoV from classic bit-pipes into smart, goal-oriented networks for next-generation transportation.

% ========================================
% Goal-Oriented Communication in Multi-UAV Networks
% ========================================
\subsection{{Goal-Oriented Communication in Multi-UAV Wireless Networks}}
\label{subsec:uav_go_comm}

{Multi-UAV wireless networks represent a natural and demanding application domain for goal-oriented communication, as they tightly couple sensing, communication, and control under stringent resource and reliability constraints. In such systems, UAVs must coordinate to accomplish tasks such as area coverage, target tracking, search and rescue, and environmental monitoring, while operating over time-varying and bandwidth-limited wireless links. These characteristics make it impractical to transmit all sensed data, thereby necessitating communication strategies that prioritize task-relevant information~\cite{2026:OJ-ComSoc}.}

{From a goal-oriented perspective, multi-UAV applications highlight the importance of explicitly quantifying the utility of information~\cite{2026:WCL_UAV}. In target tracking and surveillance tasks, communication should be driven by the reduction of uncertainty about critical states, while in coverage and exploration tasks, transmissions should occur only when local observations significantly improve global situational awareness. These requirements naturally align with metrics such as the VoI and the CoIL, which provide measures of how communication decisions affect task performance. Recent studies have begun to incorporate such metrics into UAV coordination, enabling more efficient use of limited communication resources in dynamic environments; see, for example, \cite{2025:ToC_UAV} and references therein.}

{Recent advances in learning-based coordination have enabled significant progress in this direction. In particular, multi-agent reinforcement learning (MARL) has been widely adopted to jointly learn control and communication policies in decentralized UAV networks~\cite{Riku2023, e25020221}. Through mechanisms such as message compression, attention, and gating, agents learn to exchange only information that improves collective performance. More recent works further integrate communication with wireless resource management, where channel allocation, transmission scheduling, and power control are adapted based on the estimated contribution of messages to the mission objective. For example, task-oriented channel allocation frameworks leverage VIB principles and reinforcement learning to prioritize transmissions that are most beneficial for coordination, rather than those that simply maximize throughput. Despite these advances, most existing approaches rely on implicit notions of task relevance learned through reward signals, rather than explicit semantic or information-theoretic formulations. This can limit robustness, and generalization, particularly in safety-critical or rapidly changing environments. As a result, there is growing interest in integrating learning-based coordination with semantic and goal-oriented communication frameworks, where the relevance and utility of information are explicitly modeled. Such integration has the potential to improve both communication efficiency and decision quality in large-scale UAV systems.}

{Overall, multi-UAV wireless networks serve as a key testbed for bridging theoretical advances in semantic and goal-oriented communication with practical challenges in distributed, resource-constrained systems, and will play a central role in the development of next-generation intelligent communication architectures for 6G and beyond.}

% ========================================
% Distributed SLAM with goal‑oriented communication
% ========================================
\subsection{Distributed SLAM with goal‑oriented communication}
\label{subsec:slam}

Distributed Simultaneous Localization and Mapping (SLAM) is fundamental to the deployment of multi-robot systems in real-world environments. However, traditional multi-robot SLAM architectures face significant challenges related to communication bandwidth and computation costs. In many systems, robots either exchange large, detailed sensor data, like full 3D point clouds or complete maps, at high frequencies, which is costly and redundant, or avoid sharing altogether, leading to degraded localization and inconsistent maps~\cite{ZOU:2019}. 

One promising approach in distributed SLAM is to reduce the size and frequency of the data that robots share, by transmitting only the most useful and compact information. That is, robots can share compressed and relevant summaries of what they observe. These summaries are often called \emph{lightweight feature descriptors}, and they capture the most relevant geometric or semantic features of the environment, such as corners, edges, or recognizable objects. For example, the DCL-SLAM framework~\cite{2024:DCL-SLAM} uses specialized descriptors built from LiDAR data that allow each robot to detect whether it has visited a place before—or whether another robot has—without needing to share the full scan. This saves significant bandwidth while still supporting accurate localization and map merging. Another efficient method involves dividing a robot's map into smaller, more manageable pieces called 2.5D submaps. These represent the local environment using height information, like a terrain map. Robots can align and compare these submaps with those from their teammates to determine how their positions relate. When a match is found (called a loop closure), the robots can update their maps accordingly. These updates are often performed using a sliding window approach, meaning the robot focuses on recent or relevant data only, which reduces computational cost while keeping the map consistent~\cite{2024:mdpi-slam}. Studies have shown that this selective communication strategy can reduce data exchange volume by approximately 30\% while still maintaining global map consistency and accurate localization~\cite{2024:mdpi-slam}.

Another line of work focuses on robust loop closure detection and prioritization. For example, DOOR‑SLAM~\cite{DOOR-SLAM:2020} introduces distributed pairwise consistency maximization to allow less conservative place‑recognition thresholds, yielding more valid inter‑robot matches without exchanging raw data. Using less stringent front-end thresholds while filtering outliers downstream improves both map accuracy and loop closure yield under constrained bandwidths. In~\cite{2022:distributedSLAM}, key matches are selected based on pose graph structure, proximity, and point-cloud features for loop closure prioritization for large-scale multi-robot exploration, reducing computational load by roughly 50–75\% while improving mapping accuracy.

One prominent system demonstrating this approach is \emph{Kimera-Multi}~\cite{2022:Kimera-Multi}, a fully distributed SLAM framework that allows multiple robots to collaboratively build both geometric and semantic maps. Kimera-Multi operates through peer-to-peer communication, meaning robots connect directly without relying on a central server. It integrates robust \emph{place recognition} and \emph{pose-graph optimization}, which aligns the relative positions of robots and observed landmarks. These capabilities enable the system to build globally consistent, annotated 3D mesh maps in real-time while keeping communication minimal. The result is an efficient and scalable system that supports real-world deployment scenarios, such as exploration, search and rescue, and infrastructure inspection, even under tight bandwidth constraints.

Collaborative SLAM has been effectively adapted for use in teams of UAVs, allowing them to map and navigate large environments together~\cite{IROS:2021ethz,ICRA:2023ethz,2023:fastmulti-UAV}. One practical approach involves equipping each UAV with a monocular camera and short-range wireless ranging devices. When two UAVs come close to one another, they can combine their views to form a temporary stereo vision setup. This setup allows the team to better estimate distances and improve the accuracy of the map, especially in challenging environments where scale is hard to determine from a single camera alone. Importantly, this collaboration is done in a distributed way, in which each UAV processes its own data locally and only shares essential information with its neighboring UAVs. This keeps communication fast and efficient, with minimal delay, making the system well-suited for real-time tasks like exploration, inspection, or search and rescue. By working together in this manner, UAVs can build more accurate and consistent maps while avoiding the need to send large amounts of raw sensor data over the network.

%{}

Overall,  these innovations represent a shift toward more \emph{goal-oriented communication} in multi-robot mapping.  These strategies help robots collaborate in real-time while improve both efficiency and robustness in dynamic, bandwidth-limited environments. By sharing only what’s needed, when it’s needed, they embody the principles of goal-oriented communication in SLAM. Experimental benchmarks across diverse environments (e.g. subterranean, outdoor exploration, GPS-denied settings) consistently show that such task-driven schemes outperform dense or static communication approaches in both map accuracy and bandwidth efficiency.

Looking forward, embedding explicit goal-awareness into SLAM communication policies will be critical, including learning to predict when a loop closure will significantly reduce global uncertainty, compressing semantics (e.g., object landmarks, semantic segmentation maps) rather than raw geometry, and designing adaptive coordination strategies that balance exploration versus sharing, especially in heterogeneous agent teams or dynamic environments. Such directions align with the broader vision of goal-driven communication: agents communicate not just to share data, but to support shared objectives with minimal cost.

\begin{figure*}
  \includegraphics[width=\textwidth]{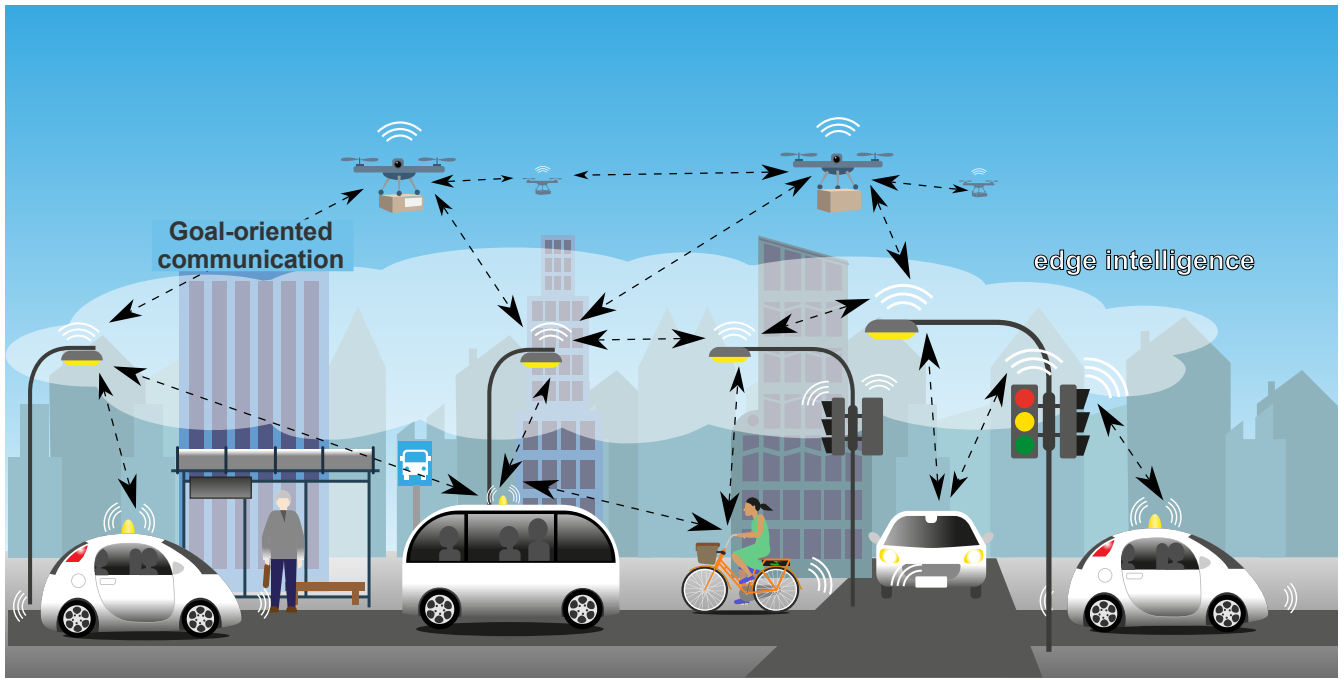}
  \caption{A unified scenario illustrating goal-oriented communication in MASs: autonomous ground vehicles, cyclists, and aerial drones collaboratively operate in a smart city environment. Agents share semantically relevant, goal-oriented information for coordinated navigation, surveillance, and human interaction, demonstrating key application domains such as distributed perception, edge intelligence, and safety-critical coordination under bandwidth constraints.}
\end{figure*}

% ========================================
% Benchmarking and Standardized Evaluation
% ========================================
\subsection{Federated Learning and Edge Intelligence}\label{subsec:FLandEI}

Another important domain where goal-oriented communication can be a key enabler are AI/ML services, relying on federated learning (FL) and edge intelligence. Here, resource and communication efficiency are pivotal and unfortunately, traditional paradigms usually rely on message exchanges that do not consider if the transmitted data effectively contributes to the global learning goal. As a remedy, goal-oriented communication can address redundant transmissions by activating devices only when their information advances the global learning objective. Considering that edge environments are characterized by heterogeneous resource availability and dynamic network conditions, the benefits of goal-oriented communication have been presented in various recent works.

{First, the role of goal-oriented communications in improving the performance of FL has been demonstrated in several studies.} The paper by Pandey \textit{et al.} \cite{10293926} introduces a risk-aware participation strategy for FL, treating device selection as a distributed goal-oriented communication problem. In greater detail, devices independently evaluate whether participating in a particular training round will benefit the global model while considering their local costs and the value of their contributions. To support this process, a parameter server is used, providing semantic feedback to each device, facilitating them to calculate the utility of its update and in case the marginal benefit is insufficient it will remain silent. This process results in significant reduction of unnecessary transmissions, while avoiding the propagation of low-quality updates. Performance evaluation demonstrates that integrating risk-awareness and semantic feedback into the participation decisions, a $1.4\times$ improvement in communication efficiency over baseline FL alternatives is achieved, without degrading model accuracy. A relevant use case for this strategy can be smart healthcare systems, where mobile and wearable devices must optimize power usage, data privacy, and relevance to the global model before contributing to collaborative diagnostics.

Then, Costafreda and Elayoubi~\cite{10922012} proposes a hybrid optimization approach, adopting goal-oriented communication in the scheduling and resource allocation of FL in mobile networks. More specifically, the authors formulate a policy selection problem to optimize the device sampling probabilities to balance resource usage and data utility, in terms of the processing capacity of each device and the quality of its data. In this setting, a hybrid genetic algorithm is presented, dynamically adapting the transmission strategies to prioritize clients whose updates maximize model accuracy under radio resource constraints. Contrary to traditional FL, where all available clients might contribute equally, this hybrid approach ensures that only clients with high data utility and low transmission cost will participate. This strategy leads to a communication-efficient FL process, achieving higher performance with a reduced number of updates. A relevant area where this method can be employed involves vehicular networks, where edge devices in cars must appropriately schedule their participation in federated perception models, adhering to strict latency and bandwidth constraints.

Furthermore, Wei \textit{et al.} in \cite{10622021} develop a novel federated semantic learning (FedSem) architecture, considering the IB principle into a multi-user semantic communication setting for FL. Their system focuses on transmitting compressed but semantically rich representations to construct semantic knowledge graphs from raw images, allowing for efficient and task-specific learning. Here, instead of treating all semantic features equally, the IB principle filters out non-essential content, only keeping task-relevant semantics for the inference objective. The proposed federated training process jointly updates device-side semantic encoders and a base station-side decoder. Results indicate important gains in rate-distortion efficiency and faster convergence under dynamic channel conditions. Architectures, such as FedSem can be well-suited to applications related to collaborative aerial surveillance with drones, where only mission-critical semantic features, such as detected threats or anomalies must be communicated to improve resource efficiency of the learning task.

Then, Wei \textit{et al.}~\cite{10225550} focuses on combining goal-oriented communication with FL, developing a resource allocation framework tailored to CPS. Their solutions departs from conventional throughput-maximization, maximizing instead the information utility gain, i.e. the marginal improvement in CPS task objectives stemming from data transmission. Thus, they model FL as a CPS task, where devices are selectively provided communication resources, based on how their data contributes to system-level goals. The resource allocation problem is formulated as a knapsack problem and solved via a divide-and-conquer greedy algorithm. Simulations highlight that optimizing for task-related information over data throughput offers high efficiency gains in FL performance. This approach can be applied to IIoT deployments, such as predictive maintenance in manufacturing plants, where sensors must prioritize energy-efficient transmissions that improve the accuracy fault prediction models.

Below we discuss two works that consider semantics-aware metrics in FL. More specifically, the work in \cite{CSCN23} proposes a scheduling policy that combines Age of Update (AoU) concepts and the data Shapley metric. This policy considers the freshness and value of received parameter updates from individual data sources and real-time channel conditions to enhance the operational efficiency of FL. Furthermore, the work in \cite{ICASSPW24} introduces the concept of Version Age of Information (VAoI) to FL. Unlike traditional AoI metrics, VAoI considers both timeliness and content staleness. Each client’s version age is updated discretely, indicating the freshness of information. VAoI is incorporated into the client scheduling policy to minimize the average VAoI, mitigating the impact of outdated local updates and enhancing the stability of FL systems.

{Regarding the use of goal-oriented communications in the field of edge intelligence there have been various works presenting significant performance enhancements. The paper by Yang \textit{et al.} \cite{9979702} provides a semantic communication-based framework for edge intelligence, enabling efficient, goal-oriented information exchange in 6G networks. More specifically, the authors highlight the limitations of Shannon-based systems and propose an edge-enabled semantic communication-based architecture where semantic extraction and background knowledge sharing take place at the network edge, reducing latency and improving privacy. At the same time, FL is integrated where edge servers collaboratively update semantic extraction models without raw data exchange, while an edge-sharing knowledge graph enhances semantic management and computational efficiency. In addition, the paper investigates semantic-aware edge intelligence, where collaborative DRL agents exploit semantic relatedness, selecting cooperative partners and applying semantic model compression for efficient distributed learning. Experiments include a case study focusing on semantic-aware resource allocation in wireless-powered IoT, demonstrating high communication savings while maintaining semantic fidelity.}

{The paper by Utkovski \textit{et al.} \cite{10364645} revisits the theoretical foundations of semantic communication, studying its impact when designing edge intelligence systems. Here, the authors introduce two complementary perspectives, i.e., goal-oriented communication and semantic interoperability to unify the data acquisition, communication, and control processes in distributed edge networks. The proposed framework leverages information-theoretic tools, including IB and VoI to quantify data relevance and perform adaptive communication between edge devices and servers. Then, a semantic-aware edge device co-inference architecture is given where on-device neural networks extract task-relevant features, while variational IB optimization minimizes communication overheads without degrading inference accuracy. Experimental validation on ther Neuromorphic-MNIST dataset shows that neuromorphic semantic encoding exhibits improved accuracy and energy efficiency over joint source-channel coding scheme, and conventional separate source-channel coding schemes.  Finally, the paper discusses cross-layer semantic networking mechanisms, allowing context-driven resource allocation and protocol adaptation in collaborative robotics, health monitoring, and autonomous control systems.}

{The paper by Diao \textit{et al.} \cite{10960403} introduces a unified communication framework, aligning reconstruction-oriented and task-oriented paradigms for edge intelligence. The main motivation for this study is to address issues of traditional systems aiming to reconstruct raw data at the receiver without considering task-specific requirements of AI-driven applications. So, the authors extend the IB theory to jointly minimize task-relevant loss while preserving the original data structure through an information reshaper module. In this manner, semantic (goal-oriented) and conventional communication are integrated, balancing data fidelity and task utility. Moreover, a variational optimization approach is adopted, handling mutual information in high-dimensional neural representations, and a compatible joint source–channel coding modulation scheme is provided, guaranteeing interoperability with existing digital infrastructures. Experimental validation for edge-based autonomous driving scenarios, using the CARLA simulator shows a 99.19\% reduction in bits per service, over classical codecs, thus demonstrating the framework’s potential for highly efficient and goal-driven edge inference.}

{In the paper by Binucci \textit{et al.} \cite{Binucci2022} presents a dynamic system and algorithmic framework enabling goal-oriented communications for edge inference. Focusing on image classification, the authors exploit a bank of convolutional encoders at the user devices to extract and transmit only the features relevant to the inference task, while adapting feature size to channel and computational conditions. Leveraging IB, the proposed solution jointly optimizes communication and computation using Lyapunov stochastic optimization, balancing energy, latency, and accuracy performance. In this respect, resource management strategies are developed, adjusting transmission rates, compression factors, and CPU cycles at the user devices and edge server. Simulation results highlight that adaptive goal-oriented feature extraction can achieve high classification accuracy even under extreme compression levels, significantly enhancing resource-efficient edge intelligence in real-time visual analytics and mission-critical autonomous systems.}

{Next, the work by Merluzzi \textit{et al.} \cite{9815737} investigates goal-oriented communication for energy-efficient edge inference in Multi-access Edge Computing (MEC)-assisted beyond-5G networks. Towards that end, an edge intelligence mechanism is given, performing system performance assessment not by classical reliability metrics, such as bit error rate but through goal effectiveness, corresponding to the inference task accuracy under predefined energy constraints. In this setting, multiple multi-access edge hosts (MEHs) cooperate and conduct either standalone or ensemble inference, optimizing communication reliability and computational load. Simulation results show that relaxing communication reliability requirements by one decimal digit can offer goal effectiveness above 84\%, while reducing device radio energy consumption by around 23\%. Furthermore, it is highlighted that ensemble inference across MEHs can improve both energy efficiency and inference accuracy when favorable backhaul conditions exist in the network.}

{The study by Binucci \textit{et al.} \cite{10639178} develops goal-oriented deep neural network (DNN) splitting towards achieving energy-efficient and low-latency cooperative edge inference. Here, a dynamic optimization strategy is introduced, adjusting the splitting point between user devices and edge cloud servers, while adapting communication and computation resources to satisfy specific latency and inference accuracy requirements. Moreover, new trade-offs, considering the accuracy degradation as a function of the splitting point selection and wireless channel conditions are investigated. Simulation results indicate that the proposed method substantially reduces energy consumption without undermining edge accuracy, even under noisy channel conditions.}

{The paper by Bijanrostami \textit{et al.} \cite{10615907} presents a goal-oriented semantic communication holistic model to improve task inference and resource utilization in IoT deployments. In this field, a three-layer architecture is presented where users exploit computing resources at access point nodes for semantic communication purposes prior to offloading their data to higher levels. Then, a DNN-based resource allocation problem is formulated, maximizing the average inference accuracy under quality-of-service (QoS) and resource constraints. The resulting mixed-integer nonlinear programming (MINLP) problem is solved via a heuristic and linear programming algorithm. Performance evaluation demonstrates that the incorporation of semantic communication improves the network admission rate by approximately 15\%, compared to conventional IoT deployments.}

{Finally, the paper by Shao \textit{et al.} \cite{10258036} proposes a task-oriented communication framework with a temporal entropy model (TOCOM-TEM) for edge video analytics, relying on camera-equipped devices, transmitting only compact and task-relevant features to an edge server instead of full video streams. Based on the deterministic IB principle, the device-side encoder removes spatial and temporal redundancy to reduce bandwidth consumption, maintaining discriminative information for downstream analytics tasks. As a way to enhance rate efficiency, TEM exploits feature correlations across consecutive frames, reducing the bitrate, and a spatial–temporal fusion module at the server integrates present and past features, increasing inference accuracy. Experiments highlight that TOCOM-TEM results in improved rate–performance trade-off, when compared to reconstruction-oriented and data-centric alternatives.}

% ========================================
%
% Open Challenges and Research Directions
%
% ========================================
\section{Open Challenges and Future Directions} \label{sec:challenges}

% ========================================
% Unifying Information-Theoretic and Learning-Based Approaches
% ========================================
\subsection{Unifying Information-Theoretic and Learning-Based Approaches} \label{subsec:unification}

A key open challenge in goal-oriented communication for MASs lies in bridging the gap between \emph{information-theoretic formulations} and \emph{learning-based communication strategies}. Classical information theory provides a rigorous framework for quantifying data relevance and efficiency using tools such as mutual information, rate-distortion theory, and channel capacity~\cite{shannon1948mathematical, cover2006elements}. More recent extensions, including goal-oriented rate-distortion~\cite{ISIT:2021RDF,TCOM:2022RD,isit:2022Photis} and semantic communication models~\cite{JSAC:2023Gunduz} aim to describe information sources that have semantic aspects and proposed corresponding rate distortion problem formulations for characterizing the amount of information content of such semantic sources. However, these methods often assume known models and distributions, which are rarely available in large-scale, real-world MASs. In contrast, learning-based approaches, especially those rooted in DRL and MARL, offer flexible mechanisms for discovering communication policies through learning. Agents can learn to determine what, when, and with whom to communicate by optimizing task-level objectives directly, often using attention mechanisms or graph-based message passing~(see relevant discussion is Section~\ref{sec:scheduling_routing}).

Despite notable advancements, a gap remains between information-theoretic and learning-based approaches to communication. Learning methods provide flexible, data-driven mechanisms for discovering effective communication strategies, yet they often offer no formal guarantees of optimality. In contrast, information-theoretic models yield frameworks for quantifying relevance and efficiency, but they typically rely on idealized assumptions and do not readily adapt to complex task-driven settings. Bridging this gap calls for the development of hybrid approaches that embed theoretical structure into learning-based models. For example, one strategy is to incorporate information relevance (formulated through rate-distortion or utility-based criteria) into the learning objectives of communication systems. By treating communication as a constrained optimization problem over both message content and utility, it becomes possible to design learning systems that are not only effective but also grounded in formal principles of informational value. Ultimately, the overarching objective is to design learning-based communication systems that are not only empirically effective but also theoretically grounded. These systems should be capable of learning what to communicate, when, and to whom according to a goal (individual or joint). Achieving this vision necessitates a deeper integration of information significance metrics with modern machine learning paradigms. 

Additionally, learning-to-communicate frameworks will need to be extended to support compositional reasoning, semantic abstraction, and causal inference to ensure generalization and interpretability (see discussion in Section~\ref{subsec:safety_reliability}). As multi-agent deployments become increasingly prevalent in real-world domains, ranging from autonomous vehicles to environmental monitoring and disaster response, learning-based goal-oriented communication will be a key enabler of efficient, scalable, and intelligent collective behavior.

%=====================================
% Scalability in Large Multi-Agent Systems
%=====================================
\subsection{Scalability in Large Multi-Agent Systems} \label{subsec:scalability}

As the number of agents in a MAS increases, communication, coordination, and learning become significantly more complex. Scalability is a critical challenge for goal-oriented communication, particularly in systems involving dense, heterogeneous, or spatially distributed agents. In such settings, naive communication strategies, such as all-to-all broadcasting or uniform message passing, become impractical due to excessive bandwidth usage, latency, and processing overhead.

To enable efficient coordination at scale, communication must become increasingly selective, structured, and decentralized. One promising direction involves exploiting \emph{topological sparsity} through graph-based message passing frameworks, where agents only communicate with a subset of their neighbors based on learned attention weights, proximity, or role-based criteria. These approaches allow for adaptive communication graphs that evolve in response to task demands and environmental dynamics, significantly reducing communication overhead without sacrificing performance. Another key aspect of scalability is the emergence of \emph{hierarchical architectures}, where agents are organized into clusters or roles (e.g., leaders, followers, relays), and intra-cluster and inter-cluster communication are treated differently. This enables localized decision-making while maintaining global coordination, and aligns naturally with goal-oriented communication models. In addition, \emph{multi-resolution} or \emph{compressed representation learning} approaches offer further scalability benefits by enabling agents to exchange task-specific features rather than raw state observations. Such techniques are particularly useful when combined with importance-aware scheduling, where agents selectively transmit only high-value information according to metrics such as CoIL, VoI or AoI (see discussion in Section~\ref{sec:WNCSs}).

{From a computational perspective, scalability is also fundamentally constrained by the exponential growth of joint state, action, and communication spaces with the number of agents, which affects both learning and communication design. \cite{TOCD:2025} highlights that this challenge can be mitigated by explicitly incorporating task relevance into the communication design and by decomposing the overall problem into tractable sub-components. In particular, the authors show that learning value-based representations of observations and designing communication policies based on such representations can significantly reduce complexity, while maintaining task performance. Moreover, they demonstrate that communication strategies learned in small-scale systems can be transferred to larger networks, effectively decoupling the scalability of the learning phase from the number of agents.}

Achieving scalable goal-oriented communication requires the joint design of communication protocols, learning algorithms, and network architectures. This includes developing sparsity-aware learning objectives, integrating hierarchical coordination schemes, and leveraging the structure of real-world environments to reduce the effective communication complexity. As the size and diversity of MASs continue to grow, the ability to scale communication in a goal-relevant manner will be essential for enabling efficient, robust, and interpretable collective intelligence.
% -----

Achieving scalable goal-oriented communication therefore requires the joint design of communication protocols, learning algorithms, and network architectures. This includes developing sparsity-aware learning objectives, integrating hierarchical coordination schemes, and leveraging the structure of real-world environments to reduce the effective communication complexity. As the size and diversity of MASs continue to grow, the ability to scale communication in a goal-relevant manner will be essential for enabling efficient, robust, and interpretable collective intelligence.

%=====================================
% Safety, Reliability, and Interpretability of Emergent Protocols
%=====================================
\subsection{Safety, Reliability, and Interpretability of Emergent Protocols}
\label{subsec:safety_reliability}

As communication protocols in multi-agent systems become increasingly learned and emergent, questions of safety, reliability, and interpretability take on critical importance. While learning-based communication policies enable adaptivity and task-awareness, they also introduce unpredictability and potential fragility, especially in safety-critical applications such as autonomous driving, industrial automation, or collaborative robotics. In such contexts, agents must ensure that their messaging strategies do not inadvertently degrade system safety or violate task constraints. For example, decisions about whether or not to transmit information can directly impact state estimation, control performance, and ultimately, system stability. This calls for communication protocols that are not only goal-oriented but also robust to uncertainties, resilient to adversarial perturbations, and transparent in their behavior.

A particularly important concern in safety-critical control is the guarantee of \emph{set invariance}, i.e., ensuring that the system's state remains within a predefined safe set over time. Various \emph{safety filters}, such as those based on control barrier functions or predictive filtering, have been developed to detect and correct unsafe control inputs in a minimally invasive manner. However, these typically require safety constraints to be explicitly defined in the state space, which limits their applicability to geometrically grounded conditions like collision avoidance. For agents operating in complex or human-centric environments, safety often extends beyond geometric constraints to encompass semantic, context-dependent rules, referred to as \emph{semantic safety constraints}~\cite{AngelaTUM:2025RAL}. Designing communication and control protocols that respect such implicit constraints remains an open challenge. It requires integrating high-level knowledge, reasoning, and shared intent into the communication loop.

In parallel, the interpretability of emergent protocols is essential for system verification, human–robot collaboration, and transferability across tasks and domains. Understanding what information is communicated, why it is shared, and how it affects collective decision-making is crucial for establishing trust in autonomous multi-agent systems. 

Ultimately, ensuring that learned communication policies remain safe, reliable, and interpretable requires combining design with adaptive learning. It calls for the development of formal tools to analyze and constrain emergent behaviors, along with empirical validation in realistic, high-stakes scenarios. As goal-oriented communication continues to evolve, these properties will be key enablers of real-world deployment.

%=====================================
% Safety, Reliability, and Interpretability of Emergent Protocols
%=====================================
\subsection{Extensions of GIB}
\label{subsec:GIBextensions}

GIB offers a foundation for designing communication protocols in MASs. 
%However, realizing its full potential in real-world applications requires further extension along multiple directions. 
We discuss some promising possible research directions that can significantly expand the applicability and impact of GIB-based methods.

First, extending GIB to \emph{temporal domains} is a crucial step toward supporting agents that operate in dynamic, sequential environments. Many MAS tasks, such as cooperative exploration, real-time traffic coordination, or adaptive monitoring, demand communication protocols that take into account the history of observations and interactions over time. A temporal extension of GIB would allow agents to encode and transmit compressed yet informative summaries of their trajectories, decisions, and evolving internal states. These models could be implemented via recurrent GNNs, graph-attention mechanisms, or temporal message-passing frameworks, capturing dependencies not only across space (agents) but also across time. Furthermore, temporal GIB could help mitigate memory bottlenecks and redundancy by filtering out outdated or irrelevant historical data, focusing instead on predictive and decision-critical features.

Second, the development of \emph{hierarchical GIB architectures} holds promise for improving scalability and modularity in large-scale MASs. In many practical settings, agents are naturally organized into subgroups, either by spatial proximity, function, or communication constraints. Hierarchical GIB models can support multi-level abstraction, where bottlenecks are enforced not only at the individual agent level but also across clusters, regions, or roles. For example, intra-group communication could involve fine-grained, high-frequency updates, while inter-group communication could be based on coarse, abstract representations. This separation of concerns allows the system to scale gracefully and remain robust under variable communication budgets or partial agent failure. Additionally, hierarchical models align well with concepts in hierarchical reinforcement learning and decentralized planning, enabling structured communication that mirrors task hierarchies.

Third, integrating GIB with tools from \emph{causal inference} can improve the interpretability, generalization, and reliability of learned communication policies. Traditional information bottleneck methods rely on statistical dependencies, which may not always correspond to causal influences on task outcomes. By incorporating causal reasoning into the bottleneck objective, GIB can help agents to generalize across different environments and task configurations, improve robustness to distribution shifts, and facilitate debugging and verification of communication strategies. 

Together, these extensions point toward a unified vision of GIB as a flexible and theoretically grounded framework for \emph{goal-oriented, scalable, and interpretable} communication in complex multi-agent environments. By integrating temporal modeling, hierarchical structure, and causal reasoning, future GIB models can better accommodate the rich demands of real-world MASs applications.

% ========================================
% Cross-Domain Synergies and Open Interfaces
% ========================================
\subsection{Cross-Domain Synergies}

While this overview has concentrated on goal-oriented communication within MASs, the concepts discussed here resonate strongly with several adjacent research areas. These domains not only present new application frontiers but also introduce complementary tools and challenges that can refine and extend goal-oriented communication paradigms. In particular, three cross-domain interfaces stand out as both promising and underexplored.

First, \emph{multi-modal perception}~\cite{Wang_2024_CVPR} integrates heterogeneous sensory streams (namely, vision, language, audio, tactile feedback) into unified representations that capture richer context than any single modality alone. Embedding goal-oriented communication within multi-modal pipelines enables agents to transmit distilled, semantically aligned summaries rather than raw sensor data, thereby reducing bandwidth and processing requirements without sacrificing task-critical content. For example, instead of streaming entire video frames, an autonomous vehicle could share compact symbolic descriptors of relevant objects and their predicted trajectories. The central challenge lies in developing multi-modal semantic encoders and decoders whose distortion measures are explicitly tied to downstream task success, rather than modality-specific reconstruction fidelity.

Second, \emph{large-scale distributed AI}~\cite{SNT2023:distributedAI}, including federated learning, distributed optimization, and cooperative inference, pushes communication constraints to the forefront. In such systems, agents often operate under stringent bandwidth, latency, and privacy requirements while jointly optimizing a shared model or decision policy. Integrating goal-oriented principles into federated decision-making could allow agents to exchange only those gradients, feature embeddings, or variables that provide maximal utility for global performance. This opens new research challenges: quantifying the value of such updates via information-theoretic or causal measures, designing adaptive communication topologies that dynamically select optimal peer interactions, and ensuring robustness under adversarial or non-stationary conditions.

Third, \emph{causal inference}~\cite{2023:JSAIT_Kurisummoottil} offers a foundation for interpretability and robustness in goal-oriented messaging. By augmenting transmitted content with causal structure (e.g., explanations of why a message is relevant or how an observation affects task outcomes) agents can improve coordination, detect misleading updates, and anticipate the impact of actions in safety-critical settings. This is especially valuable for human–machine teaming, where trust hinges on transparency and accountability. Open research directions include formalizing causal semantics within rate–distortion and information bottleneck frameworks, incorporating them into real-time communication protocols, and exploring how causal annotations can enhance resilience to distribution shifts or deceptive behavior.

Bridging goal-oriented communication with these domains will require new theoretical models, joint learning–communication optimization strategies, and cross-domain interfaces. Achieving this integration promises a new class of cyber–physical and AI systems capable of exchanging \emph{only} the most task-relevant, interpretable, and semantically rich information across sensing, computation, and control layers, thus minimizing overhead while maximizing effectiveness, transparency, and safety.

\subsection{{Integrating Goal-Oriented Communication to 6G}}

{The classical Open Systems Interconnection model enforces a strict separation among network layers, enabling abstraction, interoperability, and modular system evolution. However, such a layered architecture becomes increasingly restrictive in the context of goal-oriented communications, where transmission strategies must adapt to application-level objectives rather than solely to bit-level fidelity. In emerging 6G applications such as autonomous driving, industrial automation, and extended reality, the usefulness of transmitted data depends directly on how it contributes to a specific task. As a result, communication, computation, and control are tightly coupled, and the assumptions underlying strict layer separation are no longer fully valid.}

{This motivates a rethinking of network architecture. In \cite{strinati24goals}, it is proposed to extend the conventional design by introducing an additional functional plane that operates alongside the user plane and control plane. The purpose of this new plane is to represent the goals of the application and to guide how information is processed and transmitted across the network. Instead of treating all data equally, the network can use this information to prioritize data that is more relevant for achieving the desired task. In this way, communication decisions are influenced by the intended outcome, rather than only by channel conditions or throughput considerations. This idea is closely related to semantic communications \cite{qin2025deep, 10038657, JSAC:2023Gunduz}, but it goes one step further by explicitly linking communication to task performance and decision-making~\cite{11393542}.}

{At the same time, the benefits of such integration must be carefully balanced against the well-established advantages of modular system design. Layered architectures allow independent development, testing, and upgrading of network components, thereby reducing complexity and improving scalability. In contrast, tightly coupled designs may introduce unintended interactions between layers, making the system more difficult to analyze, maintain, and standardize. As discussed in \cite{kawadia05cautionary}, cross-layer approaches can even lead to performance degradation when dependencies are not properly managed.}

{For this reason, integrating goal-oriented communication into 6G will require careful consideration to determine whether the projected performance improvements justify the redesign efforts and architectural changes. One possible approach is to maintain the layered structure while allowing high-level goals to influence lower-layer operations through well-defined interfaces. In this context, the additional functional plane can serve as a coordinating entity that translates application requirements into communication and resource allocation strategies without fully coupling all layers. Technologies such as software-defined networking and network function virtualization provide suitable tools to support this type of flexible and adaptive control.}

% ========================================
%
% Conclusion
%
% ========================================
\section{Concluding remarks}\label{sec:conclusions}

Goal-oriented communication is emerging as a foundational paradigm in the design of intelligent and efficient multi-agent systems. By shifting the focus from transmitting data for the sake of fidelity to communicating information that is \emph{relevant for achieving shared objectives}, this paradigm provides a way to reduce bandwidth, improve coordination, and enhance task performance in resource-constrained environments. This overview has outlined the theoretical frameworks, mentioning some real-world applications of goal-oriented communications, offering a unified lens that bridges information theory and learning-based strategies in the context of MASs. We have highlighted how classical models such as rate-distortion theory and mutual information serve as building blocks for more advanced constructs like goal-oriented and {SRD} frameworks. These approaches provide formal metrics to quantify the utility of information with respect to tasks, offering critical insights for communication scheduling and prioritization in multi-agent settings. At the same time, learning-based techniques have demonstrated the potential to discover efficient, decentralized communication policies directly from data. As multi-agent systems grow in scale, complexity, and heterogeneity, the importance of designing communication protocols that are \emph{task-aware, adaptive, and scalable} will continue to grow. We have discussed the key strategies to meet these needs and emphasized the critical importance of safety, reliability, and interpretability in learned communication protocols, especially for real-world deployment in high-stakes domains such as autonomous driving, robotics, and industrial automation.

Despite significant progress, numerous open challenges remain. The integration of information-theoretic structure into neural architectures, the design of benchmarks for evaluating goal-oriented communication systems, and the development of robust methods for reasoning under semantic and temporal uncertainty are all fertile grounds for future research. There is also a pressing need for tools that can formally analyze and constrain the behavior of emergent communication protocols to ensure safe and predictable operation.

The vision outlined in this paper is one where communication in MASs becomes a purposeful, context-aware act, driven not by the quantity of data, but by its value to a collective goal. This overview can serve as a roadmap for researchers and practitioners seeking to build the next generation of communication-aware multi-agent systems that are not only efficient, but also intelligent, robust, and aligned with their operational objectives.

% ========================================
%
% References
%
% ========================================
%\section*{References}
%\def\refname{\vadjust{\vspace*{-1em}}} %Please don't do this in a real paper.
\bibliographystyle{IEEEtran}
\bibliography{bibliographia}

%\clearpage
% ========================================
%
% IEEEbiography
%
% ========================================
\begin{IEEEbiography}[{\includegraphics[width=1in,height=1.25in,clip,keepaspectratio]{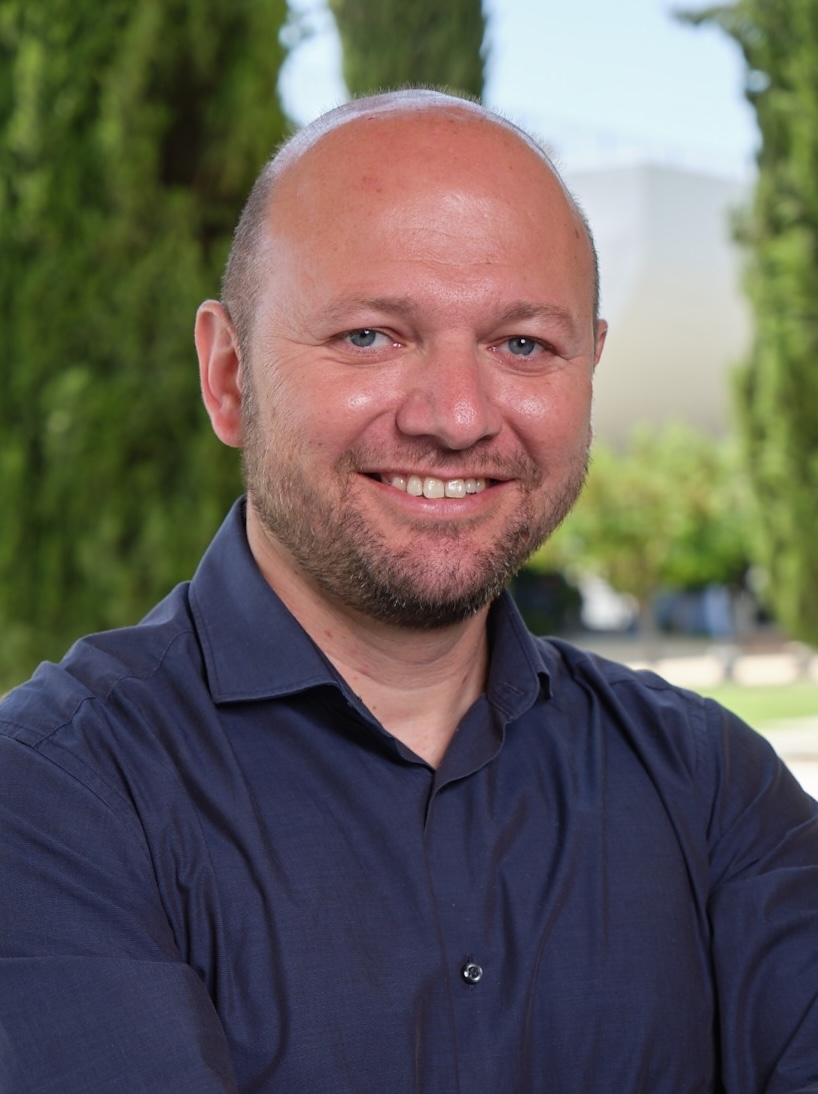}}]{Themistoklis Charalambous}~(Senior Member, IEEE) eceived his BA (First Class Honours) and M.Eng (Distinction) in Electrical and Information Sciences from Trinity College, University of Cambridge. He completed his PhD at the Control Laboratory of the Engineering Department, University of Cambridge, in 2009.
Following his PhD, he joined the Human Robotics Group at Imperial College London as a Research Associate (2009–2010). He then served as a Visiting Lecturer at the Department of Electrical and Computer Engineering, University of Cyprus (2010–2011). From 2012 to 2015, he was a Postdoctoral Researcher at the Division of Decision and Control Systems, KTH Royal Institute of Technology. He subsequently held a postdoctoral position at Chalmers University of Technology in the Communication Systems group (2015–2016). 
In 2017, he joined the Department of Electrical Engineering and Automation at the School of Electrical Engineering of Aalto University as an Assistant Professor. In September 2018, he was awarded the title of Academy Research Fellow by the Academy of Finland, and in July 2020, he was promoted to Associate Professor.
In September 2021, he continued his academic career at the University of Cyprus, where he joined the Department of Electrical and Computer Engineering as a tenure-track Assistant Professor. In May 2022, he was awarded an European Research Council (ERC) Consolidator Grant for the research project “MINERVA: Emerging Cooperative Autonomous Systems: Information for Control and Estimation.” From May 2026, he was appointed as a tenured Associate Professor at the University of Cyprus. He also remains affiliated with Aalto University as a Visiting Professor and, since April 2023, serves as a Visiting Professor at the FinEst Centre for Smart Cities.
Prof. Charalambous is an Associate Editor of \textsc{IEEE Transactions on Automatic Control}
\end{IEEEbiography}

\begin{IEEEbiography}[{\includegraphics[width=1in,height=1in,clip,keepaspectratio]{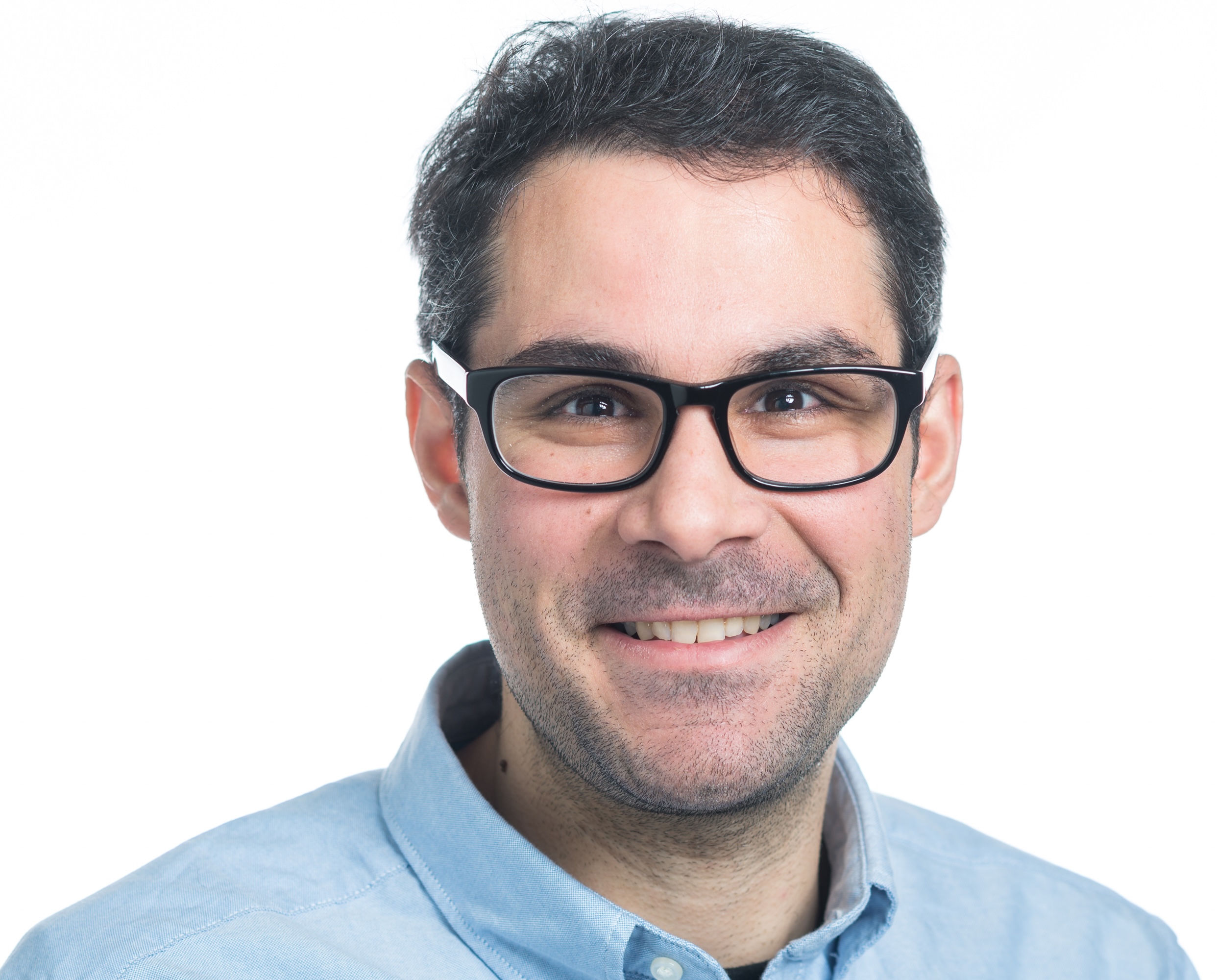}}]{Nikolaos Pappas}~(Senior Member, IEEE) received the first B.Sc. degree in computer science, the second B.Sc. degree in mathematics, the M.Sc. degree in computer science, and the Ph.D. degree in computer science from the University of Crete, Greece, in 2005, 2012, 2007, and 2012, respectively. From 2005 to 2012, he was a Graduate Research Assistant with the Telecommunications and Networks Laboratory, Institute of Computer Science, Foundation for Research and Technology-Hellas, Heraklion, Greece; and a Visiting Scholar with the Institute of Systems Research, University of Maryland at College Park, College Park, MD, USA. From 2012 to 2014, he was a Post-Doctoral Researcher with the Department of Telecommunications, CentraleSupec, France. He is currently an Associate Professor with the Department of Computer and Information Science, Link\"oping University, Link\"oping, Sweden. His main research interests include the field of wireless communication networks, with an emphasis on semantics-aware communications, energy harvesting networks, network-level cooperation, age of information, and stochastic geometry. He has served as the Symposium Co-Chair for the IEEE International Conference on Communications in 2022. He was the general chair for the 23rd International Symposium on Modeling and Optimization in Mobile, Ad hoc, and Wireless Networks (WiOpt 2025). He is an Area Editor of the \textsc{IEEE Open Journal of the Communications Society} and an Expert Editor of invited papers of the \textsc{IEEE Communications Letters}. He is Associate Editor for four IEEE Transactions journals.
\end{IEEEbiography}

\begin{IEEEbiography}[{\includegraphics[width=1in,height=1.25in,clip,keepaspectratio]{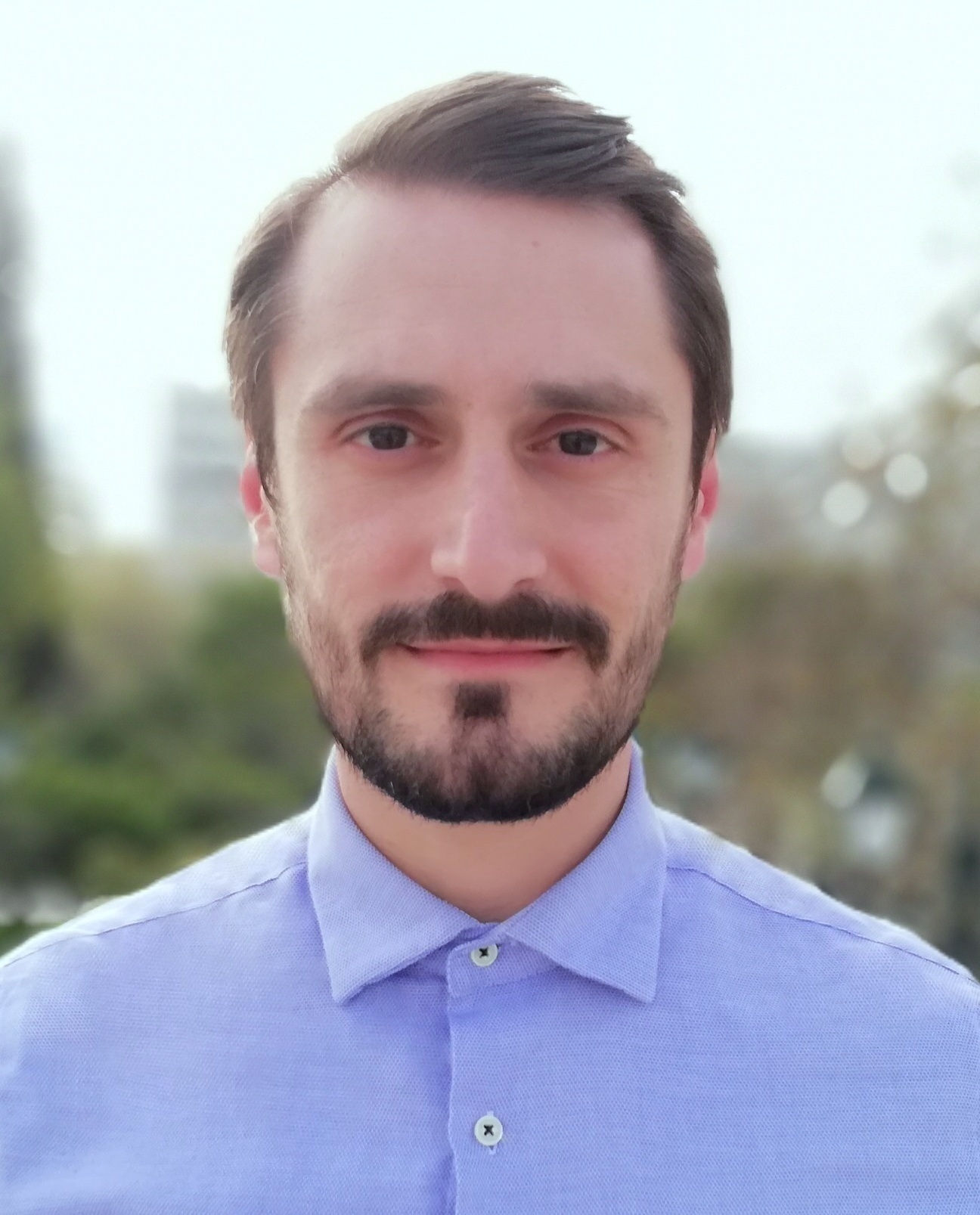}}]{Nikolaos Nomikos}~(Senior Member, IEEE) received the Diploma in electrical engineering and computer technology from the University of Patras, Patras, Greece, in 2009, and the M.Sc. and Ph.D. degrees from the Information and Communication Systems Engineering Department, University of
the Aegean, Samos, Greece, in 2011 and 2014, respectively. Since 2025, he has been an Assistant Professor of Mobile and Satellite Communications Systems, Department of Information and Communication Systems Engineering, University of the Aegean, Samos, Greece. Moreover, he is a Project Manager with Four Dot Infinity P.C. His research interests include cooperative communications, non-orthogonal multiple access, non-terrestrial networks, and machine learning for wireless networks optimization. Prof. Nomikos is an Editor of IEEE TRANSACTIONS ON COMMUNICATIONS and Associate Editor for Frontiers in Communications and Networks. He is a Member of the IEEE Communications Society and the Technical Chamber of Greece.
\end{IEEEbiography}

\begin{IEEEbiography}
    [{\includegraphics[width=1in,height=1.25in,clip,keepaspectratio]{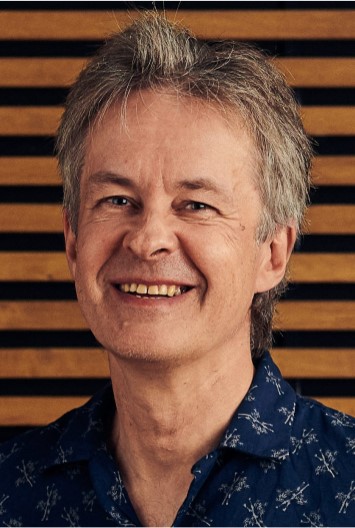}}]
    {Risto Wichman}~(Senior Member, IEEE)
 received his M.Sc.  and D.Sc. (Tech) degrees in digital signal processing from Tampere University of Technology, Finland, in 1990 and 1995, respectively.  From 1995 to 2001, he worked at Nokia Research Center as a senior research engineer.  In 2002, he joined the Department of Information and Communications Engineering, Aalto University School of Electrical Engineering, Finland, where he has been a full professor since 2008.  His research interests include signal processing techniques for wireless communication systems.
\end{IEEEbiography}

% ========================================
%
% That's all OLD folks! ;-)
%
% ========================================
\end{document}